\documentclass[twocolumn,tighten,twocolappendix]{aastex631}

\usepackage{comment}

\newcommand{\SExtractor} {\textsc{SExtractor}}

\newcommand{\code}[1]	{\textsf{\small #1}}

\defcitealias{Sheen_2012}{S12}
\defcitealias{oh18}{O18}

\received{2023 December 8}
\revised{2024 February 25}
\accepted{2024 March 10}
\published{2024 April 30}
\submitjournal{ApJ}

\shorttitle{Distribution of merging and post-merging galaxies in nearby galaxy clusters}
\shortauthors{Kim et~al\@.}
\graphicspath{{./}{figures/}}

\begin{document}

\title{Distribution of merging and post-merging galaxies in nearby galaxy clusters}

\correspondingauthor{Duho Kim}
\email{duhokim81@gmail.com}

\author[0000-0001-5120-0158]{Duho Kim}
\affiliation{Department of Astronomy and Space Science, Chungnam National University, Daehak-ro 99, Yuseong-gu Daejeon 34134, Republic of Korea}
\affiliation{Korea Astronomy and Space Science Institute, Daedeokdae-ro 776, Yuseoung-gu Daejeon 34055, Republic of Korea}

\author[0000-0002-3211-9431]{Yun-Kyeong Sheen}
\affiliation{Korea Astronomy and Space Science Institute, Daedeokdae-ro 776, Yuseoung-gu Daejeon 34055, Republic of Korea}

\author[0000-0003-2150-1130]{Yara L. Jaff\'e}
\affiliation{Departamento de F\'isica, Universidad T\'ecnica Federico Santa Mar\'ia, Avenida Espa\~na 1680, Valpara\'iso, Chile}

\author[0000-0002-8130-3593]{Kshitija Kelkar}
\affiliation{Departamento de F\'isica, Universidad T\'ecnica Federico Santa Mar\'ia, Avenida Espa\~na 1680, Valpara\'iso, Chile}
\affiliation{Instituto de F\'isica y Astronom\'ia, Universidad de Valpara\'iso, Avda. Gran Breta\~na 1111 Valpara\'iso, Chile}

\author[0000-0001-9882-1576]{Adarsh Ranjan}
\affiliation{Space Telescope Science Institute, 3700 San Martin Drive, Baltimore, MD 21218, USA}

\author[0009-0008-0197-3337]{Franco Piraino-Cerda}
\affiliation{Departamento de F\'isica, Universidad T\'ecnica Federico Santa Mar\'ia, Avenida Espa\~na 1680, Valpara\'iso, Chile}
\affiliation{Instituto de F\'isica y Astronom\'ia, Universidad de Valpara\'iso, Avda. Gran Breta\~na 1111 Valpara\'iso, Chile}

\author[0000-0002-9810-1664]{Jacob P. Crossett}
\affiliation{Departamento de F\'isica, Universidad T\'ecnica Federico Santa Mar\'ia, Avenida Espa\~na 1680, Valpara\'iso, Chile}
\affiliation{Instituto de F\'isica y Astronom\'ia, Universidad de Valpara\'iso, Avda. Gran Breta\~na 1111 Valpara\'iso, Chile}

\author[0000-0002-4393-7798]{Ana Carolina Costa Louren\c co}
\affiliation{Instituto de F\'isica y Astronom\'ia, Universidad de Valpara\'iso, Avda. Gran Breta\~na 1111 Valpara\'iso, Chile}
\affiliation{European Southern Observatory (ESO), Alonso de Cordova 3107, Santiago, Chile}

\author[0000-0003-2939-8668]{Garreth Martin}
\affiliation{Korea Astronomy and Space Science Institute, Daedeokdae-ro 776, Yuseoung-gu Daejeon 34055, Republic of Korea}
\affiliation{Steward Observatory, University of Arizona, 933 N. Cherry Ave, Tucson, AZ, USA}

\author[0000-0002-7356-0629]{Julie B. Nantais}
\affiliation{Facultad de Ciencias Exactas, Departamento de Ciencias F\'isicas, Instituto de Astronom\'ia,
Fern\'andez Concha 700, Edificio C-1, Piso 3, Las Condes, Santiago, Chile}

\author[0000-0003-3921-2177]{Ricardo Demarco}
\affiliation{Institute of Astrophysics, Facultad de Ciencias Exactas, Universidad Andr\'es Bello, Sede Concepci\'on, Talcahuano, Chile}

\author[0000-0001-7568-6412]{Ezequiel Treister}
\affiliation{Instituto de Astrof{\'{\i}}sica, Facultad de F{\'{i}}sica, Pontificia Universidad Cat{\'{o}}lica de Chile, Campus San Joaquín, Av. Vicu{\~{n}}a Mackenna 4860, Macul Santiago, Chile, 7820436}

\author[0000-0002-4556-2619]{Sukyoung K. Yi}
\affiliation{Astronomy Department and Yonsei University Observatory, Yonsei University, 50 Yonsei-ro, Seodaemun-gu,  Seoul 03722, Republic of Korea}



\begin{abstract}

We study the incidence and spatial distribution of galaxies that are currently undergoing gravitational merging (M) or that have signs of a post merger (PM) in six galaxy clusters (A754, A2399, A2670, A3558, A3562, and A3716) within the redshift range, 0.05$\lesssim$\,$z$\,$\lesssim$0.08. To this aim, we obtained Dark Energy Camera (DECam) mosaics in $u^{\prime}$, $g^{\prime}$, and $r^{\prime}$-bands covering up to $3\times R_{200}$ of the clusters, reaching 28\,mag/arcsec$^2$ surface brightness limits. We visually inspect $u^{\prime}$$g^{\prime}$$r^{\prime}$ color-composite images of volume-limited ($M_r < -20$) cluster-member galaxies to identify whether galaxies are of M or PM types. We find 4\% M-type and 7\% PM-type galaxies in the galaxy clusters studied. By adding spectroscopic data and studying the projected phase space diagram (PPSD) of the projected clustocentric radius and the line-of-sight velocity, we find that PM-type galaxies are more virialized than M-type galaxies, having 1--5\% point higher fraction within the escape-velocity region, while the fraction of M-type was $\sim$10\% point higher than PM-type in the intermediate environment. Similarly, on a substructure analysis, M types were found in the outskirt groups, while PM types populated groups in ubiquitous regions of the PPSD. Adopting literature-derived dynamical state indicator values, we observed a higher abundance of M types in dynamically relaxed clusters. This finding suggests that galaxies displaying post-merging features within clusters likely merged in low-velocity environments, including cluster outskirts and dynamically relaxed clusters.

\end{abstract}



\setcounter{page}{2}

\section{Introduction} \label{sec:intro}

In the prevailing $\Lambda$CDM paradigm, galaxies are thought to assemble much of their mass through hierarchical bottom-up mergers \citep{white&rees78,Fall1980,Bundy2009,Kaviraj2015}. These bottom-up mergers are understood to play a key role in driving changes in the global galaxy properties, such as galaxy structure \citep{Toomre1977,Dekel2009,Conselice2009,Naab2014,Fiacconi2015,Graham2015,Gomez2017,Martin2018,Jackson2022} or the quenching and enhancement of star formation rates \citep{Schweizer1982,Mihos1996,Schawinski2014,Pontzen2017,Martin2017,Martin2021,tanaka2023}. Past mergers also leave behind a morphological record in the form of various \textit{post-merging} features, which can be the result of major mergers -- typically the case for plumes \citep[e.g.][]{Lauer1988} and tidal tails \citep[e.g.][]{Pfleiderer1963,Toomre1972} -- or minor mergers, which usually produce streams \citep[e.g.][]{Johnston1999} or shells \citep[e.g.][]{Malin1983,Quinn1984}. By examining these signatures, it is possible to infer a galaxy's recent interaction history and thereby disentangle the role of various merging processes governing galaxy properties \citep[e.g.][]{Johnston2008,Martinez-Delgado2009,Spavone2020,VeraCasanova2021}.

The importance of mergers in driving the evolution of galaxies within clusters, where effects such as ram-pressure stripping \citep{Gunn1972} and tidal interactions \citep{Jones2000,Gnedin2003} begin to play an increasingly important role, is a matter of discussion. Although clusters represent particularly dense environments, the incidence of mergers among their members is not expected to be high due to high relative velocities. Analytically, the energy input and dynamical friction of a collision scale with the inverse square of the relative velocity of the interacting galaxies \citep{binney&tremaine87}, such that fast encounters, which become preponderant in denser cluster environments, are less likely to result in a merger than slow encounters, which are more common in the field \citep{makino97,mihos03}. Studies such as \citet{Sheen_2012} (\citetalias{Sheen_2012} hereafter) and \citet{oh18} (\citetalias{oh18} hereafter) report, indeed, that only around four percent of the early-type galaxies (ETGs) found in cluster environments appear to be actively undergoing major mergers. \citet{oh18} report only a marginal change in the fraction of ongoing mergers, which increases from 4\% to 7.7\% when going from $R/R_{200}$\,$<$2 to $R/R_{200}$\,$>$2 in clustocentric distance, concluding that merger processes are unlikely to significantly impact galaxy evolution within the virialized region of a cluster. \citet{McIntosh2008} found also negative correlation of the major merging frequency of massive satellite galaxies with the host halo mass. Their sample of SDSS satellite galaxies with a stellar mass $M_{star}$\,$\geq$\,$5\times10^{10}$M$_{\odot}$ dropped their major merging frequency from 3\% in halos with their mass around $M_{halo} \simeq 10^{13.5} M_{\odot}$ to almost 0\% in cluster-scale halos with $M_{halo} \simeq 10^{15} M_{\odot}$.

Given the apparently negligible merger rates in cluster environments, the high proportion of cluster galaxies that exhibit \emph{post-merging} features is therefore unexpected. Deep optical surveys of galaxies in dense cluster environments show a high fraction of \emph{post-merging} features ($\sim$25\% of galaxies brighter than $-$20 in the $r$ band in the red sequences in \citetalias{Sheen_2012}; $\sim$20\% of galaxies with similar brightnesses in \citetalias{oh18}), with abundances comparable to that of the field \citep[see e.g.][]{vandokkum05,Kado-Fong2018,Bilek2020,Sola2022,Trujillo2022,Martin2022,Valenzuela2022}. \citetalias{Sheen_2012} attribute this result to migration of galaxy populations that have undergone mergers prior to entering the cluster \citep[pre-processing;][]{Fujita2004}. The work of \citet{yi13} and \citet{ji14} lend support to this argument, showing that, merger features continue to be detectable for about 4\,Gyr at a limiting surface brightness of 28\,mag\,arcsec$^{-2}$ in their hydrodynamic simulations of major mergers between Sa- and Sb-type galaxies. Evidence of \emph{pre-processing} is also seen at higher redshifts. For instance, \citet{vandokkum99} find that merging galaxies are found preferentially in the outskirts of galaxy clusters at $z\sim0.8$, and \citet{Olave-Rojas2018} find high quenching efficiencies in galaxies at large distances from the cluster centers at $z\sim0.4$. 

Besides, clusters are consistently accreting groups. \citet{das23} found a higher level of star formation (SF) and bluer colors in low-mass SDSS galaxy pairs in filaments and sheets around galaxy clusters. \citet{Kleiner2014} found more asymmetric galaxies in the inner region of A1664 where a merging group most likely passed the pericenter, and they argued that a merger between halos of a group and a cluster enhanced galaxy-galaxy interactions \citep[see also][]{Vijayaraghavan2013}. \citet{mihos03} also stressed that the infalling galaxy groups through the ``cosmic web'' would allow galaxies to interact with each other strongly by slow encounters. \citet{Lokas2023} also found galaxy-galaxy interactions in a galaxy group prior to cluster infall in the IllustridTNG100 simulation. Analyses of substructures in/outside of clusters seem to be necessary to study galaxy-galaxy interactions in merging clusters \citep[i.e.,][]{Olave-Rojas2023}.

In this paper, we study the abundances of gravitationally interacting or post-merger galaxies using deep ($\gtrsim$28 mag arcsec$^{-2}$)\footref{sb} and wide ($\gtrsim$3\,deg$^2$; $R/R_{200}$\,$\gtrsim$3) photometric data sets (in $u^{\prime}$, $g^{\prime}$, and $r^{\prime}$) of six Abell clusters: A754, A2399, A2670, A3558, A3562, and A3716 in a redshift range, 0.05$\lesssim$\,$z$\,$\lesssim$0.08 collected using the Dark Energy Camera \citep[DECam;][]{DePoy2008}.  In combination with public spectroscopic data, we study the prevalence of these features in substructures and in regions of the projected phase-space diagram. This is done to understand the evolution of galaxies over the course of their accretion onto the clusters. 

The rest of the article is divided into seven separate sections as follows: In Section~\ref{sec:data}, we present an overview of the data we use in this study. In Section~\ref{sec:sam}, we outline how we select our sample of galaxies in each cluster. In Section~\ref{sec:vis}, we explain how we classify our sample as merging or post-merging galaxies. In Section~\ref{sec:dyn}, we explain how we adopt and use the dynamical state indicator values from the literature for our sample galaxy clusters. In Section~\ref{sec:sub}, we describe how we identify substructures in each galaxy cluster. In Section~\ref{sec:res}, we show our results on the relations between merging and post-merging rates and the dynamical states of clusters, the substructures, and the projected phase-space diagram. In Section~\ref{sec:dis}, we present our discussion and conclusions.

In this article, we use AB magnitudes \citep{Oke74,Oke83}, and a Standard $\Lambda$CDM cosmology for the paper with a Hubble constant $H_0$\,$=$\,67.4\,km\,s$^{-1}$\,Mpc$^{-1}$, present matter density $\Omega_m$\,$=$\,0.315, and dark energy density $\Omega_{\Lambda}$\,$=$\,0.684 from \citet{planck18} throughout. 

\section{Data} \label{sec:data}

\subsection{Photometric measurements} \label{sec:obs}

\begin{figure}
 \gridline{ 
            \fig{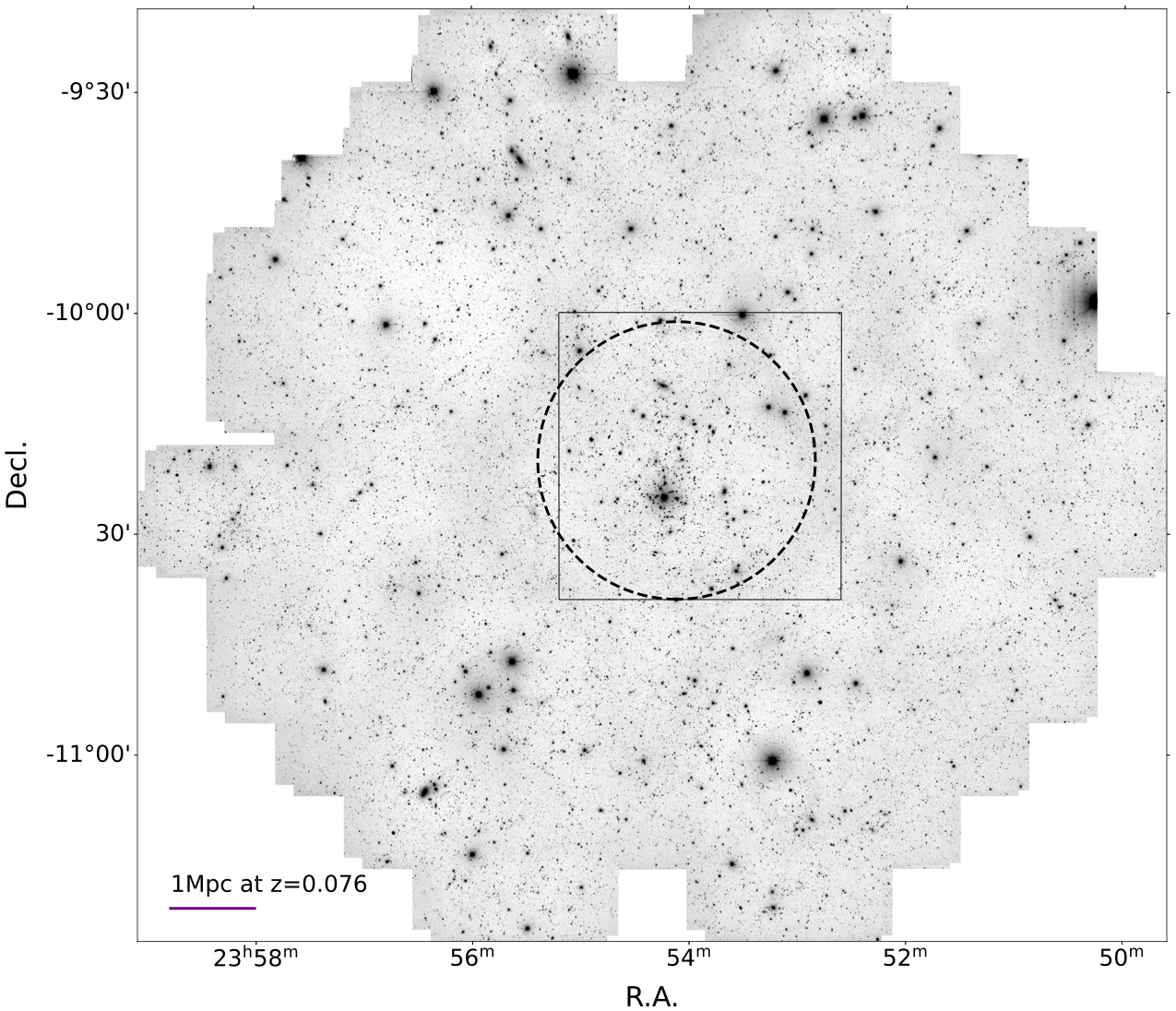}{0.5\textwidth}{}
          }
\caption{\emph{Combined} DECam mosaic images of A2670 in the $r^{\prime}$ band. The dashed circle corresponds to R$_{200}$ derived from the SDSS spectroscopic data, which is 1.69\,Mpc from the cluster center (exact celestial coordinates given in Table~\ref{tab:clusters}). The FoV of MOSAIC {\footnotesize II} used by \citetalias{Sheen_2012} is shown as a solid square, for comparison.
\label{fig:mosaics1}}
\end{figure}

Optical mosaics in $u^{\prime}$, $g^{\prime}$, and $r^{\prime}$ bands of the field of views (FoVs) around seven Abell galaxy clusters were taken using the Dark Energy Camera \citep[DECam; e.g.,][]{decam} installed on the Blanco 4m Telescope at CTIO from April 10th to 11th in 2013 (Proposal ID: 2013A-0612, PI: Y. Sheen) and August 19th to 22nd in 2014 (Proposal ID: 2014B-0608, PI: Y. Jaff\'e), as presented in Table~\ref{tab:clusters} (for A3574 and A3659, visit the website\footnote{\url{https://data.kasi.re.kr/vo/DECam_catalogs/}\label{website}}). DECam features a $\sim$3\,deg$^2$ FoV consisting of sixty-two 2k $\times$ 4k pixel$^2$ CCDs, with a total of 520 million pixels for each FITS\footnote{Flexible Image Transport System \citep{FITS1,FITS2}} image. Figure~\ref{fig:mosaics1} shows an example of the \emph{combined} FITS image of A2670 from multiple exposures. The characteristic R$_{200}$ radius of the cluster is drawn as a dashed circle centered at the A2670's center coordinate given in Table~\ref{tab:clusters}. The FoV of MOSAIC{\footnotesize II} which was used by \citetalias{Sheen_2012} is also overlaid as a thin solid square, for comparison. MOSAIC{\footnotesize II} is a predecessor of the DECam which was mounted on the Blanco telescope.

\begin{deluxetable*}{cccccccc}
\tablecaption{CTIO DECam Observations Summary \label{tab:clusters}}
\tablewidth{0pt}
\tablehead{
\colhead{Cluster ID} & \colhead{R.A.$^a$} & \colhead{Decl.$^a$} & \colhead{Redshift$^a$} & \colhead{Filter} & 
\colhead{$t_{exp}$$^b$} & \colhead{Date$^b$} & \colhead{MW Extinction$^c$} \\
\colhead{} & \colhead{(J2000)} & \colhead{(J2000)} & \colhead{} & \colhead{} & 
\colhead{(s)} & \colhead{} & \colhead{mag}
}
\startdata
A754    & 09:09:08.4    & $-$09:39:58   & 0.0542   & $u^{\prime}$ & 300/3600  & 2013 Apr 11/11 & 0.308 \\
        &               &               &           & $g^{\prime}$ & 300/7200  & 2013 Apr 11/11 & 0.240 \\
        &               &               &           & $r^{\prime}$ & 300/10800 & 2013 Apr 10/10 & 0.166 \\
A2399   & 21:57:25.8    & $-$07:47:41   & 0.0579   & $u^{\prime}$ & 300/6000  & 2014 Aug 20/20,21 & 0.159 \\
        &               &               &           & $g^{\prime}$ & 60/4200   & 2014 Aug 19/22 & 0.124 \\
        &               &               &           & $r^{\prime}$ & 300/10500 & 2014 Aug 19/19,20 & 0.086 \\
A2670   & 23:54:13.7    & $-$10:25:08   & 0.0762   & $u^{\prime}$ & 300/6300  & 2014 Aug 21/21,22 & 0.188 \\
        &               &               &           & $g^{\prime}$ & 60/4200   & 2014 Aug 21/22 & 0.146 \\
        &               &               &           & $r^{\prime}$ & 300/9000  & 2014 Aug 19/19--21 & 0.101 \\
A3558   & 13:27:57.5    & $-$31:30:09   & 0.0480   & $u^{\prime}$ & 300/3600  & 2013 Apr 11/11 & 0.212 \\ 
        &               &               &           & $g^{\prime}$ & 300/7200  & 2013 Apr 11/11 & 0.165 \\
        &               &               &           & $r^{\prime}$ & 300/10800 & 2013 Apr 10/10 & 0.114 \\
A3716   & 20:51:16      & $-$52:41.7    & 0.0462   & $u^{\prime}$ & 300/3300  & 2014 Aug 21/21,22 & 0.157 \\ 
        &               &               &           & $g^{\prime}$ & 300/2700  & 2014 Aug 21/21,22 & 0.122 \\
        &               &               &           & $r^{\prime}$ & 300/4800  & 2014 Aug 19/19--22 & 0.085\\
\enddata
\tablecomments{\emph{(a)} Source: NASA/IPAC Extragalactic Database (\url{https://ned.ipac.caltech.edu/}) 
\emph{(b)} The first number and date is for the \emph{best} single-exposure mosaic, and the second number and date is for the combined mosaic. \emph{(c)} Extinction by dust in the Milky Way from \url{https://irsa.ipac.caltech.edu/applications/DUST/} \citep{mwext} towards the line of sight.
} 
\end{deluxetable*}

We make use of pipeline-reduced science, weight, and data quality map (DQM) FITS image data of the clusters for both \emph{single}-exposure and \emph{combined} mosaics from the NOIRLab Astro Data Archive\footnote{\url{http://astroarchive.noirlab.edu}}. The pipeline reduces raw DECam data to \emph{single}-exposure mosaics by removing instrument signature, applying world coordinate system (WCS) and photometric calibration, and re-projecting to a common grid. The \emph{combined} mosaics are then stacked and/or tiled in a region of sky derived from multiple, spatially overlapping \emph{single}-exposures in the same filter band. Detailed information about the reduction process can be found in \citet{noao}. Basic information for the observed clusters and their FITS mosaics that we used is summarized in Table~\ref{tab:clusters}. Multiple standard-star fields were also observed each observing night and used for the flux calibration. Each field was observed at least twice with a large enough time intervals to measure  atmospheric extinction at different airmasses (refer to the website\footref{website} for more information). 

\begin{figure}
 \gridline{ 
            \fig{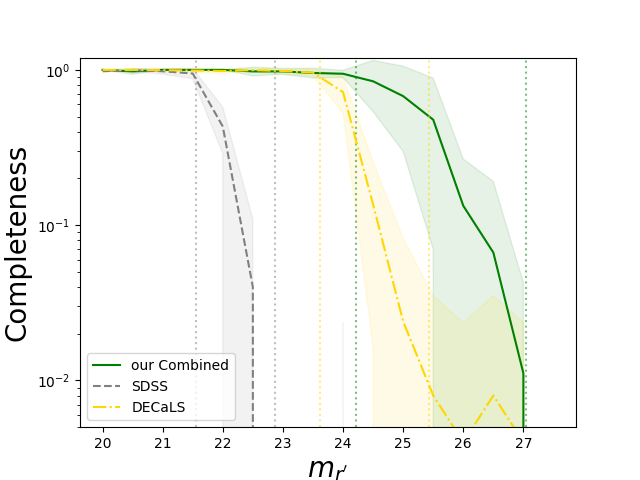}{0.5\textwidth}{}
          }
\caption{Fractions of mock sources, which was added on the $r^{\prime}$ band images of A2670, that are recovered by \SExtractor. Green solid, grey dashed, and yellow dot-dashed lines are for our combined DECam, SDSS, and DECaLS images, respectively. The mean and standard deviation among 9, 25, and 25 tiles of three surveys, respectively, are plotted. Vertical dotted lines are the magnitudes the fraction fall below 90\% and 1\% for each survey in matched color.
\label{fig:completeness}}
\end{figure}

The detection of faint merger features, such as tidal tails and shells, is depth-dependent. To measure the depth of our photometric data, we added mock sources with various magnitudes at random locations in the images using a 2D Gaussian model\footnote{\url{https://docs.astropy.org/en/stable/api/astropy.modeling.functional_models.Gaussian2D.html}}. The 2D Gaussian models are commonly used to model the point spread function (PSF) of astronomical images. Those are also relatively easy to generate and control the parameters especially for measuring depths of data from multiple sources. We then determined the completeness as a function of limiting magnitude by counting the number of injected sources detected by \SExtractor (see Figure~\ref{fig:completeness}). We also downloaded and measured depths of two overlapping public surveys: the Sloan Digital Sky Survey \citep[SDSS;][]{sdssdr16} and the DECam Legacy Survey \citep[DECaLS;][]{decals}. The magnitude where the detection rate drops below 90\% ($m_{lim,90\%}$) was around 24\,mag in $g^{\prime}$- and $r^{\prime}$-band of our \emph{combined} DECam mosaics similar to that of DECaLS. The $m_{lim,90\%}$ of SDSS was around 22\,mag in the same bands. The fainter magnitude limit where the detection rate drops below 1\% ($m_{lim,1\%}$), on the other hand, was around 27\,mag in our data in $g^{\prime}$- and $r^{\prime}$-band which was $\sim$1\,mag (1.5 mag in the case of Figure~\ref{fig:completeness}) deeper than that of DECaLS. The $m_{lim,1\%}$ of SDSS in $g^{\prime}$- and $r^{\prime}$-band was around 23\,mag. We share the result for all sample clusters in three bands on the website\footref{website}. 1-$\sigma$ surface brightness limits\footnote{\url{https://www.gnu.org/software/gnuastro/manual/html_node/Surface-brightness-limit-of-image.html}\label{sb}}, using the median of the standard deviation calculated in moving meshes in the $r^{\prime}$ band, are tabulated in the website\footref{website}. The values are $\mu \gtrsim 28$\,mag arcsec$^{-2}$ for the data we use in this study.

\subsection{Spectroscopy} \label{sec:spec}

Five out of seven DECam cluster fields that we observed overlap with public spectroscopic surveys. Table~\ref{tab:spec} lists the survey programs and the basic properties of each cluster that we derived from the combined spectroscopic catalogs. We combined sources by cross-matching them within a 1\arcsec\ radius. Using the combined spectroscopic data from multiple surveys, we measured the central velocity, $v_r$, and the velocity dispersion, $\sigma_{v_r}$, using \code{curve\_fit}\footnote{\url{https://docs.scipy.org/doc/scipy/reference/generated/scipy.optimize.curve_fit.html}} with the Gaussian model function `\code{Gauss}'.

\begin{figure*}
\center
\includegraphics[width=\textwidth]{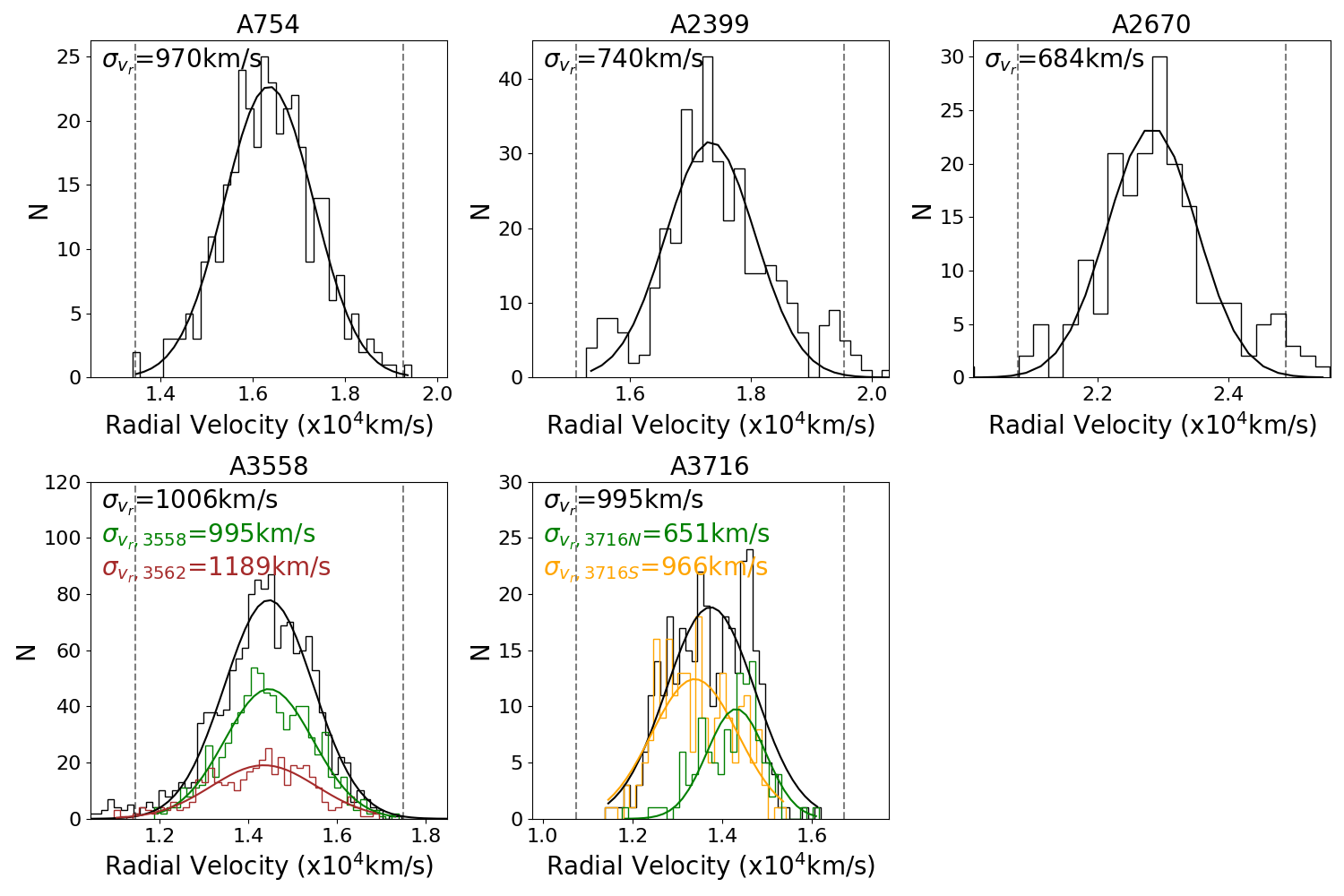}
\caption{Fitted Gaussian curves from the radial velocity histograms of galaxies in five clusters. We combined multiple spectroscopic survey data and measured the central, $v_r$, and the standard deviation, $\sigma_{v_r}$, of the radial velocities in $km/s$. Vertical dashed lines indicate a range of $\pm3\times\sigma_{rv}$ from the center, which we adopt as a cluster membership cut. A3558 and A3716 have two X-ray peaks coinciding with two substructures found by \code{mclust} (see Figure~\ref{fig:rosat2}). Colored histograms and Gaussian curves are for the substructures with same-color symbols in Figure~\ref{fig:rosat2}. \label{fig:rad_vel}}
\end{figure*}

Figure~\ref{fig:rad_vel} shows a histogram of the radial velocities from the spectroscopic surveys and the fitted 1D Gaussian curves. The A3558 mosaic also includes a second cluster, A3562. We therefore split the spectroscopic catalog of the A3558 field into two subgroups `a3558' and `a3562' (see Section~\ref{sec:sub} and the result for A3558 in Section~\ref{sec:res}), and fit a Gaussian curve separately (see the bottom left panel in Figure~\ref{fig:rad_vel}). We derived the size and mass of each cluster using the fitted $\sigma_{v_r}$ following the method outlined by \citetalias{Sheen_2012} (see their Equation~4).

We evaluated the completeness of the public spectroscopic surveys by comparing number counts within red-sequence strips (see Appendix~\ref{sec:speccomp}). The spectroscopic surveys covered around half of the bright, $M_{r^{\prime}}$\,$<$\,$-$20, cross-matched photometric sources on the red-sequence strips. Appendix~\ref{sec:speccomp} describes the procedure in detail.

\begin{deluxetable*}{cccccccc}
\tablecaption{Spectroscopic Properties of the Clusters\label{tab:spec}}
\tablewidth{0pt}
\tablehead{
\colhead{Cluster} & \colhead{$v_r^a$} & \colhead{$\sigma_{v_r}^b$} & \colhead{$R_{200}^c$} & \colhead{$M_{200}^d$} & \colhead{N$_{spec}^e$} & \colhead{N$_{member}^f$} &  \colhead{Reference} \\
\colhead{} & \colhead{$10^4$\,km\,s$^{-1}$} & \colhead{km\,s$^{-1}$} & \colhead{Mpc} & \colhead{10$^{14}M\odot$} & \colhead{} & \colhead{} & \colhead{}
}
\startdata
A754    & 1.64  & 970$\pm$41    & 2.40  & 15.8  & 339   & 336   &  W$^g$ + OW$^h$  \\
A2399   & 1.73  & 739$\pm$60    & 1.83  & 7.0   & 372   & 357   &  W$^g$ + OW$^h$ + SDSS$^i$ \\
A2670   & 2.28  & 683$\pm$61    & 1.69  & 5.5   & 208   & 189   & SDSS$^i$ \\
A3558   & 1.45  & 1005$\pm$16   & 2.49  & 17.6  & 1475  & 1367  & OW$^h$ + SS$^j$ \\
a3558   & 1.45  & 994$\pm$40    & 2.46  & 16.2  &       & 791   & green \code{mclust} comp2 in Fig~\ref{fig:rosat2} \\
a3562   & 1.44  & 1188$\pm$75   & 2.94  & 19.4  &       & 396   & brown \code{mclust} comp5 in Fig~\ref{fig:rosat2} \\
A3716   & 1.37  & 995$\pm$87    & 2.46  & 15.7  & 327   & 327   & OW$^h$    \\
A3716N  & 1.43  & 651           & 1.61  & 10.6  &       & 115   & green \code{mclust} comp2 in Fig~\ref{fig:rosat2} \\
A3716S  & 1.34  & 965           & 2.39  & 17.7  &       & 212   & orange \code{mclust} comp1 in Fig~\ref{fig:rosat2} 
\enddata
\tablecomments{\emph{(a)} The radially receding central velocity of member galaxies in the cluster.
\emph{(b)} The velocity dispersion of member galaxies in the cluster.
\emph{(c)} The clustocentric radius within which the density becomes 200\,$\times$\,$\rho_c$, being $\rho_c$ the critical density of the Universe, which is derived from $\sigma_{v_r}$.
\emph{(d)} The halo mass enclosed inside the $R_{200}$ derived from $\sigma_{v_r}$.
\emph{(e)} Number of galaxies with spectroscopic receding velocity information from multiple spectroscopic surveys.
\emph{(f)} Number of spectroscopic member galaxies which are inside the $\pm$3\,$\sigma$ range in Figure~\ref{fig:rad_vel}.
\emph{(g)} WINGS \citep{cava09}
\emph{(h)} OmegaWINGS \citep{moretti17}
\emph{(i)} SDSS DR12 \citep{alam15}
\emph{(j)} Shapley Supercluster \citep{Quintana2020} 
} 
\end{deluxetable*}

\section{Galaxy Sample} \label{sec:sam}

In this Section, we describe how we constructed a sample of galaxies that were visually inspected for the existence of ongoing- or post-merger features. We first select cluster members having radial velocities within a $\pm$3\,$\sigma$ range (vertical dashed lines in Figure~\ref{fig:rad_vel}). Then, we cross-match them to the photometric sources. We use photometric sources, extended (CLASS\_STAR from \SExtractor\,$<$\,0.5), cross-matched in at least two bands, and not spurious (saturated pixels or cosmic rays) (visit the website\footref{website} for more information). Figure~\ref{fig:cmd} shows the color-magnitude diagram of the photometric sources, with the $M_r$\,$=$\,$-20$ cut shown as vertical dashed lines on the left. The cross-matched sources are our final galaxy sample marked as teal-colored circles on the left in Figure~\ref{fig:cmd}.

\section{Identification of merging and post-merging galaxies} \label{sec:vis}

Accurate estimation and subtraction of background levels is crucial in the process of galaxy stamp generation because the faint tidal features around galaxies can easily be erased by either over- or under-subtraction of the background. In this work, we use the background subtraction of \code{Galapagos-2}, which best eliminates background flux while preserving faint tidal features in the galaxy images (see Appendix~\ref{sec:back} for a comparison of several background fitting techniques). \code{Galapagos-2} is an improved version of \code{Galapagos} \citep{galapagos}, an \code{IDL}\footnote{\href{https://www.l3harrisgeospatial.com/Software-Technology/IDL}{Interactive Data Language}} code based on \code{GALFIT} \citep{galfit}. \code{Galapagos-2} incorporates multi-band capabilities and allows for precise measurement of the sky background by performing a multi-source 2D fitting based on the \SExtractor\ output catalog. We also measured the background RMS in $r^{\prime}$ band images and computed surface brightness limits\footref{sb}. The results are online\footref{website}. For our sample of galaxy clusters, the surface brightness limit, $\mu$, was in the range 28.0--28.7\,mag\,arcsec$^{-2}$ with a moving mesh size of 1000x1000 pixel$^2$.

\subsection{Visual Inspection Tool}\label{sec:vis_ins_tool}

A set of stamp images was generated for each galaxy: 1) assorted stretches of the \emph{combined} DECam $u^{\prime} g^{\prime} r^{\prime}$-color images; 2) the result of the \code{Galapagos-2} running, the original, the model, and the model-subtracted original $r^{\prime}$-band stamps. We developed a GUI named \code{GIVIT} to effectively show these stamps to inspectors.\footnote{\url{https://github.com/DuhoKim/visual_inspection_tool}}  Users are able to label multiple morphological and feature types based on the multiple postage stamp images of each galaxy. 

\subsection{Sample classification}\label{sec:class}

\begin{figure}
 \gridline{ \fig{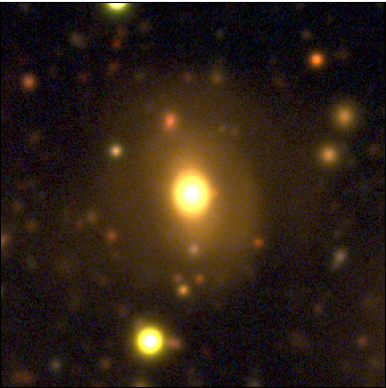}{0.23\textwidth}{}
          \fig{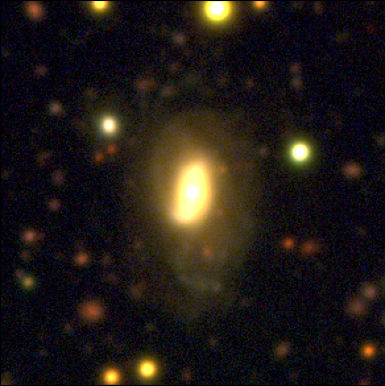}{0.23\textwidth}{}
          }
\gridline{ \fig{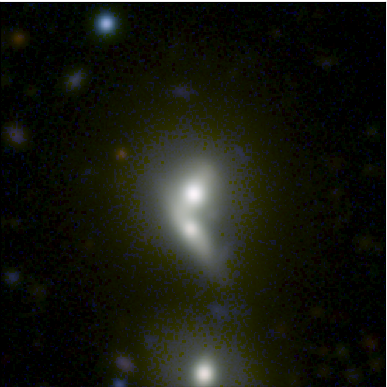}{0.232\textwidth}{}
          \fig{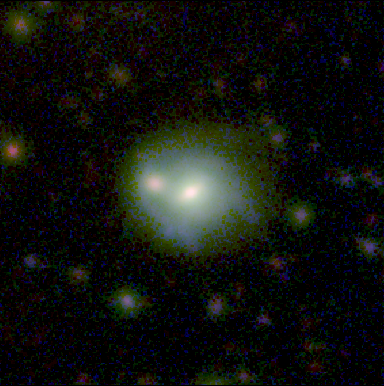}{0.23\textwidth}{}
          }
\caption{Example color stamps of galaxies selected as `Post-Merger', PM-types (top), showing faint tidal features on the left and asymmeric structures on the right, and `ongoing-Merger', M-types (bottom), which are accompanying companions with a distorted disk (left) or arms (right).\label{fig:example}}
\end{figure}

Four inspectors visually inspected the full sample and classified whether a galaxy is merging with another galaxy (M) or has recently merged, showing post-merging features (PM). We defined M and PM types as follows:
\begin{itemize}
\item M type: galaxies displaying signs of ongoing gravitational interactions connected with one or more nearby companions. This includes connecting bridges, tital tails or asymmetrical morphological features that are clearly linked to another galaxy;
\item PM type: galaxies displaying signs of past gravitational interactions, but without any current evidence of an ongoing interaction with a companion. This includes features like tidal tails, shells, or asymmetrical morphological features not connected to another galaxy.
\end{itemize}
\noindent The key distinction between M and PM types lies in the presence of companions associated with tidal features.
Inspectors opted to select M type, PM type, both, or none.
We assigned each galaxy a M or PM type if more than two inspectors came to an agreement.
Out of 872 galaxies inspected in five clusters, we identified 26 M types ($\sim$3\%) and 40 PM types ($\sim$4.5\%). Two galaxies received equal votes for both M and PM classification; these were ultimately assigned as M type. Figure~\ref{fig:example} shows representative examples of galaxies classified as M or PM type.

\begin{figure}
 \gridline{ \fig{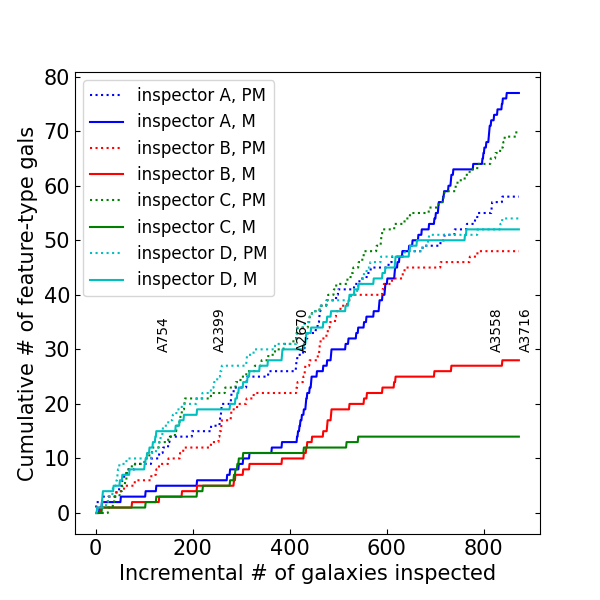}{0.5\textwidth}{}
          }
\caption{Cumulative number counts for M and PM types voted by four inspectors (A,B,C,D). The cluster names are shown as vertical texts at the position of last member from each cluster. The PM-type classifications (dotted line), more or less, agree with each other, while the M-type classification deviate from each other, especially in A3558. We use M and PM types voted by at least two inspectors.\label{fig:agree}}
\end{figure}

Figure~\ref{fig:agree} shows the cumulative counts of M and PM-type votes for each inspector. The name of a cluster is annotated as a vertical text where the last member places in the whole sample. The classification for PM types seems more robust than M types in terms of agreement. In the case of M types, the vote counts suggest inspectors have different thresholds for what should count as an ongoing merger. However, it also shows that some inspectors are not necessarily more conservative than others. Inspectors A, B, and C generally agree on each other on casting votes for the M types for galaxies in all clusters but A3558, while Inspector D seems to disagree with other inspectors but agrees with Inspector B in A3558. The disagreement rate in A3558 was $\sim$31\% slightly higher than the average. Inspector A was most generous while Inspector C was most strict. Clumpy galaxies could have contributed to the disagreement in M-type classification because inspectors were not told to include/exclude minor mergers, so some inspectors might have considered clumps as the remains of minor mergers, whereas the others thought clumps were SF regions in galaxies.


\section{Dynamical state of galaxy clusters}\label{sec:dyn}

To study the relationship between the fraction of M ($f_M$) and PM ($f_{PM}$) types and the dynamical state of the galaxy clusters, we used the result of a study of Chandra and XMM-Newton X-ray images of 1844 galaxy clusters by \citet{yuan22}. Their catalog includes five out of six cluster samples (Table~\ref{tab:dyn}). A3716 is included in their previous catalog from \citet{yuan20}, but we decided not to use it because 1) for the consistency, 2) authors say they excluded clusters with poor data quality in their most recent catalog in \citet{yuan22} (personal comm.). They derived a set of parameters indicating the dynamical state of the galaxy clusters using the X-ray archival data: 1) the concentration index, $c$, defined as the ratio of X-ray luminosities within circular regions with radii of 100\,kpc and 500\,kpc; 2) the centroid shift, $\omega$, defined as the standard deviation of the projected separation between the X-ray brightness peak and the model-fitted center; 3) the power ratio, $P_3/P_0$, dimensionless morphological parameters from the 2D multipole expansion of the projected gravitational potential of clusters within 500\,kpc; 4) the profile parameter $\kappa=(1 + \epsilon) / \beta$, ($\beta$ and $\epsilon$ are the power-law index and the ellipticity for the fitted $\beta$-model); 5) the asymmetry factor, $\alpha$, defined as the fraction of the X-ray flux which is asymmetrical; and 6) the morphology index, $\delta$, $\delta = 0.68\log_{10}(\alpha) + 0.73\kappa + 0.21$.

1) to 3) are parameters that are widely used as indicators for the dynamical state of galaxy clusters, which are calculated in a circular central region with a fixed radius of 500\,kpc. `Disturbed' clusters have lower $c$, higher $\omega$, and higher $P_3/P_0$ values than `relaxed' ones \citep[see Figure~5 in][]{yuan22}. \citet{yuan20} derive two adaptive parameters 4) to 5) from the best fitted elliptical region and defined 6), and they argue it can be excellent indicator for dynamical state by having higher $\delta$ as galaxy clusters disturbed.
Table~\ref{tab:dyn} shows the values for our sample from \citet{yuan22}.
\begin{deluxetable}{ccccccc}
\tablecaption{Dynamical status parameter values from \citet{yuan22}\label{tab:dyn}}
\tablewidth{0pt}
\tablehead{
\colhead{Cluster} & \colhead{c$^a$} & \colhead{$\omega^b$} &  \colhead{$(P_3/P_0)^c$} & 
\colhead{$\kappa^d$} & \colhead{$\alpha^e$} & \colhead{$\delta^f$} \\
& [$\log_{10}$] & [$\log_{10}$] &  [$\log_{10}$] & & [$\log_{10}$] & \\
}
\startdata
A754    & {\bf $-$1.17} & {\bf $-$1.50} & {\bf $-$5.42} & {\bf 2.62}    & $-$0.30       & {\bf 1.92} \\
A2399   & {\bf $-$0.89} & {\bf $-$1.88} & {\bf $-$5.90$^g$} & {\bf 2.97}    & $-$0.85       & {\bf 1.80} \\
A2670   & $-$0.48       & {\bf $-$1.92} & $-$7.04$^h$       & 1.11          & {\bf $-$1.18} & {\bf 0.22} \\
A3558   & {\bf $-$0.67} & {\bf $-$1.94} & $-$8.38       & {\bf 1.78}    & {\bf $-$1.09} & {\bf 0.77} \\
A3562   & $-$0.54       & {\bf $-$2.25} & {\bf $-$6.38} & {\bf 1.53}    & {\bf $-$1.16} & {\bf 0.54} \\
\enddata
\tablecomments{\emph{(a)} The concentration index, $c$. \emph{(b)} The centroid shift, $\omega$. \emph{(c)} The power ratio, $P_3/P_0$. \emph{(d)} The profile parameter, $\kappa$. \emph{(e)} The asymmetry factor, $\alpha$. \emph{(f)} The morphology index, $\delta$. 
Uncertainty ranges were $\pm$0.01 for all values except \emph{(d)}, \emph{(g)}, and \emph{(h)}. These ranges were calculated using the Jacknife method on X-ray pixel data and error propagation (personal communication). Uncertainty for \emph{(d)} was not provided, while \emph{(g)} and \emph{(h)} had uncertainties of 0.03 and 0.15, respectively. Boldfaced values indicate the values that `disturbed' clusters generally exhibit \citep[see][Figure~5]{yuan22}.
} 
\end{deluxetable}

\section{Identification of substructures} \label{sec:sub}

As \citet{binney&tremaine87} and \citet{mihos03} point out, slow encounters between galaxies in groups are more likely to end in a merger than fast encounters in clusters. Taking advantage of our large FoV data, we use two complementary techniques: \code{mclust}\footnote{\url{https://cran.r-project.org/web/packages/mclust/index.html}} (\code{R} package for normal mixture models) and an improved version of the traditional \citet{ds} test \citep[\code{DS+}\footnote{\url{https://github.com/josegit88/MilaDS}}][]{DS+} to find galaxy groups in and around galaxy clusters. We input R.A., DEC., and the redshift from Section \ref{sec:spec}. We use all available spectroscopic sources, the $N_{spec}$ column in Table~\ref{tab:spec}, as input for these algorithms.

The \code{mclust} package uses an expectation-maximization (EM) algorithm for normal mixture models \citep{mclust} to find an optimum number of components representing a given multidimensional data. \code{mclust} calculates the Bayesian Information Criterion (BIC) values for up to nine components having 14 covariance structures. The BIC penalizes models with more components more heavily than the Akaike Information Criterion (AIC) and less likely to overfit the data. \code{mclust} is better at separating large/clear substructures while \code{DS+} can identify smaller substructures. We select a model with the highest BIC score for each galaxy cluster with more than two components to separate the substructures from the main cluster (see Figure~\ref{fig:rosat1}--\ref{fig:rosat2}). 

For A3558, we select a model with the 7th BIC score, a VVV--having the covariance matrix with varying volume, shape, and orientation--a model with 6 components, which we consider to have two components that match a3558 and a3562 best (see green and brown components in the top left panel in Figure~\ref{fig:rosat2}). For other clusters, we define an extended component near the central position on the X-ray map as the main cluster (colored orange in A754 and A2670 and brown in A2399). In the case of A3716, two components with equivalent sizes were located north and south, matching double X-ray peaks, so we exclude A3716 from the substructure analysis based on the \code{mclust} result because we cannot identify which is the main and the other is the substructure. We use the double X-ray peaks in A3558 and A3716 and the velocity dispersion of galaxies in two \code{mclust} components, each coinciding with the X-ray peaks separately for the projected phase-space diagram analysis (see Figure~\ref{fig:rosat2}). 

The original \code{DS test} measures the local (closest 11 galaxies) deviation of the line-of-sight velocities from the mean velocity of a galaxy cluster in order to determine whether a cluster is relaxed or has substructures by summing the local deviation values. \code{DS+} has an additional feature, which is returning membership information of a galaxy in a background (main cluster) or in a substructure. Unlike the \code{DS+}, \code{mclust} does not define a main cluster so that all galaxies are assigned to one of the substructures. We set a parameter `nsims', the number of simulations, as 1000, and `Plim\_P', the minimum probability of the \code{DS+} selection parameter, as 0.1. We tried a grid of `Plim\_P' values, 0.01, 0.05, 0.1, 0.5, 1, 5, and 10, and found that lower than 0.1 values return a limited number of substructures while larger values shred structures into way more than physical ones. For A3558, we ran the \code{DS+} separately for a3558 and a3562 because we had an overabundance of substructures from the fitting of the whole sample in A3558.

\section{Results} \label{sec:res}

In this Section, we show the results of our study on $f_M$ and $f_{PM}$ in the cluster environment relating to 1) the dynamical states of galaxy clusters, 2) substructures, and 3) the projected phase space.

\subsection{Relation with the dynamical state of clusters} 

\begin{figure*}
 \gridline{ 
            \fig{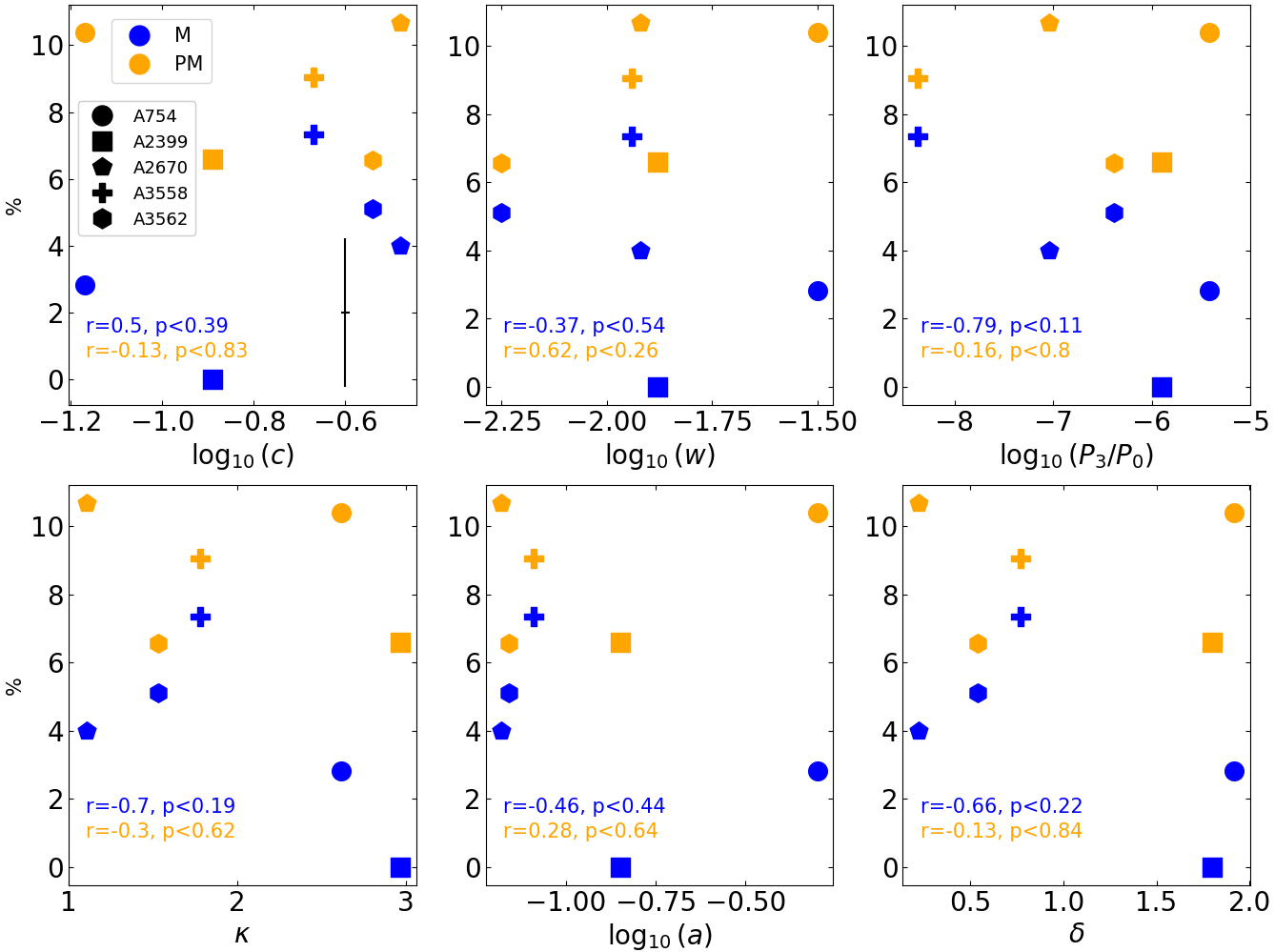}{0.9\textwidth}{}
          }
\caption{Fractions of merging ($f_M$, blue) and post-merging ($f_{PM}$, orange) galaxies within the R$_{200}$ of six galaxy clusters (represented by different symbols). These fractions are plotted against six dynamical state parameters from \citet{yuan22}. Mean errors in the x and y axes are indicated in the top left panel. Y-axis errors were derived from squared Poisson errors based on feature-type galaxy counts. As noted in Section~\ref{sec:dyn}, `relaxed' clusters generally exhibit lower parameter values (except for concentration $c$) and `disturbed' clusters clusters exhibit higher values. Pearson correlation coefficients ($r$) and associated $p$-values for each parameter and feature-type fraction are displayed at the bottom of each panel. We observe a marginal trend of higher $f_M$ in `relaxed' clusters (negative $r$ values except for $c$), though with weak statistical significance ($p$-values\,$<$\,0.11--0.54).
\label{fig:dyn}}
\end{figure*}

Figure~\ref{fig:dyn} displays $f_M$ and $f_{PM}$ within the R$_{200}$ of each cluster as a functions of their dynamical status indicators. For each indicator, Pearson correlation coefficients ($r$) and $p$-values are computed and shown at the bottom of each panel. Positive correlations approach $r$\,$=$\,1, while negative correlations approach $r$\,$=$\,$-$1. Results are considered statistically significant with $p$-values below 0.05. Note that the `relaxed' clusters generally exhibit lower indicator values (positioned towards the left), except for concentration, $\log_{10}(c)$, where they have higher values \citep[top left panel; see][Figure~5]{yuan22}. We observed a marginal trend suggesting higher fractions of merging galaxies in `relaxed' clusters. This trend correlates positively with concentration and negatively with the others. However, the correlation remains weak ($p$-value\,$<$\,0.11--0.54). No such trend was found for PM-type fractions.

\subsection{Substructures} \label{sec:sub_res}

Figures~\ref{fig:rosat1}--\ref{fig:rosat2} and Figures in Appendix~\ref{sec:dsp} show substructure identification results of galaxy clusters from \code{mclust} and \code{DS+}, respectively. On the left side we show X-ray contour maps from ROSAT All-Sky Survey 
(RASS\,3)\footnote{\url{https://skyview.gsfc.nasa.gov/}} \citep{rosat} data as transparent contours. For the central part of the X-ray maps, we use high-resolution X-ray data from XMM-Newton \citep{yuan22} (for A2399, A3558, and A3716) and Chandra \citep{yuan20} (for A754 and A2670) shown as darker contours.
On the right side, we present projected phase-space diagrams (PPSDs; refer to Section~\ref{sec:ppsd} for the definition of the alphabetical sections and their implication). 

Colors other than black represent substructures, and only the result from \code{DS+} (Figures~\ref{fig:rosat_dsp1}--\ref{fig:rosat_dsp2}) show the main cluster component in black, because \code{mclust} fits the data into multi Gaussian curves while \code{DS+} scans substructures having different line-of-sight velocities from the main cluster. 
We specify the results in detail for each cluster below:
\begin{deluxetable}{cccc}
\tablecaption{The result of substructure identification\label{tab:sub}}
\tablewidth{0pt}
\tablehead{
\colhead{Cluster} & \colhead{$N_{tot}^a$} & \colhead{\code{mclust}$^b$} &  \colhead{\code{DS+}$^c$} 
}
\startdata
A754    & 126           & {\bf 121}, 5 (1)                          & {\bf 123}, 3 (1) \\
A2399   & 115           & {\bf 52}, 21, 20, 6, 6,                   & {\bf 93}, 7, 4, 3,  \\
        &               & 5, 4, 1, 0 (8)                            & 2, 2, 2, 1, 1 (8) \\
A2670   & 171           & {\bf 76}, 35, 25, 19, 16 (4)              & {\bf 144}, 9, 7, 3, 3, 3, 2 (6) \\
A3558   & 400           & {\bf 229(a3558), 155(a3562)}              &    \\
        &               & 8, 7, 1, 0  (4)                           &    \\
a3558   & 229           &                                           & {\bf 212}, 7, 5, 3, 1, 1 (5) \\       
a3562   & 155           &                                           & {\bf 140}, 5, 4, 3, 2, 1 (5) \\
A3716   & 59            & {\bf 39(S), 20(N)} (0)                          & {\bf 59} (0) \\
\enddata
\tablecomments{\emph{(a)} Total number of sample galaxies cross-matched against spectroscopic surveys. \emph{(b)} Number of sample galaxies in each component from \code{mclust}. \emph{(c)} Number of sample galaxies in the main and substructures found by \code{DS+}. Boldfaced numbers are numbers of main clusters. Numbers in parentheses are substructure numbers found by each code. Note that some structures have single or no constituent because we exclude galaxies with brightness $M_r^{\prime}$$>$$-20$, which were included in the \code{mclust} and \code{DS+} code running.
} 
\end{deluxetable}
\begin{figure*}
 \gridline{ \fig{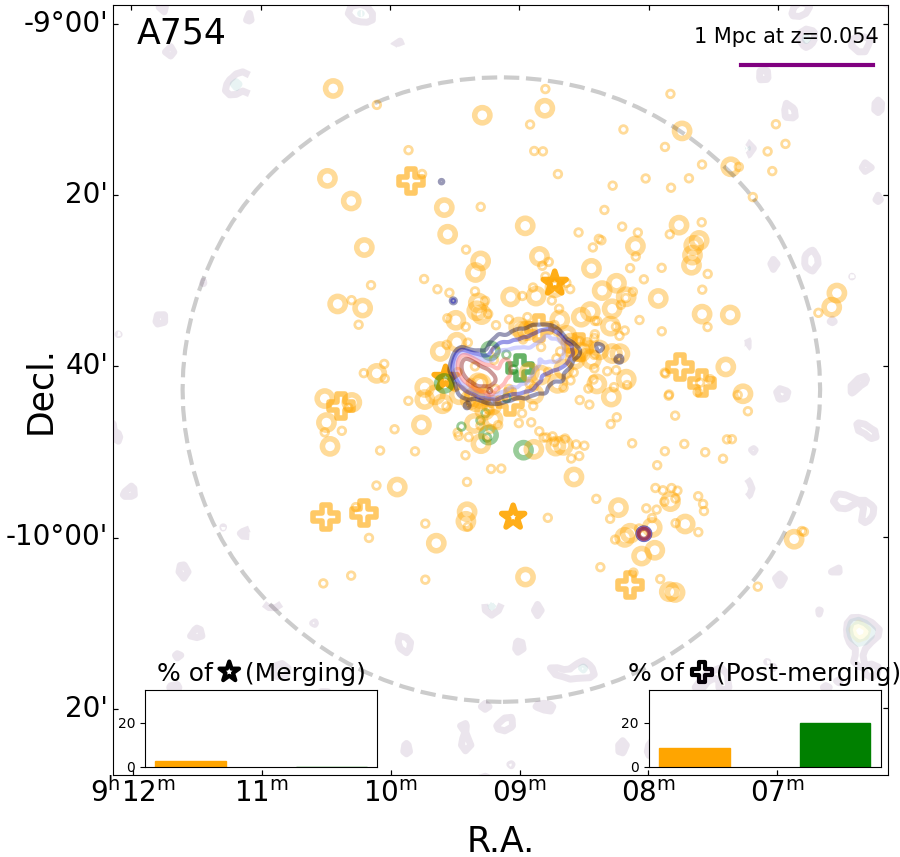}{0.38\textwidth}{}
          \fig{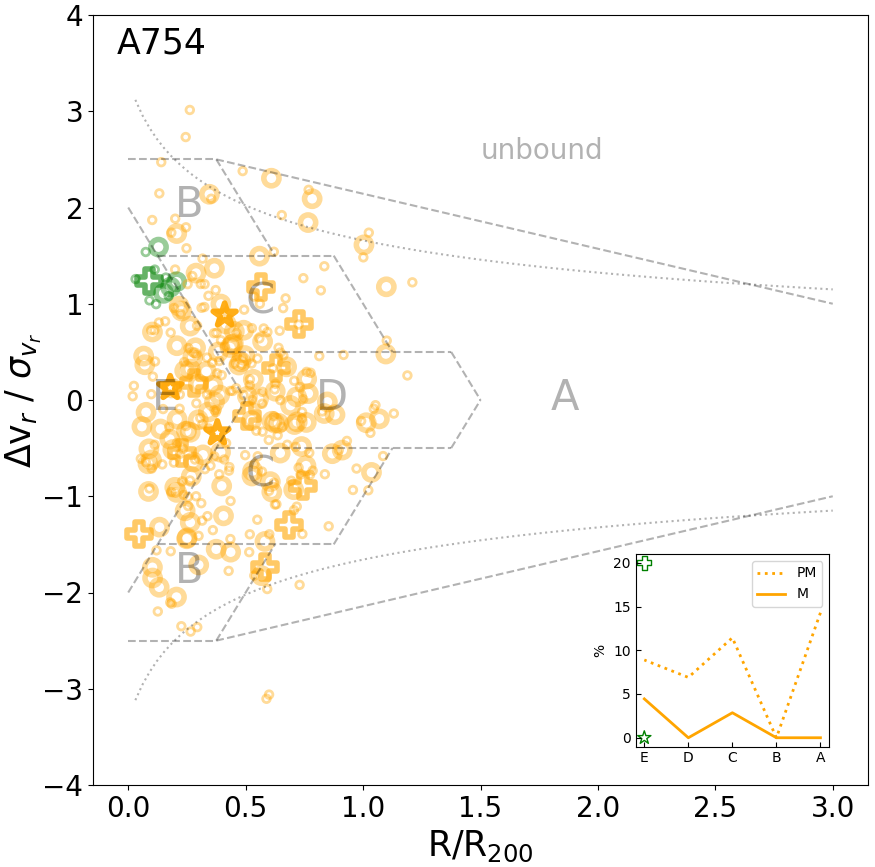}{0.38\textwidth}{}
          }
\vspace{-1cm}
\gridline{ \fig{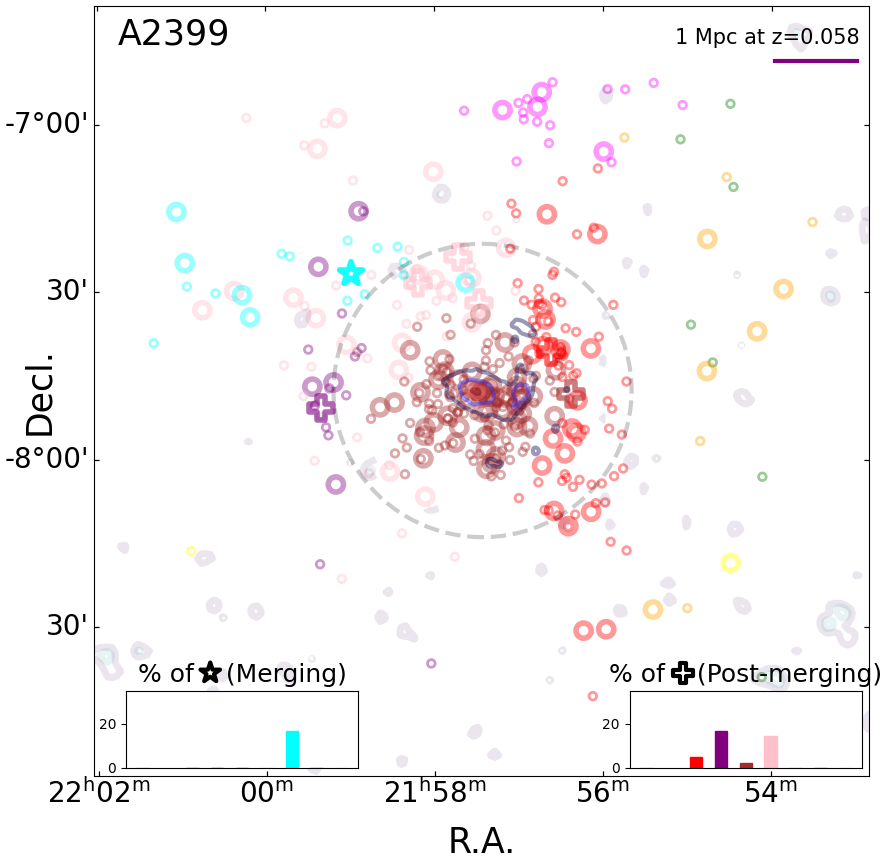}{0.38\textwidth}{}
          \fig{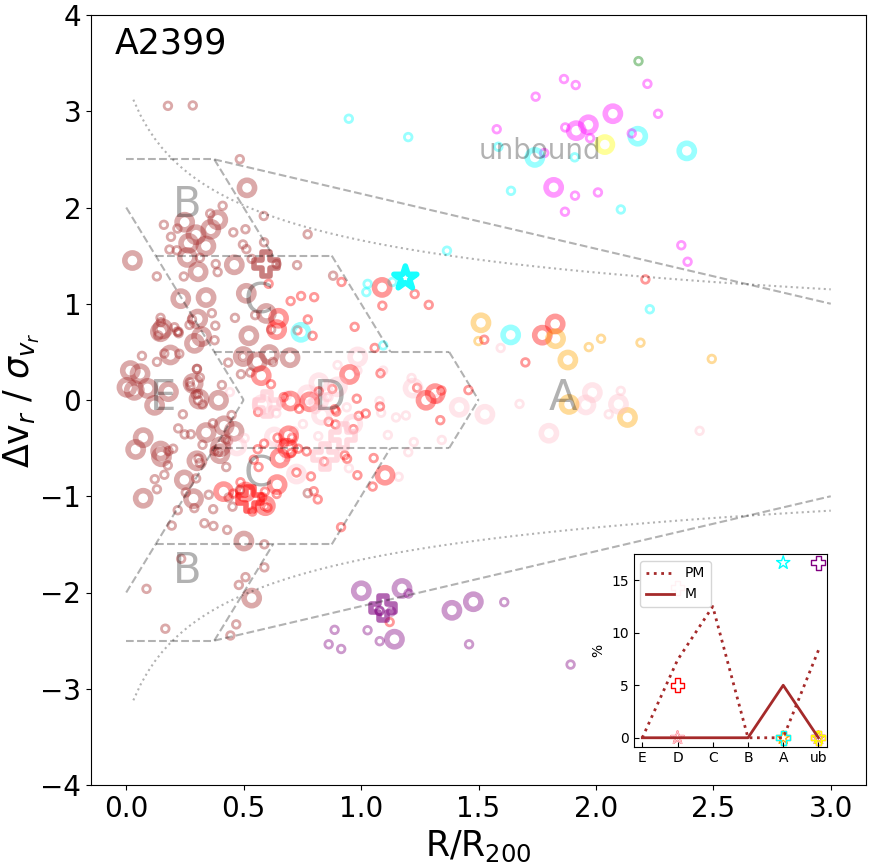}{0.38\textwidth}{}
          }
\vspace{-1cm}
\gridline{ \fig{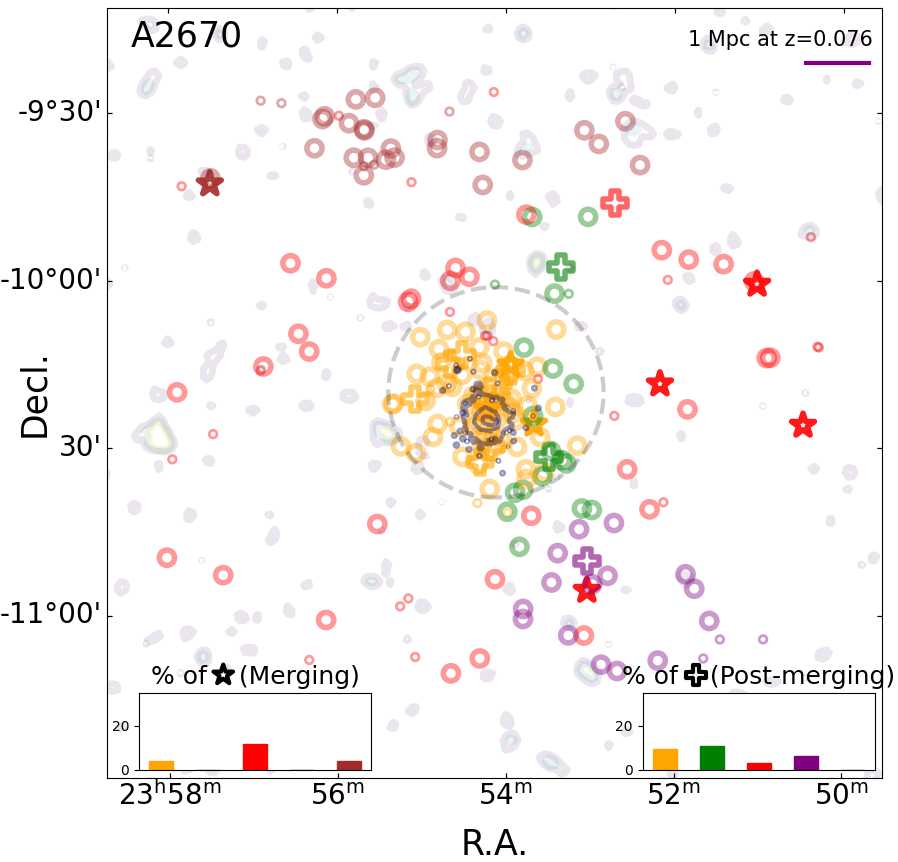}{0.38\textwidth}{}
            \fig{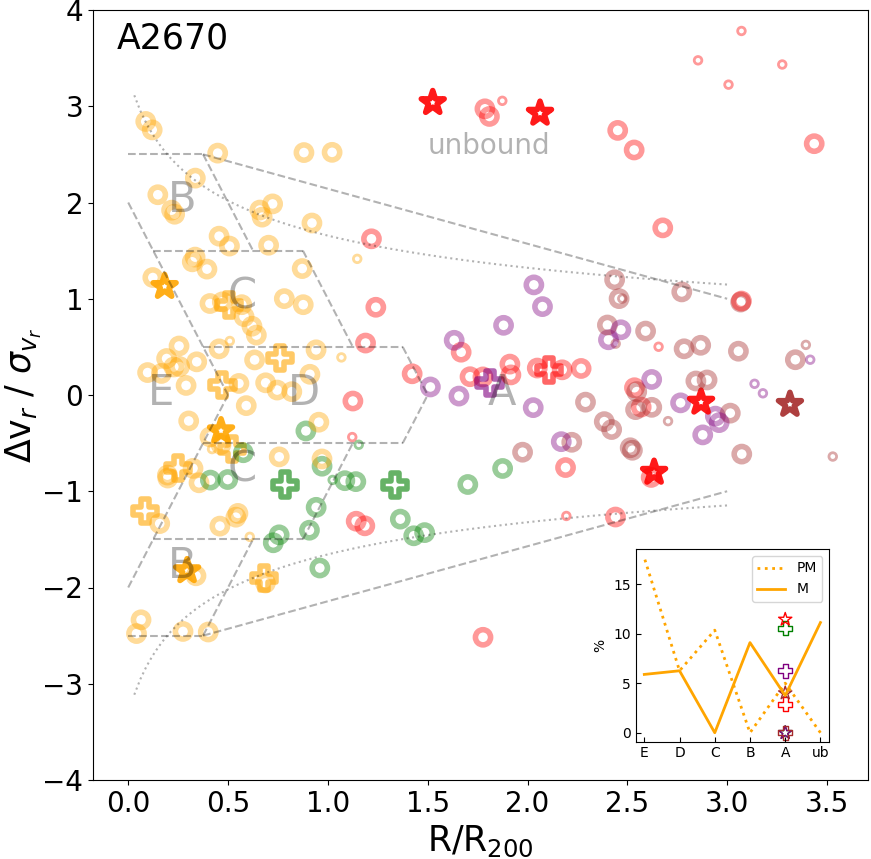}{0.38\textwidth}{}
          }
\vspace{-1cm}
\caption{\emph{(left)} Distribution of galaxies in spectroscopic surveys on the celestial coordinates. Dashed circles indicate the R$_{200}$ of each cluster. Small empty circles represent galaxies from spectroscopic catalogs used for substructure identification. These galaxies were excluded from our sample due to their low luminosity, $M_r$\,$>$\,$-$20. Sample galaxies are shown as either star, plus, or large-circular symbols based on their classification of Merging (M), Post-Merging (PM), or non-feature, respectively. Colors correspond to substructures found by the \code{mclust} model-based clustering algorithm. Background contours are from ROSAT (light gray) and XMM-Newton (central dark) X-ray data. Lower insets show the fraction of the M ($f_M$) and PM ($f_{PM}$) types inside each substructure. \emph{(right)} Distribution of the same galaxies but on the projected phase space diagram (PPSD). The straight-dotted grey lines are from \citet{Rhee2017} (see their Figure~6 and Section~\ref{sec:ppsd}). The escape velocity from the NFW halo is overplotted as curved dotted lines. `unbound' is the region out of the limit of subhalos, `A' has the largest fraction of the first infallers, `B'--`D' are in the order of the fraction of recent infallers, and `E' has the largest fraction of ancient infallers. The inset shows $f_M$ and $f_{PM}$ for the main cluster members (dotted and solid lines) and for substructures (star and empty-plus symbols) in each region on the PPSD.\label{fig:rosat1}}
\end{figure*}

\begin{figure*}
 \gridline{ \fig{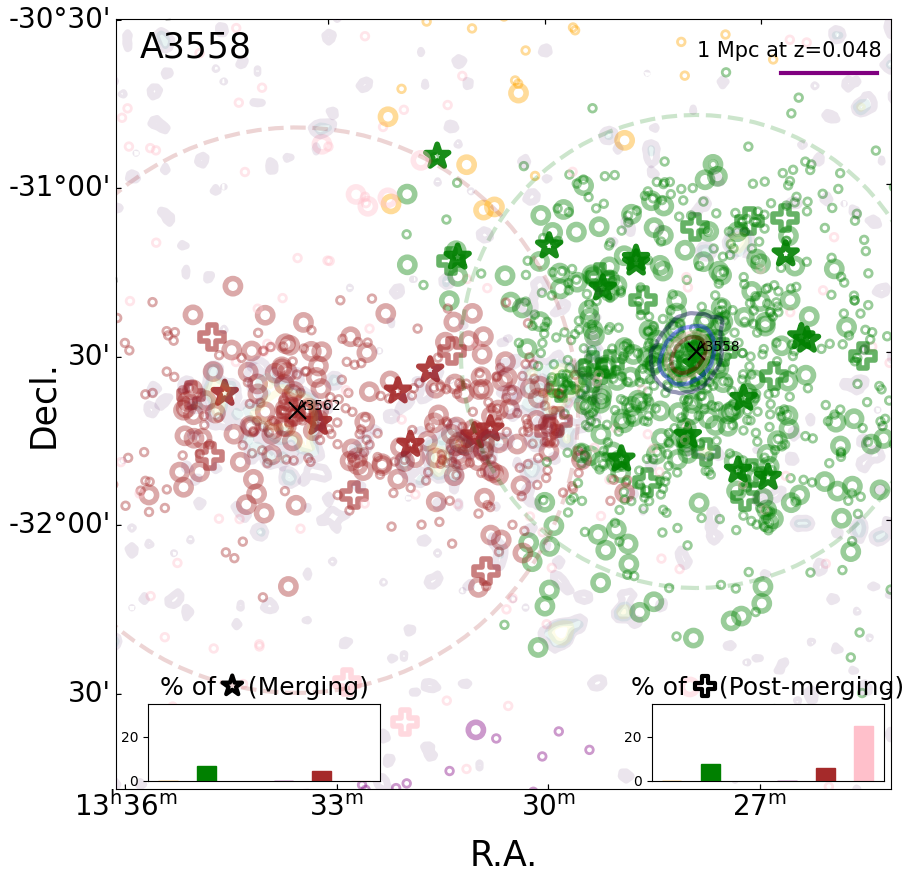}{0.4\textwidth}{}
          \fig{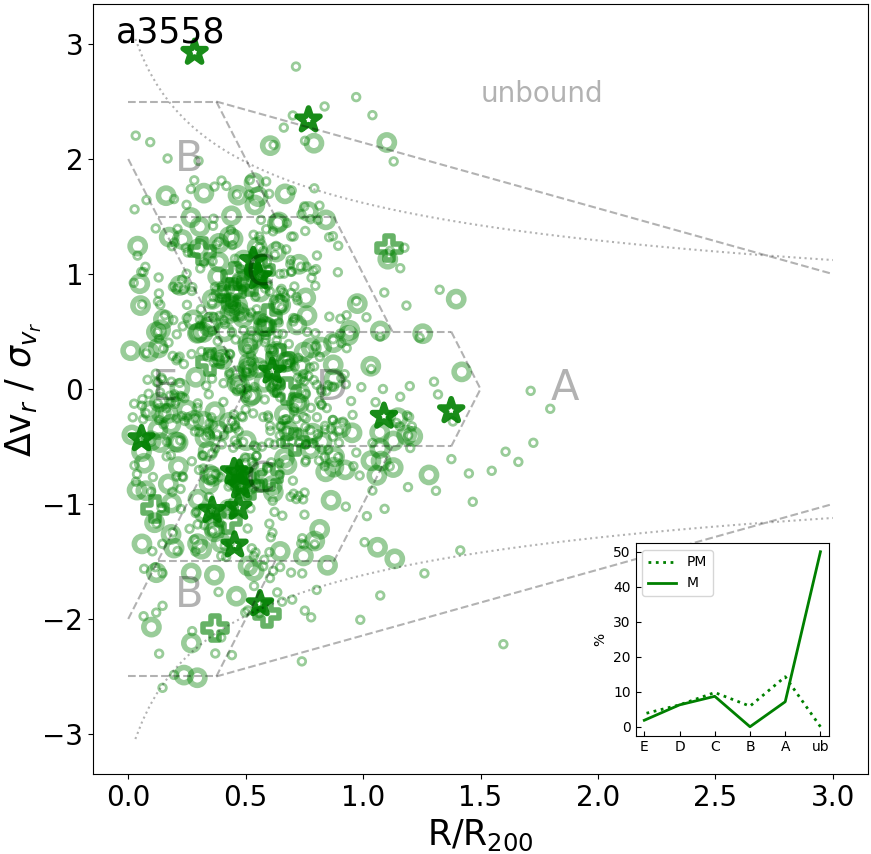}{0.4\textwidth}{}
          }
\vspace{-1cm}
\gridline{ \fig{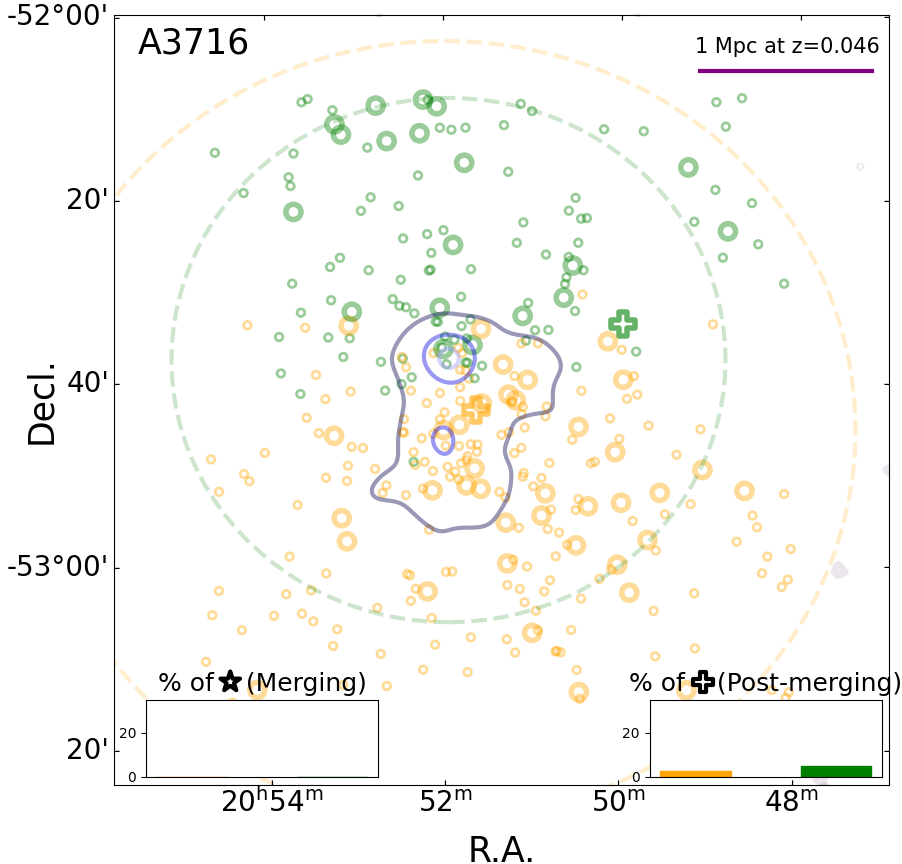}{0.4\textwidth}{}
          \fig{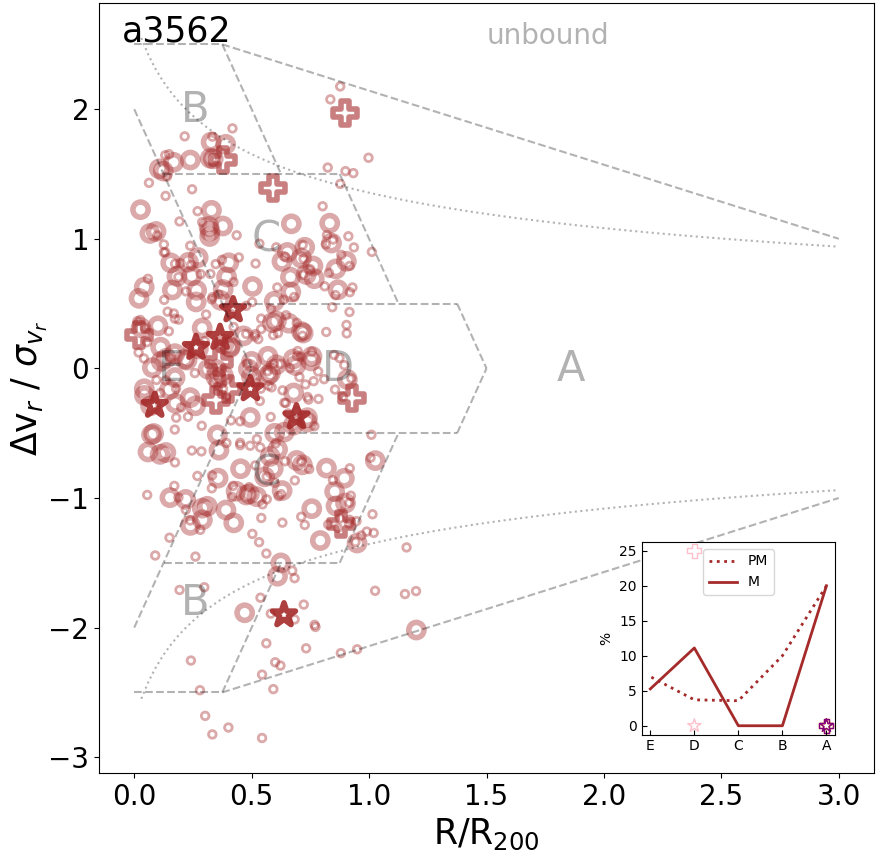}{0.4\textwidth}{}
          }
\vspace{-1cm}
\gridline{ \fig{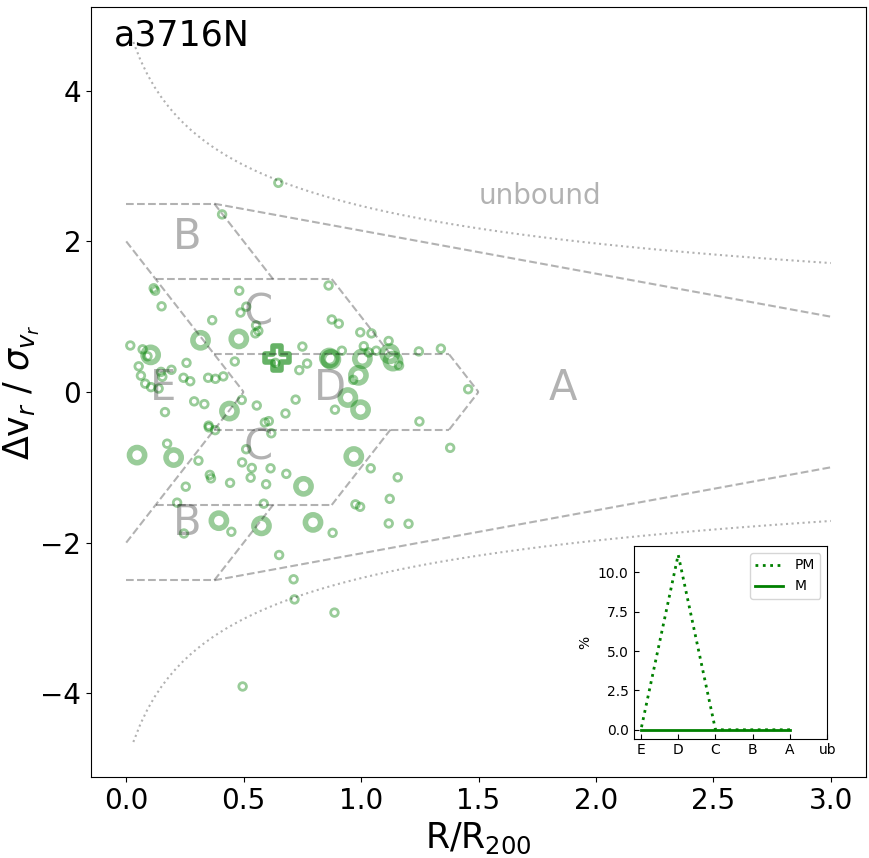}{0.4\textwidth}{}
          \fig{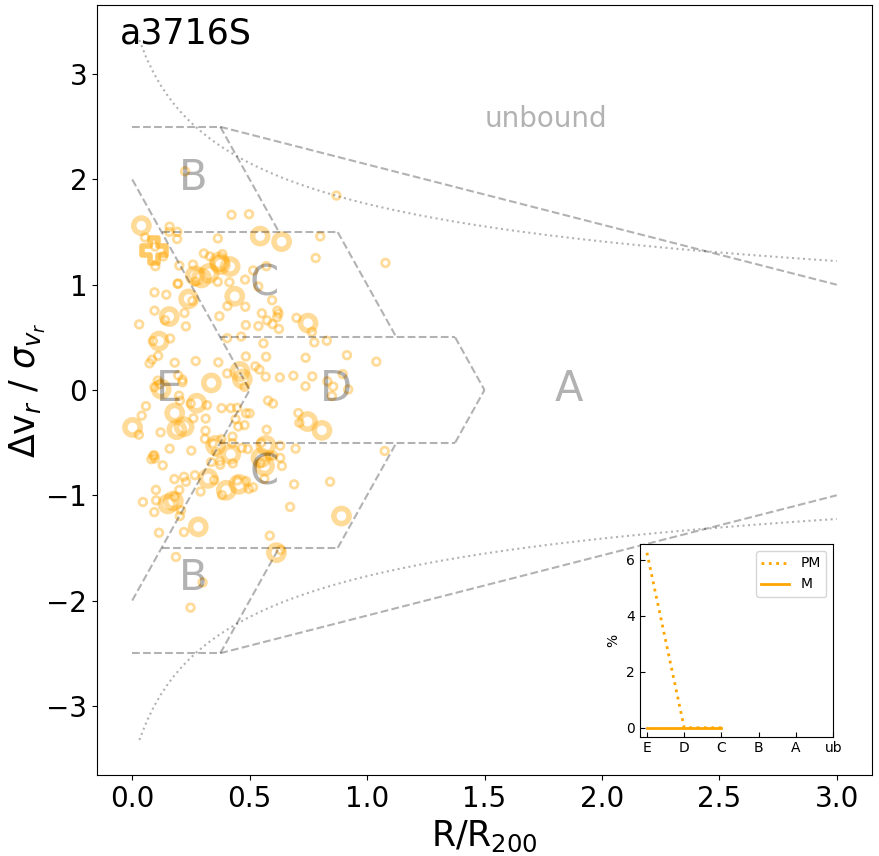}{0.4\textwidth}{}
          }
\vspace{-1cm}
\caption{continued from Figure~\ref{fig:rosat1}, but the X-ray data of A3716 is from XMM-Newton on the left. The PPSD for A3558 is separated into `a3558' and `a3562', which correspond to the green and brown substructures in the `A3558' field. Similarly, the PPSD for A3716 is separated into `A3716N' and `A3716S' which are the green and orange substructures centered in the two X-ray peaks in the `A3716' field.
\label{fig:rosat2}}
\end{figure*}

{\bf A754:} The number of components for the best \code{mclust} model was one, which means the cluster has no substructure. In spite of that, we use a model with two components because there are reports on substructures in A754 \citep{Zabludoff1995, flin06}, and we want to separate the substructures from the main cluster and compare the properties. The red X-ray contours at the center are facing a shock front in the southeast direction \citep{macario11}. A hydrodynamical merger simulation \citep{roettiger98} of two dark matter clouds, with a mass ratio of 2.5:1, produced at 0.3\,Gyr after the closest approach an X-ray temperature map that resembled the observed X-ray data. The core of the more massive main cluster coincides with the shock front. Passing the pericenter from left to right, the smaller cloud's core is where the right part of the elongated X-ray contour is. \code{DS+} found a substructure located near the right part of the X-ray core.

{\bf A2399:} We adopt the best \code{mclust} model having nine components. \citet{lourenco20} found the best \code{mclust} model with three components from similar input data sets. We select the brown component as the main cluster. \code{DS+} also found eight substructures.

{\bf A2670:} We adopt the model with the highest BIC score having 5 components and set the orange component as the main cluster. The cluster has $f_M$ 4\%$\pm$1.5 higher than the average 3\%$\pm$2.8 (cluster to cluster). \citet{piraino23} performed the \code{mclust} analysis with the spectroscopic data from the SDSS around A2670 in a wider range of 5$\times R_{200}$ and find six components. \code{DS+} found six substructures. 

{\bf A3558:} Maximum number of components, nine, were found by the \code{mclust} in the field of A3558. Still, we adopt a model with six components because the second and fifth components, colored in green and brown in Figure~\ref{fig:rosat2}, matched well with the X-ray peaks. The two X-ray peaks are the centers of two galaxy clusters, A3558 and A3562, in the Shapley Supercluster \citep{Raychaudhury90}\footnote{\url{http://www.atlasoftheuniverse.com/superc/shapley.html}\label{Shapley}} (see `X' marks in the top-left panel of Figure~\ref{fig:rosat2}). We fit the Gaussian curves for line-of-sight velocities of the two substructures separately (see Figure~\ref{fig:rad_vel}) and draw their R$_{200}$ in the upper-left panel of Figure~\ref{fig:rosat2}. We name the two clusters `a3558' and `a3562' to avoid confusion with the DECam mosaic ID `A3558', which contains two clusters, and consider both the green and brown components as the main clusters. The `a3558' is also known as `Shapley 8'\footref{Shapley} located at the core of the Shapley Supercluster. Four other components found by \code{mclust} are defined as substructures. In the case of \code{DS+}, as mentioned in Section~\ref{sec:sub}, we find substructures in `a3558' and `a3562' separately, and combine the result. `a3558' and `a3562' resulted in five substructures each. 

{\bf A3716:} A two-component model from \code{mclust} coincides with the two X-ray peaks, so we adopt it even though the single-component model had a higher BIC score. We exclude A3716 from the substructure analysis because 1) the two components divided the sample evenly, 2) no galaxy was classified as M type, 3) a PM-type galaxy was found in each of A3716N and A3716S, and 4) the substructures found by \code{DS+} do not include any visually classified sample. Still, we use the result from \code{mclust} in the projected phase-space diagram analysis by applying two X-ray peaks and the velocity dispersion values separately for A3716N and A3716S.

\begin{figure}
 \gridline{ \fig{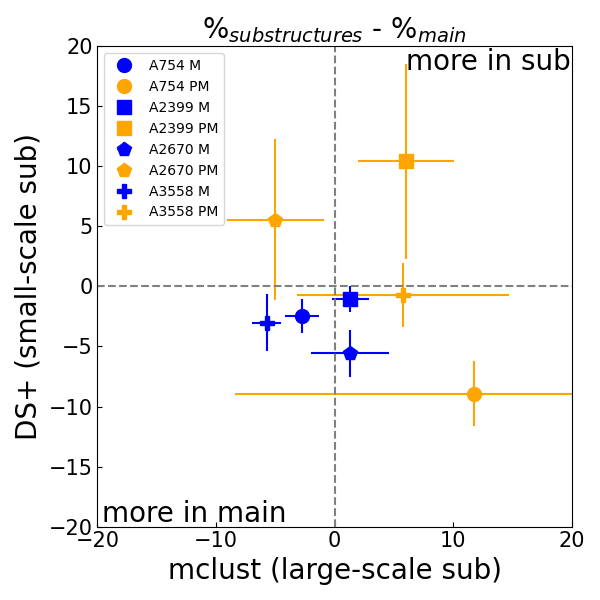}{0.45\textwidth}{}
          }
\caption{Comparison of $f_{M,PM}$ between substructures and the main cluster. Substructures were found by \code{mclust}, which finds large-scale substructures using Gaussian Mixture Modeling, (x-axis) and \code{DS+}, which finds small-scale substructures using local mean velocity deviation, (y-axis). PM-type galaxies seem to prefer the large-scale substructures to the main clusters, while M-types found 0--5\% points more in the main clusters than the small-scale substructures.\label{fig:frac_sub}}
\end{figure}

Figure~\ref{fig:frac_sub} compares $f_M$ with $f_{PM}$ in the main cluster and substructures defined by the \code{mclust} (x-axis) and \code{DS+} (y-axis) algorithms. A positive value means more feature-type galaxies in substructures.
We count the numbers of M-, PM-, and no feature-type galaxies in the main cluster and substructures and calculate the:
\begin{equation}
(\sum_{i=1}^{n} \#_{feat} / \sum_{i=1}^{n} \#_{total} - \sum_{main}^{} \#_{feat} / \sum_{main}^{} \#_{total}) * 100,
\label{eq:sub_main}
\end{equation}
where n is the number of substructures, $\#_{feat}$ is the number of M- or PM-type galaxies, and $\#_{total}$ is the total number of galaxies. Error bars are the sum of the square root of Poissonian errors on the numbers of feature-type galaxies in the main cluster and substructures. In general, PM-type galaxies were found slightly, in three out of four clusters, more in large-scale substructures than the main clusters, while M-type galaxies were found 0--5\% point more in the main clusters than in small-scale substructures. 

\subsection{Projected phase-space diagram} \label{sec:ppsd}

To study the relationship between the time since first infall into the cluster and $f_{M, PM}$, we used the regions that split up the projected phase space diagram (PPSD) in \citet{Rhee2017}. We show the regions on the PPSD on the left panels of Figures~\ref{fig:rosat1}--\ref{fig:rosat2} and \ref{fig:rosat_dsp1}--\ref{fig:rosat_dsp2}. `unbound' is the region outside of the limit of subhaloes, `A' contains more than half of `Interlopers' that are non-members of the cluster and most (around a third) of `First infallers', `B' contains most ($\sim$40\%) of `Recent infallers', `C' to `E' contain a decreasing number of `Recent infallers', and around half of the galaxies in `E' becomes the `Ancient infallers'  (see their Figure~6). The escape velocity from a Navarro-Frenk-White (NFW) halo, as computed by \citet{Gill04}, is drawn as curved-dotted lines for comparison. For that, we adopt Equations~1--3 from \citet{Rhee2017} converting to $v_{LOS} = v_{esc} / \sqrt{3}$ and $r_{proj} = \pi / 2 \times R_{3D}$. 

\begin{figure}
 \gridline{ 
            \fig{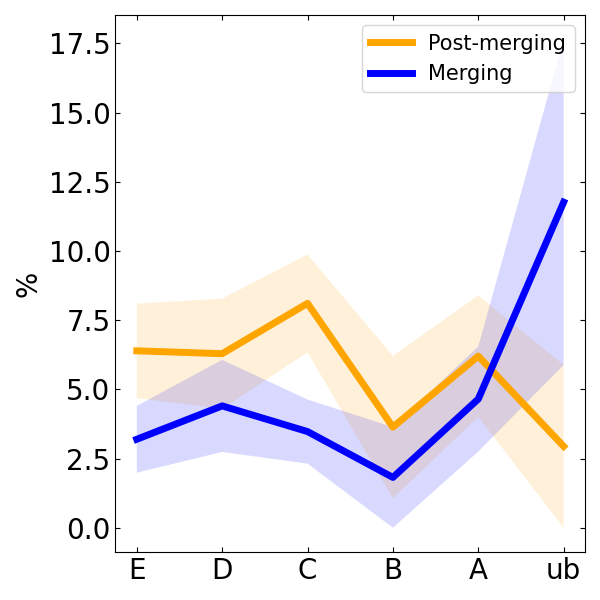}{0.5\textwidth}{}
          }
\caption{$f_{M,PM}$ in PPSD regions. The X-axis is in the order of the distance from the center of the galaxy clusters to each region of PPSD. The shaded areas correspond to the fractional range of $\pm$ Poisson errors ($\sqrt N$). The fraction of M-type galaxies peaks outside of the cluster, whereas the fraction of PM-type galaxies peaks in the middle of the clusters.\label{fig:phase}}
\end{figure}
Figure~\ref{fig:phase} shows the combined $f_{M,PM}$ of all galaxy samples on each PPSD region. We count the numbers of the feature-type galaxies in each region and calculate $f_{M,PM}$. $f_{PM}$ is $\sim$1--5\% point higher than $f_M$ inside the `A--E' cluster regions, yet in the outside of the cluster, the `unbound' region, $f_M$ is $\sim$10\% point higher than $f_{PM}$. Overall, we find that the PM galaxies are more virialized than the M population. This is also shown in substructures.

\begin{figure}
 \gridline{ 
            \fig{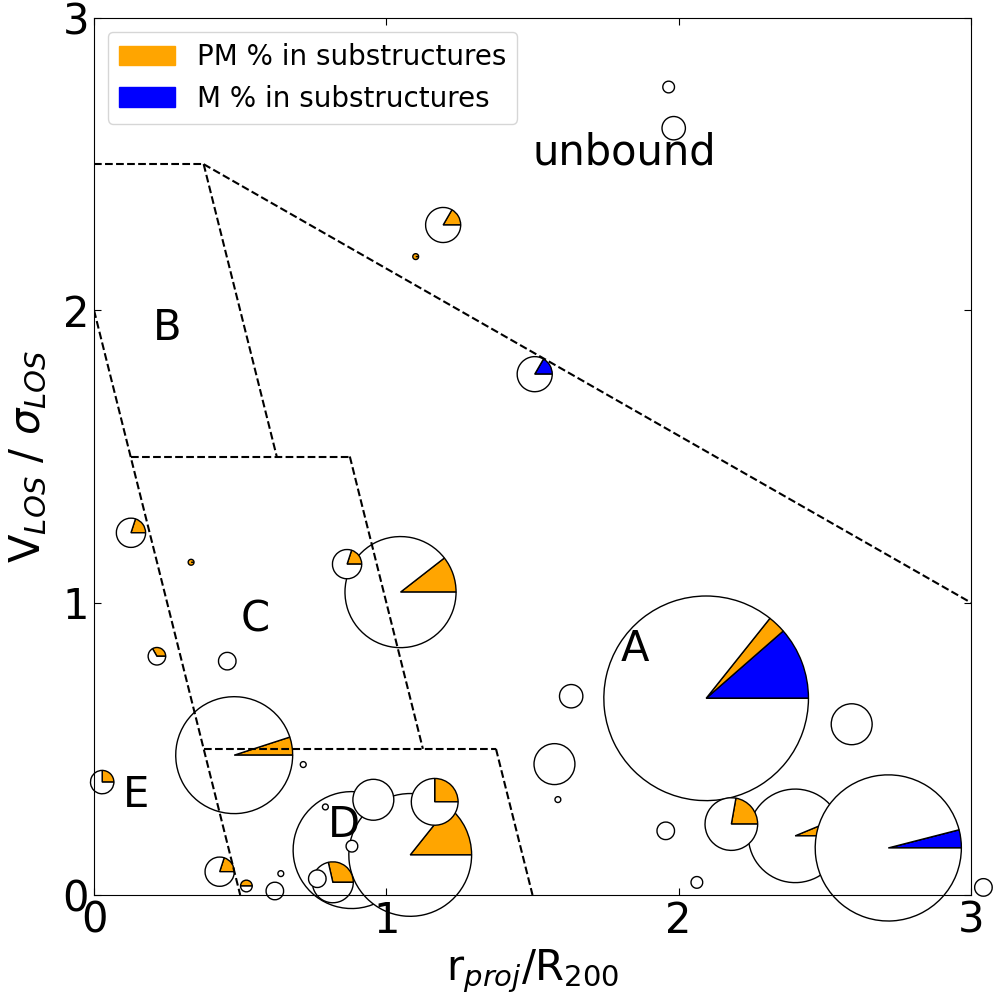}{0.5\textwidth}{}
          }
\caption{$f_{M,PM}$ in substructures as pie charts on the PPSD. The PPSD is partitioned following \citet{Rhee2017} (see Section~\ref{sec:ppsd} and the right panels of Figure~\ref{fig:rosat1}--\ref{fig:rosat2}). The size of each pie chart represents the number of constituent galaxies in each substructure. Orange and blue wedges in each pie chart show $f_{PM}$ and $f_{M}$ in each substructure, respectively. Almost all M-type galaxies in substructures were in an `A' region, whereas PM types were found in substructures in ubiquitous regions.
\label{fig:phase_grp}}
\end{figure}
Figure~\ref{fig:phase_grp} shows $f_{M,PM}$ in each substructure as a pie chart on the PPSD. The size of each pie chart depicts the number of galaxies in the substructure. Generally, \code{mclust} and \code{DS+} find large- and small-size substructures, respectively. As shown in Figure~\ref{fig:phase}, $f_{PM}$ is high, especially in small substructures found by \code{DS+}. M types are found only in the region `A' and mostly in large substructures. `Interlopers' and `First infallers' dominate the `A' region population. Despite this, M types in the `unbound' region from Figure~\ref{fig:phase} are not found in Figure~\ref{fig:phase_grp}. 

\section{Discussion and Conclusion} \label{sec:dis}

We visually inspected a volume-limited sample of spectroscopically-confirmed member galaxies in six Abell clusters in the redshift range 0.046\,$\lesssim$\,$z$\,$\lesssim$\,0.076 on the wide ($R/R_{200}$\,$<$\,3) and deep ($\mu_{r^{\prime}} \gtrsim 28$\,mag arcsec$^{-2}$) stacked DECam images, and classified galaxies according to the indication of ongoing- (M) and post- (PM) merging features. On average: (1) galaxies that are currently undergoing a merger occur with a frequency of $\sim$4\% with a decreasing frequency from the outside to the inside of the cluster; and (2) galaxies with post-merger features are found $\sim$7\% preferentially inside the escape velocity region on the PPSD and substructures. This is consistent with a scenario where galaxies undergo pre-processing outside of the galaxy clusters and accrete onto the galaxy clusters, retaining their post-merger features.

We obtain a similar ongoing merger rate to \citetalias{Sheen_2012} and \citetalias{oh18}, while \citet{McIntosh2008} reported about a dex lower rate, $\sim$0.2\%, at a dark matter halo mass range greater than $10^{14.5}M_{\odot}$ from the SDSS. Shallower images are almost certainly the origin of the main difference, as \citet{yi13} and \citet{ji14} clearly demonstrated. For PM galaxies, our observed rate of 6\%$\pm$2 (the mean and standard deviation of the PM fraction in each cluster) is approximately 15\% point lower than that reported by \citetalias{oh18} and 20\% point lower than \citetalias{Sheen_2012}. \citetalias{Sheen_2012} analyzed a sample within a red-sequence strip, using imaging data of similar depth to ours ($\mu_{r^{\prime}} \sim 28$\,mag\,arcsec$^{-2}$). \citetalias{oh18} included all Hubble types brighter than $M_{r} = -19.8$ with data approximately one magnitude shallower than ours. Consistent with \citetalias{Sheen_2012} and \citetalias{oh18}, we find higher PM rates compared to M rates. This suggest one or both of the following: 1) PM features persist for longer timescales than the total duration of the merging process itself. 2) PM galaxies migrate into the cluster after the merger has occurred outside of it.

The ratio of M- to PM-type galaxies, \emph{M/PM}, in this study goes from 0.58 (34/58; voted by at least two out of four) to 0.8 (16/20; voted by at least three out of four). \emph{M/PM} from the \citetalias{Sheen_2012} and \citetalias{oh18} studies are 0.1--0.2. N-body simulations of equal-mass gas-rich disc galaxy mergers show a merging time scale of $\sim$0.5--3\,Gyr between the first pass and the final merger of the nuclei \citep[see Table~3 in][]{Lotz08}, while \citet{yi13} and \citet{ji14} scrutinized the time scale for galaxies retaining post-merger features, reporting $\sim$4\,Gyr from the imaging data with a depth of $\mu_r$\,$=$\,28\,mag\,arcsec$^{-2}$. The range of \emph{M/PM} from the two simulation studies is 0.125--0.75, which matches the observational results. 

The galaxy-galaxy mergers in the outskirts of clusters are expected in the \emph{pre-processing} scenario where the transformation from gas-rich blue spiral galaxies to red-and-dead elliptical galaxies is happening in galaxy groups prior to their accretion onto the galaxy cluster. \citet{pasquali19} identified a peak in the average global specific star formation rate (sSFR) coinciding with a minimum in the average age of satellite galaxies. This occurred in projected phase space where the average time since satellite galaxies fell into their host halos is around 2.8 Gyr. \citetalias{Sheen_2012} also discussed that their low \emph{M/PM} values would originate in the post-merging migration of galaxies into the clusters.
 
To find a trace of the post-merging migration of galaxies from the outside to the inside of the galaxy cluster environment, we partitioned the PPSD and compared the fractions of M- and PM-type galaxies. About 10\% point more M-type galaxies were located in the outside (`unbound' region) of the galaxy clusters, while the M-type fraction was less than the PM-type fraction inside (region `A' through `E') clusters (see Figure~\ref{fig:phase}). By discerning the galaxy groups in and around galaxy clusters, we found M-type galaxies to be predominantly in the cluster outskirts (region `A'). A spectroscopically limited FoV of 1--3$\times R_{200}$ might have failed to detect substructures located further outside clusters. \citet{piraino23} expand the area of investigation of interactions around the massive interacting cluster A2670 and find a significant fraction of not only gravitational but also hydrodynamic interactions up to 5$\times R_{200}$, highlighting the importance of galaxy processing in cluster infall regions.

On the other hand, 5--10\% point more PM-types were found in galaxy groups (substructures) than in the main clusters in three out of four cases (see Figure~\ref{fig:frac_sub}), and those groups were located in ubiquitous regions on the PPSD (see Figure~\ref{fig:phase_grp}). The galaxy groups with high $f_{PM}$ that are deep inside galaxy clusters might be relics of the accreted galaxy groups harboring galaxies retaining signs of past interactions for a significant amount of time \citep[$\sim$4\,Gyr;][]{yi13, ji14}. Repeated high-speed fly-by encounters can also distort morphologies, such that a post-merger may actually have experienced multiple recent fly-bys \citep[Galaxy harassment;][]{Moore96,Moore98,Moore99}. The increase in post-merging galaxies in the cores may also be due to this process. As \citet{Lotz08,Gordon19} showed, the disturbance of an ongoing minor merger can look similar to a post-major merger. High speeds in the galaxy clusters would prohibit not only major but also minor mergers, so we can expect minor mergers to also be more frequent in the outskirts. If we move some fraction of $f_{PM}$ to $f_{M,minor}$, our result having higher $f_{M}/f_{PM}$ in the outskirt than inside of the galaxy clusters would be reinforced.

Several studies highlight the impact of cluster dynamics on SF and active galactic nuclei (AGN) activity. \citet{Ferrari06} observed enhanced SF in a colliding cluster, while \citet{hwang&lee09} detected excess SF and AGN activity within substructures of a merging binary cluster. Additionally, \citet{Sobral15} and \citet{Stroe15} linked SF enhancement in local merging clusters to shocks associated with the merging process. Contrasting mergers with relaxed clusters, \citet{Stroe21} found an even distribution of SF galaxies within 3\,Mpc of merging cluster centers but noted their rarity in relaxed clusters. Interestingly, they observed AGN primarily in relaxed cluster outskirts \citep[$\sim$1.5--3\,Mpc; see also][]{Koulouridis2018,Koulouridis2024,Koulouridis2019}. However, \citet{Wittman24} found no correlation between star-forming/AGN fractions and time since pericenter. As suggested by \citet{Stroe15}, the shock fronts in merging clusters may directly promote SF by facilitating molecular gas collapse or by fueling black holes. To investigate whether cluster mergers promote galaxy mergers, we computed Pearson correlation coefficients and corresponding $p$-values between the fraction of ongoing/post mergers within R$_{200}$ and the six dynamical state parameters from \citet{yuan22}. Overall, we observed weak correlations ($p$-values\,$<$\,0.11--54) between the $f_M$ and all six parameters, with slightly higher fractions of ongoing mergers in `relaxed' clusters. However, fractions of post-mergers showed mixed correlations. Our sample size may limit statistically robust conclusions. Notably, we lack `relaxed' cluster samples with lower (negative) morphology index ($\delta$) values. Nonetheless, the consistent correlations across all six parameter spaces warrant further investigation.

Overall, we conclude that slow encounters in the outskirts of galaxy clusters, potentially with a higher frequency in relaxed clusters, promote galaxy mergers.

\begin{acknowledgments}

We would like to thank the anonymous reviewer for his/her valuable comments and suggestions, which helped to improve this manuscript.
D.K. acknowledges support from the Korea Astronomy and Space Science Institute under the R\&D
program (Project No. 2022-1-868-04) supervised by the Ministry of Science and ICT.
D.K. also acknowledges support from the National Research Foundation of Korea(NRF) grant funded by the Korean government(MSIT) (No. NRF-2022R1C1C2004506). 
D.K. thanks Prof. Soo-Chang Rey for constructive discussion and helpful suggestions.
YKS acknowledges support from the National Research Foundation of Korea (NRF) grant funded by the Ministry of Science and ICT (NRF-2019R1C1C1010279).
Y.J. acknowledges financial support from ANID BASAL project No. FB210003 and FONDECYT Regular project No. 1230441.
This research is based on observations at Cerro Tololo Inter-American Observatory, NSF’s NOIRLab (NOIRLab Prop. ID 2013A-0612; PI: Y. Sheen and ID 2014B-0608; PI: Y. Jaff\'e), which is managed by the Association of Universities for Research in Astronomy (AURA) under a cooperative agreement with the National Science Foundation.
K.K. acknowledges full financial support from ANID through FONDECYT Postdoctrorado Project 3200139.
JPC acknowledges financial support from ANID through FONDECYT Postdoctorado Project 3210709.
J.N. acknowledges support by UNAB internal grant DI-07-22/REG.
R.D. gratefully acknowledges support by the ANID BASAL project FB210003.
ET acknowledges support from: ANID through Millennium Science Initiative Program NCN19\_058, CATA-BASAL ACE210002 and FB210003, FONDECYT Regular 1190818 and 1200495.
S.K.Y. acknowledges support from the Korean National Research Foundation (2020R1A2C3003769, 2022R1A6A1A03053472).
This research made use of the ``K-corrections calculator'' service available at \url{http://kcor.sai.msu.ru/}

\end{acknowledgments}

\vspace{5mm}
\facilities{CTIO:4m Blanco, ROSAT, Sloan, XMM-Newton, Chandra}


\software{\code{Astropy} \citep{astropy},  
        \code{wxPython},
          \SExtractor \citep{sextractor},
          \code{statsmodel},
          \code{IDL},
          \code{GALAPAGOS-2} \citep{galapagos},
          \code{mclust} \citep{mclust}, 
          \code{DS+} \citep{DS+}
          }



\appendix

\section{Spectroscopic completeness} \label{sec:speccomp}

\begin{figure*}
 \gridline{ \fig{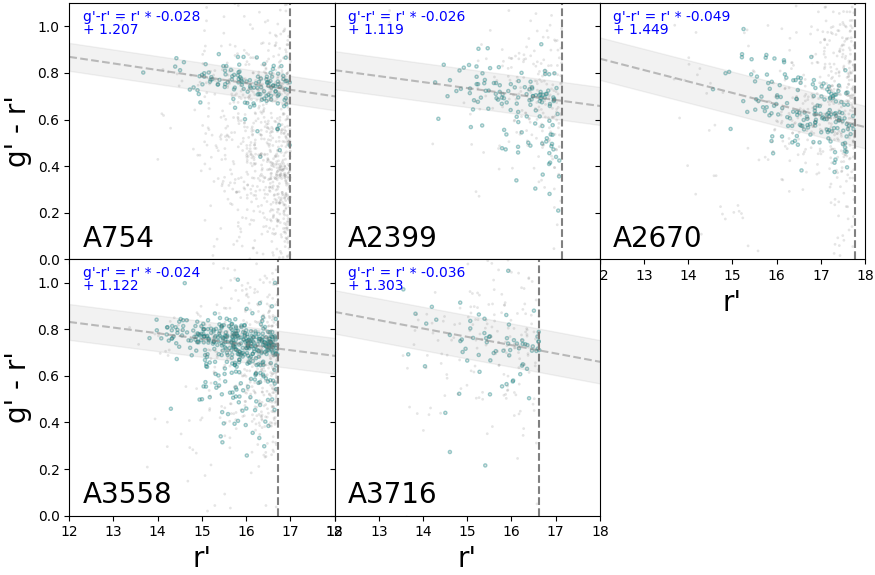}{0.4\textwidth}{}
          \fig{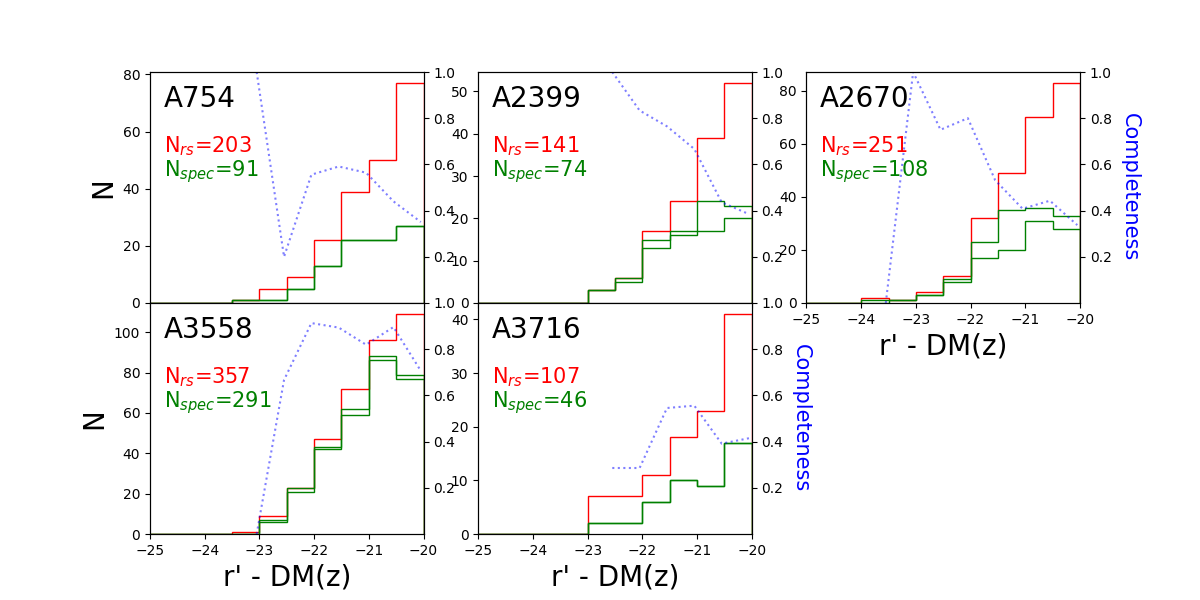}{0.6\textwidth}{}
          }
\caption{(\emph{left}) Color-magnitude diagram of sources having their absolute magnitude in $r^{\prime}$ greater than $-$20 in our photometric catalog. Large teal-colored circles are sources that are cross-matched with the spectroscopic members (see Table~\ref{tab:spec}). Vertical dashed lines are where $M_r=-$20, which is the value we adopt to set our volume-limited sample. Horizontal dashed lines are converged \code{WLS}-fitted red sequences of photometric sources inside the one-sigma scatter region (shaded region). The fitting process was repeated multiple times, excluding the sources outside of the 1~$\sigma$ range (grey shaded region), where the onset box range was 0.5\,$<$\,$g^{\prime}-r^{\prime}$\,$<$\,1.0 and $M_{r^{\prime}}$\,$<$\,$-$20. (\emph{right}) Spectroscopic completeness (dotted blue) histogram comparing numbers of photometric (red) and spectroscopic (green) sources inside a 1~$\sigma$ shaded red-sequence strip on the left plot. The spectroscopic surveys cover $>$40\% of the photometric red-sequence sources brighter than $M_r = -20$.\label{fig:cmd}}
\end{figure*}

To measure the spectroscopic completeness, we compare the numbers of photometric and spectroscopic sources inside the red sequence strips. To find the red-sequence galaxies in each cluster, we linearly fit the $g^{\prime}-r^{\prime}$ colors of sources that are cross-matched against the spectroscopic surveys (Table~\ref{tab:spec}) as a function of $r^{\prime}$. We use $K$-corrected magnitudes:
\begin{equation}
    m_{r^{\prime}, kc} = m_{r^{\prime}, MEC} + kcorr(m_{r^{\prime}, MEC}, z, m_{g^{\prime}, MEC}-m_{r^{\prime}, MEC}) ,
\end{equation}
where $m_{g^{\prime}, MEC}$ and $m_{r^{\prime}, MEC}$ are the Milky-Way extinction-corrected (see the last column of Table~\ref{tab:clusters}) magnitudes from the photometric catalog\footref{website}, and the $kcorr$ is the $K$-correction function which we calculated with a Python code downloaded from the website\footnote{\url{http://kcor.sai.msu.ru/}} \citep{kcorr1, kcorr2}. We use the \code{Weighted Least Squares (WLS)} code\footnote{\url{https://www.statsmodels.org/dev/generated/statsmodels.regression.linear_model.WLS.html}} to fit the red sequence beginning with the sources that have:
\begin{itemize}
\item $m_{g^{\prime}, kc}-m_{r^{\prime}, kc} > 0.5$,
\item $m_{g^{\prime}, kc}-m_{r^{\prime}, kc} < 1.0$, and
\item $M_r^{\prime} < -20$ , 
\end{itemize}
where 
\begin{equation}
    M_r = m_{r^{\prime}, kc} - \mathrm{DM}(z) ,
\label{eq:absmag}
\end{equation}
being DM the distance modulus, 37.002, 37.138, 37.762, 36.716, and 36.630 for A754, A2399, A2670, A3558, and A3716, respectively, as a function of the redshift. We then repeat the fitting, excluding sources outside of the 1-$\sigma$-scatter ranges until the fitted slope converges so that the change becomes less than 0.001.  The result with the 1\,$\sigma$ scatter is shown on the left in Figure~\ref{fig:cmd}. The sources inside the 1\,$\sigma$ strips are shown as red histograms on the right panels in Figure~\ref{fig:cmd}. The green histograms are the numbers of sources cross-matched with the public spectroscopic catalogs. Around half of the photometric red-sequence sources brighter than $M_{r^{\prime}} < -20$ were covered by the spectroscopic surveys.

\section{Background estimation} \label{sec:back}

\begin{figure}
\center
\includegraphics[width=0.45\textwidth]{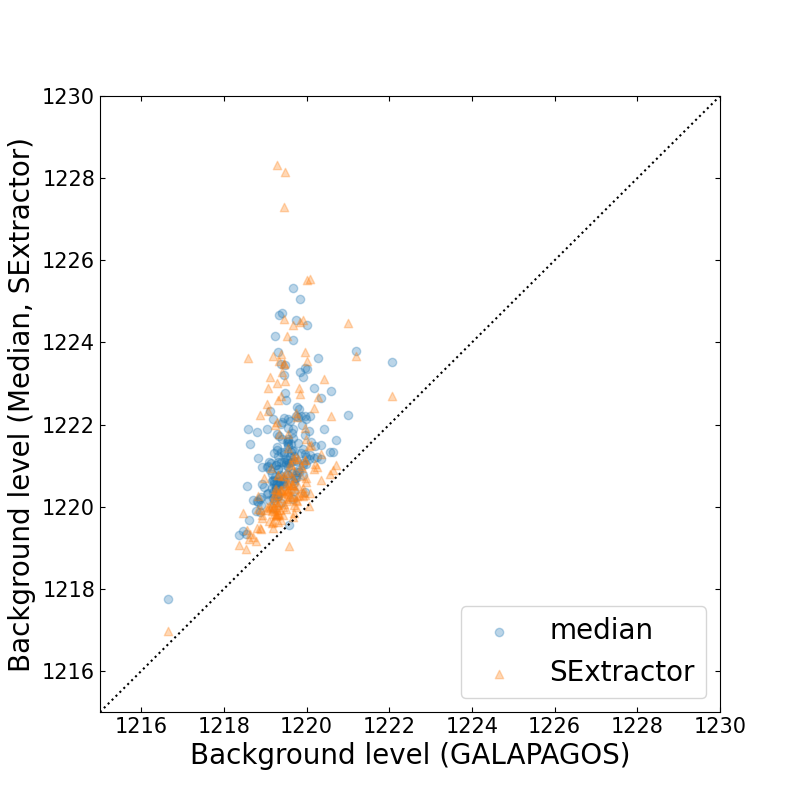}
\caption{The background values are estimated differently. The x-axis shows the values from \code{Galapagos-2}. The y-axis shows the median values in the cutout boxes (blue circles) and  the values from \SExtractor (orange triangles).\label{fig:back}}
\end{figure}

We compared the background levels from various techniques: 1) median pixel values, 2) BACKGROUND values from \SExtractor, and 3) `Sky background' values from \code{Galapagos-2}.
Figure~\ref{fig:back} shows that 1) and 2) have larger values than 3). We checked that the faint features become more prominent when we subtract the background values from \code{Galapagos-2}. Therefore, we adopted the background level from 3) when we generate galaxy stamp images for the visual inspection.

\section{\code{DS+}} \label{sec:dsp}

Figures~\ref{fig:rosat_dsp1}--\ref{fig:rosat_dsp2} are the same as Figures~\ref{fig:rosat1}--\ref{fig:rosat2} but the substructures are found by \code{DS+}.

\begin{figure*}
 \gridline{ \fig{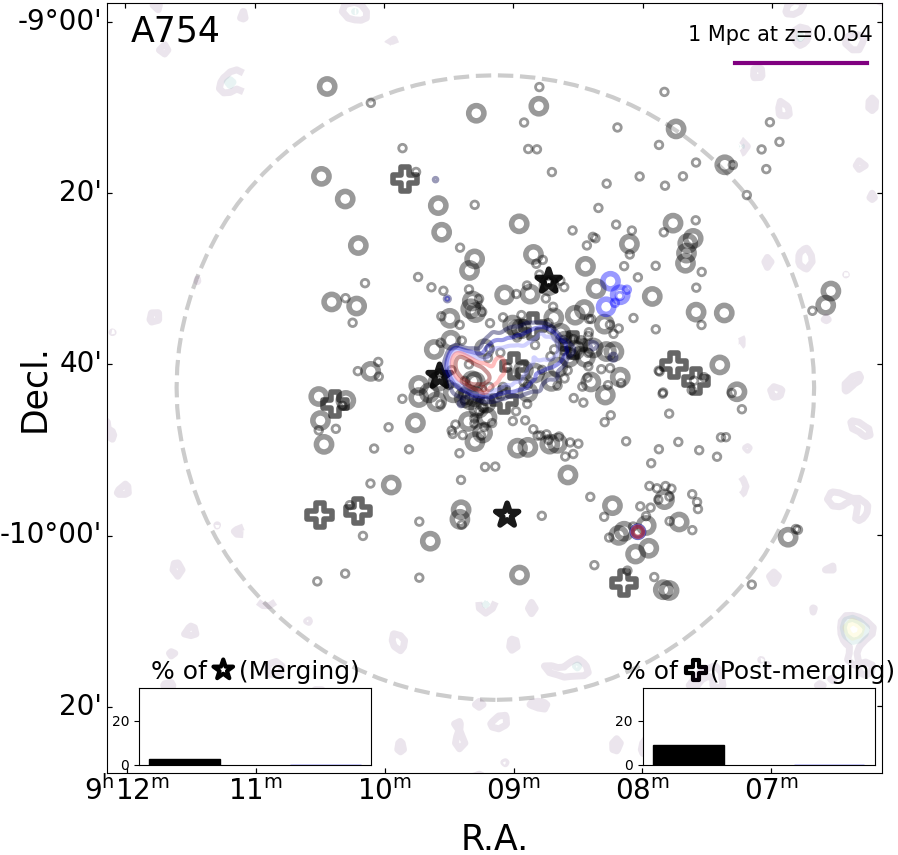}{0.45\textwidth}{}
          \fig{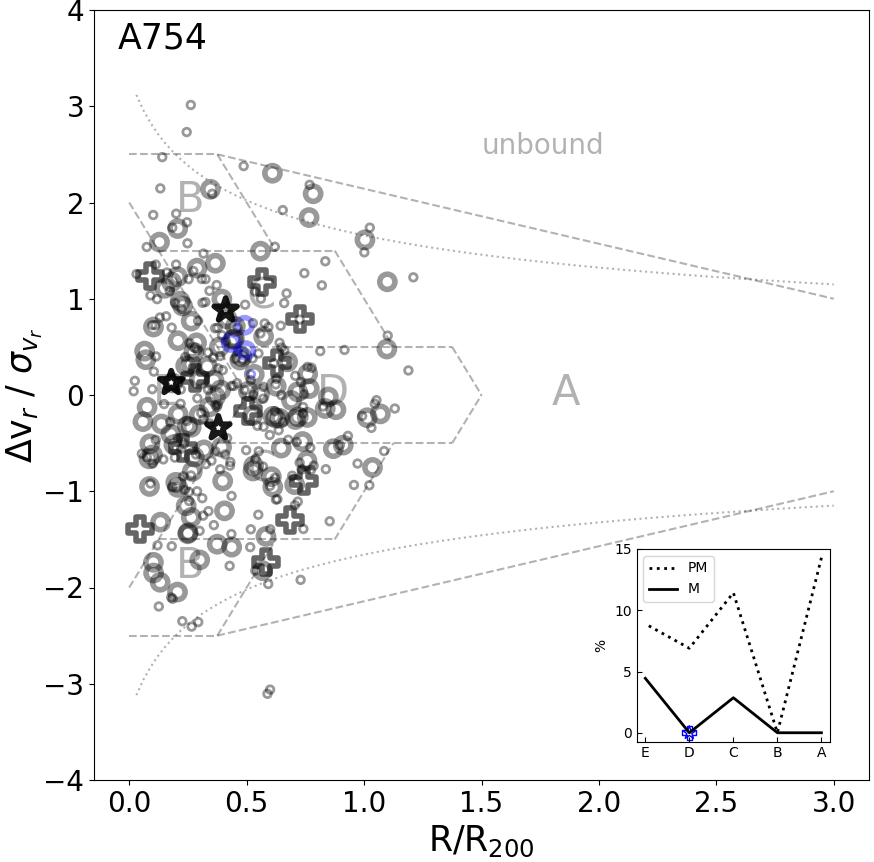}{0.45\textwidth}{}
          }
\vspace{-1cm}
\gridline{ \fig{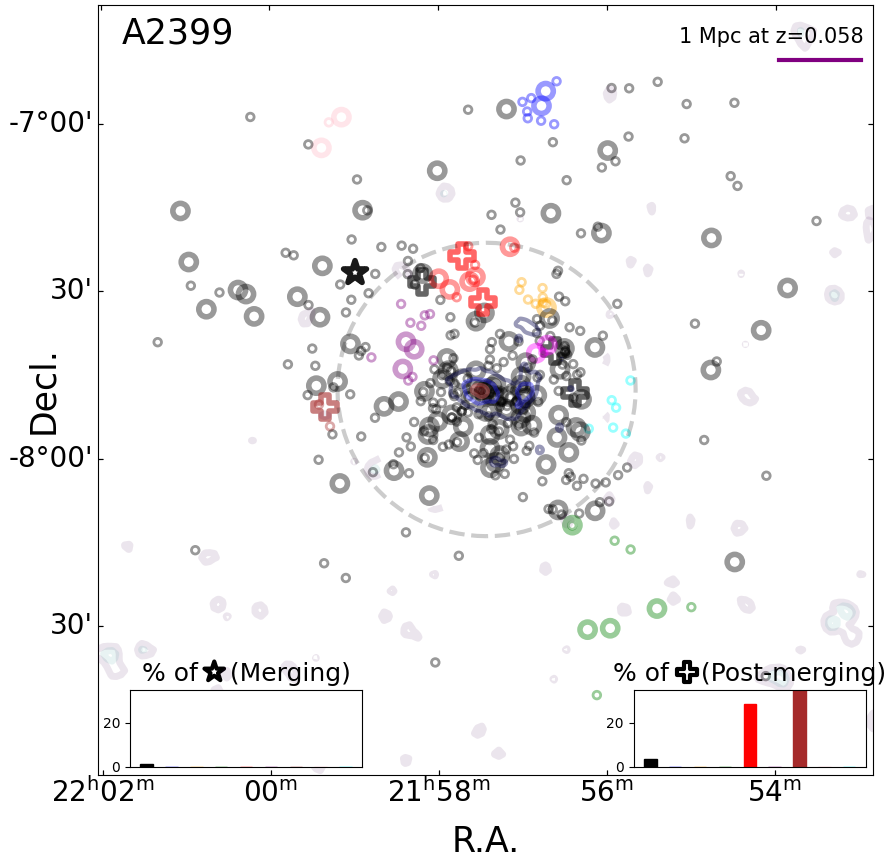}{0.45\textwidth}{}
          \fig{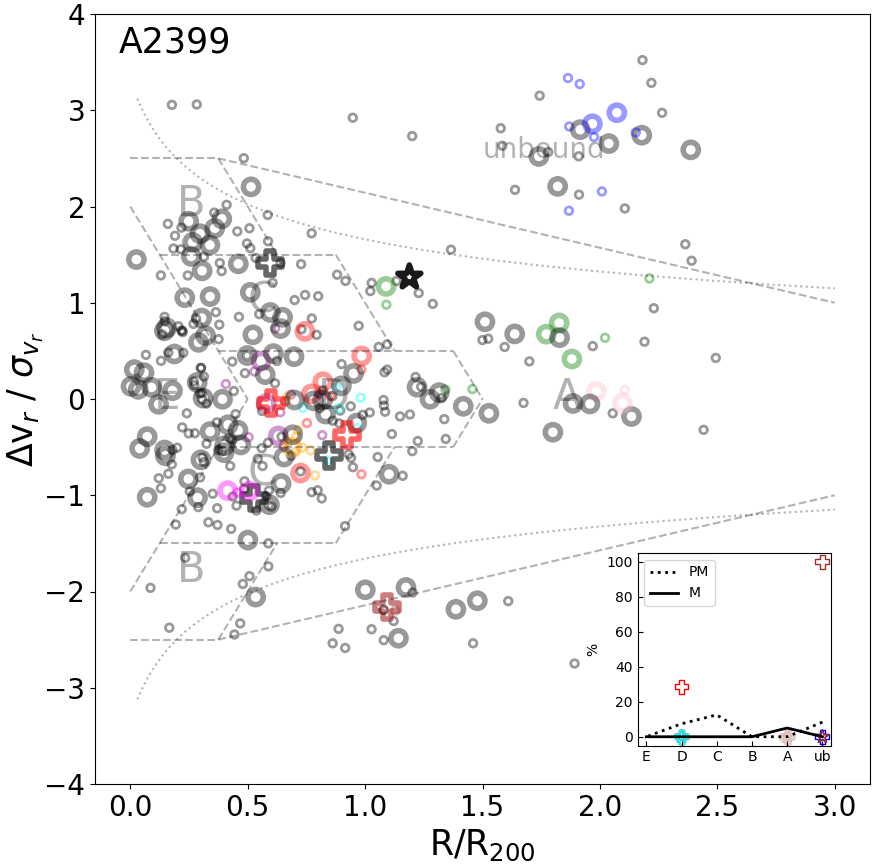}{0.45\textwidth}{}
          }
\vspace{-1cm}
\gridline{ \fig{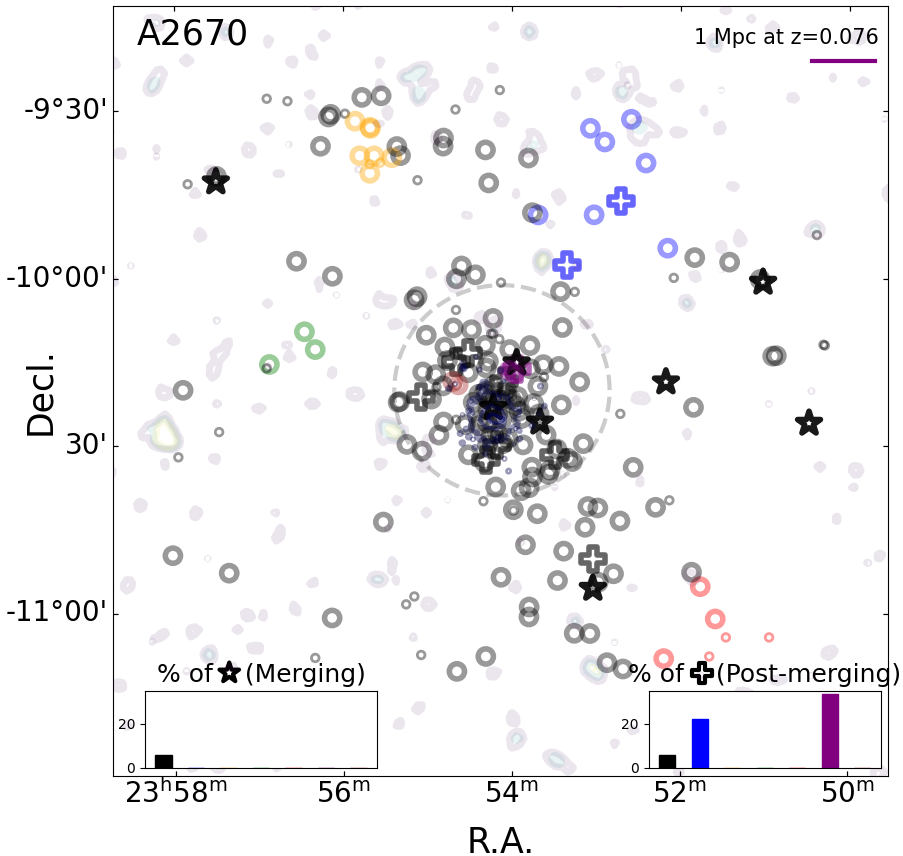}{0.45\textwidth}{}
            \fig{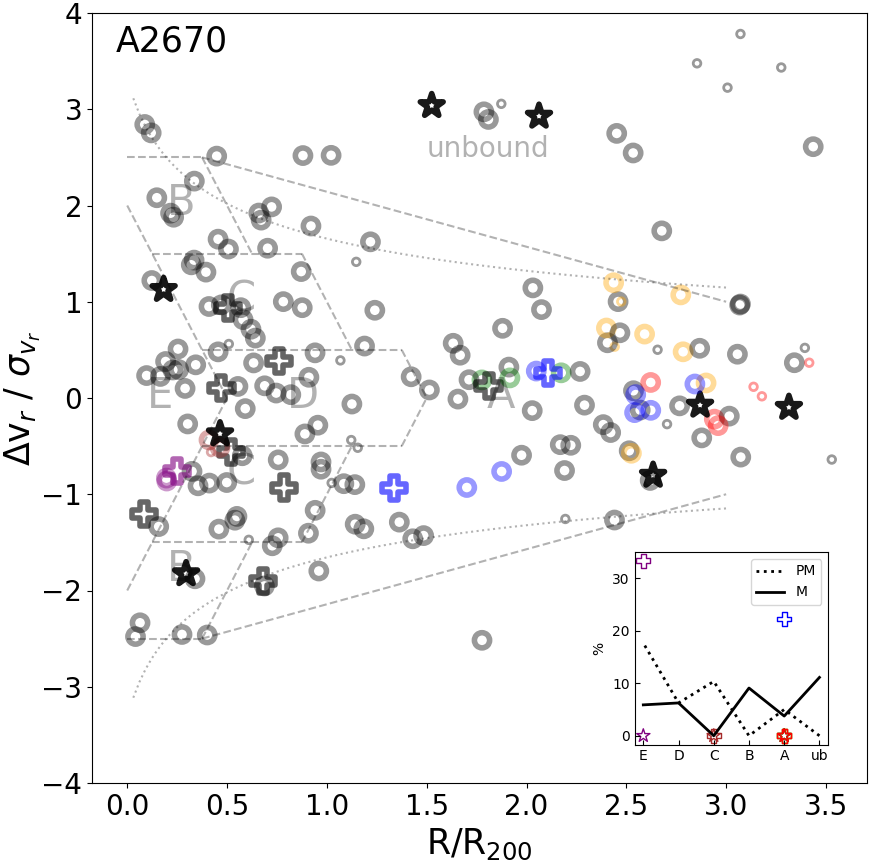}{0.45\textwidth}{}
          }
\vspace{-1cm}
\caption{Same as Figure~\ref{fig:rosat1} but substructure membership (colors) is defined by \code{DS+} algorithm which is an upgraded version of a method by \citet{ds}. Black symbols are main cluster members, and the color symbols are sources associated with the substructures when we input the value of the `Plim\_P' argument as 0.1.\label{fig:rosat_dsp1}}
\end{figure*}

\begin{figure*}
 \gridline{ \fig{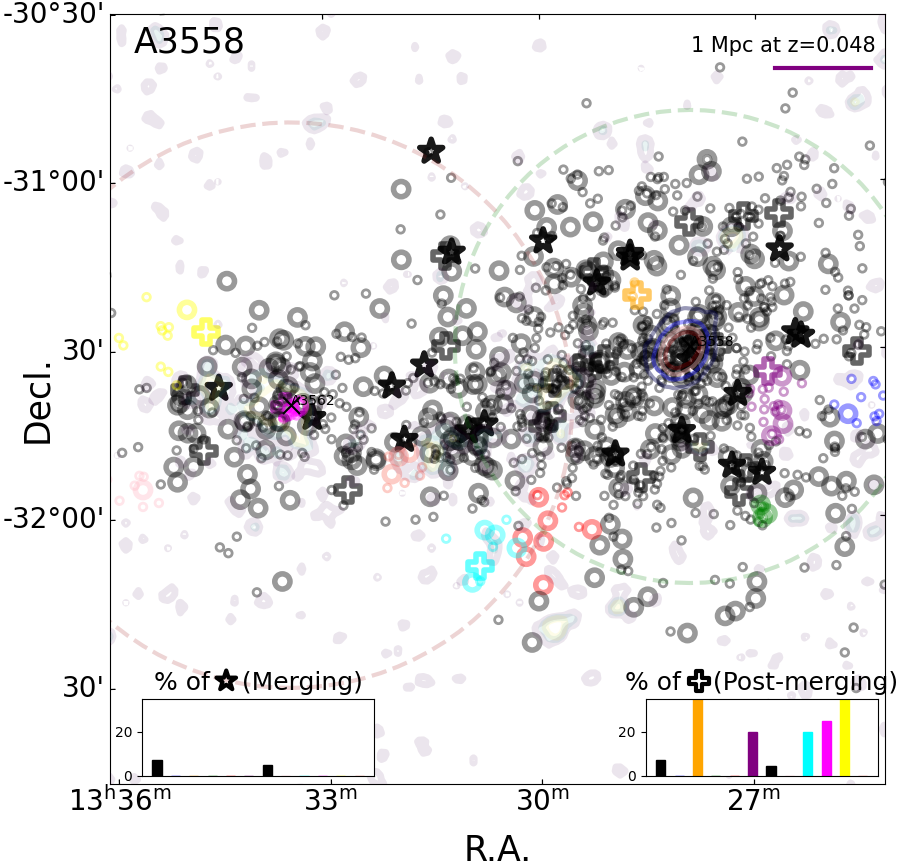}{0.45\textwidth}{}
          \fig{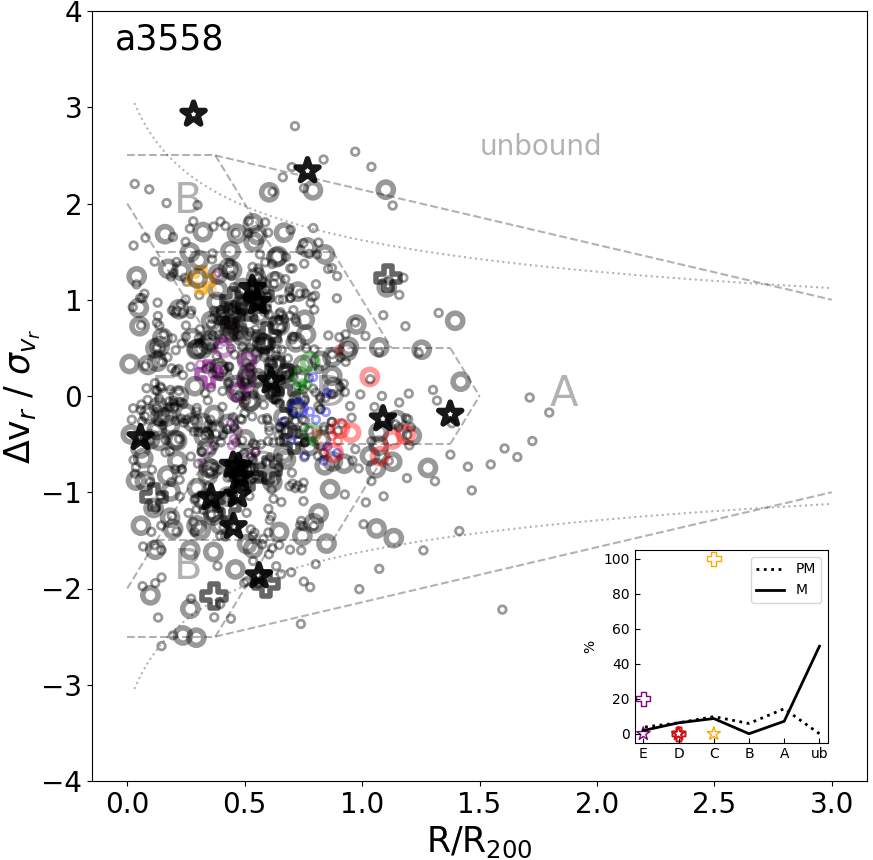}{0.45\textwidth}{}
          }
\vspace{-1cm}
\gridline{ \fig{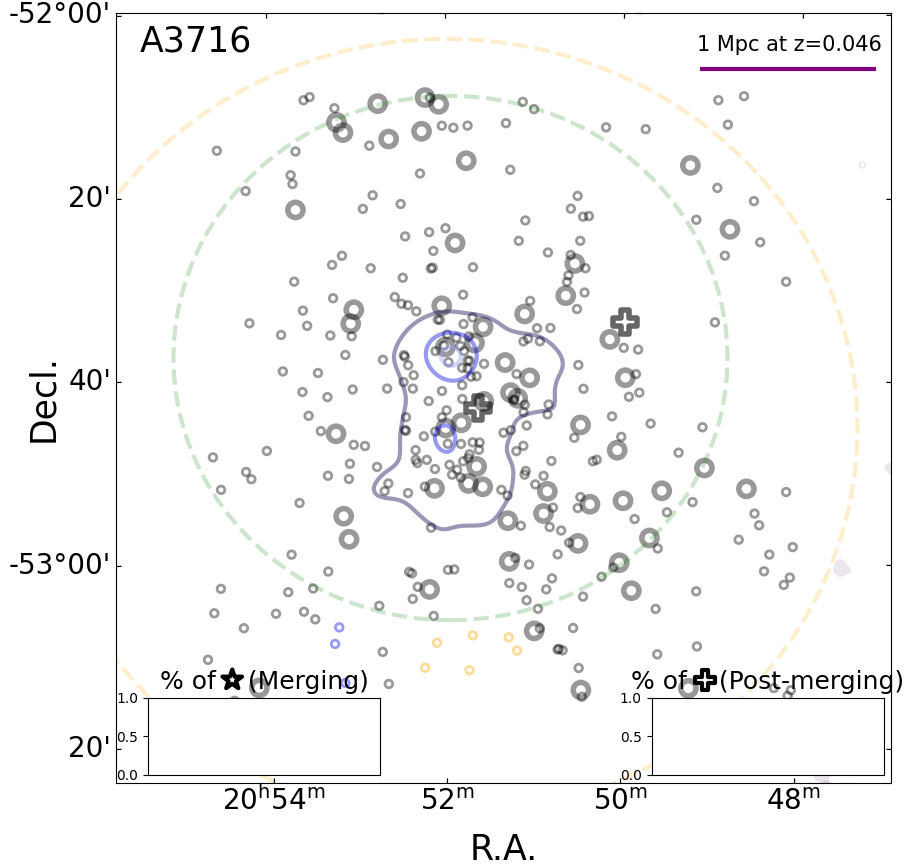}{0.45\textwidth}{}
          \fig{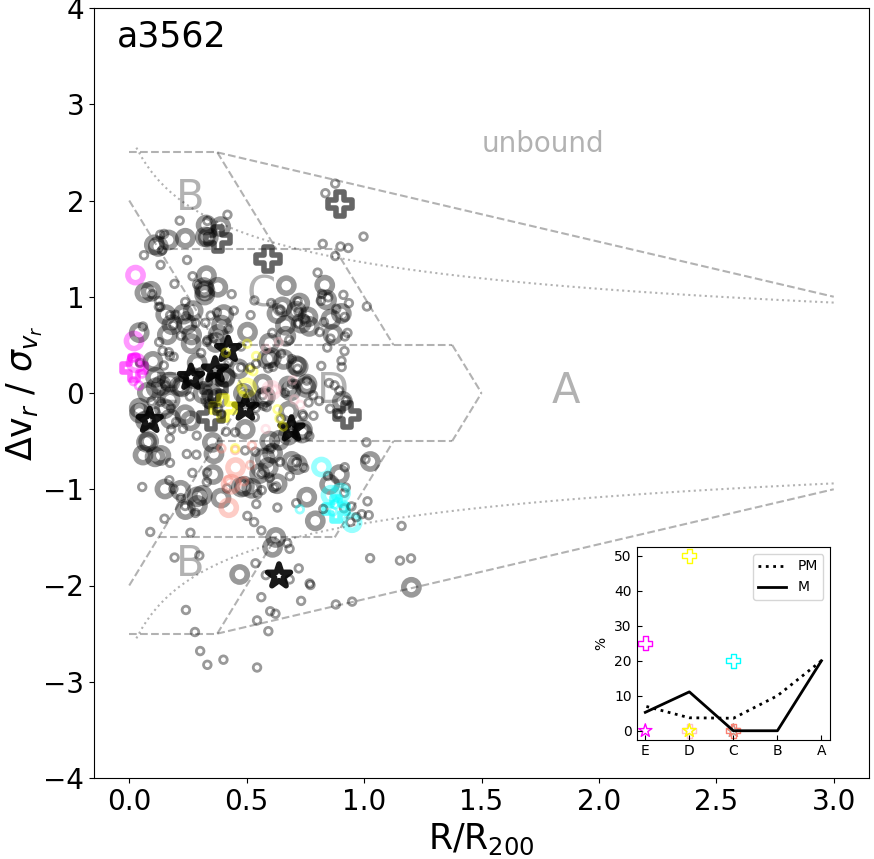}{0.45\textwidth}{}
          }
\vspace{-1cm}
\gridline{ 
          \fig{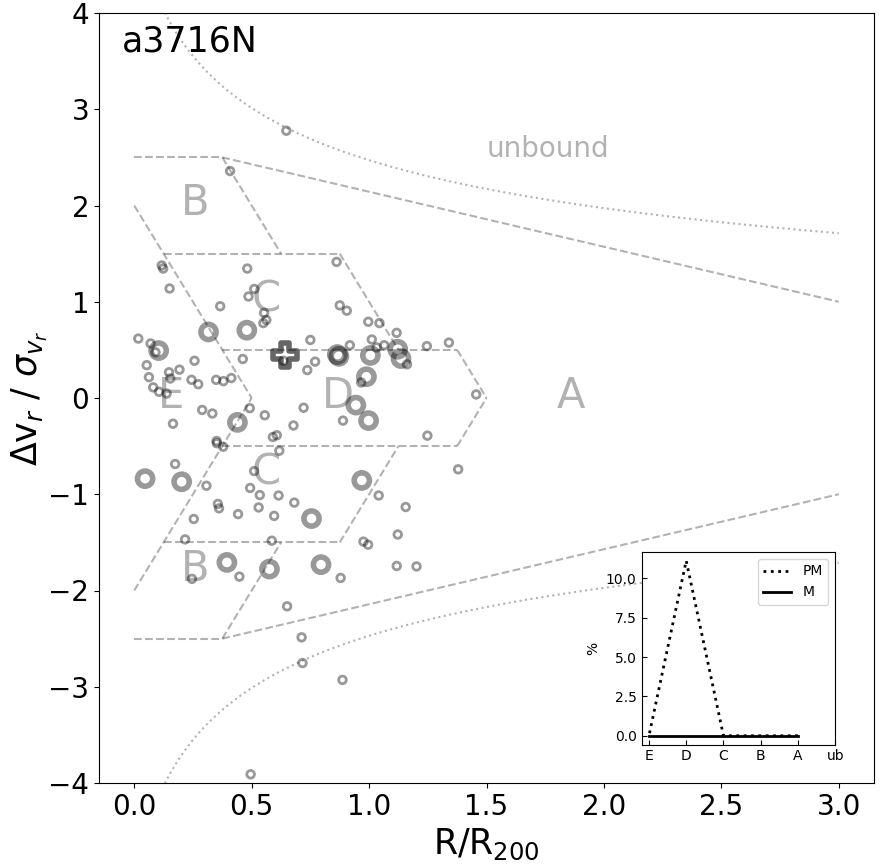}{0.45\textwidth}{}
          \fig{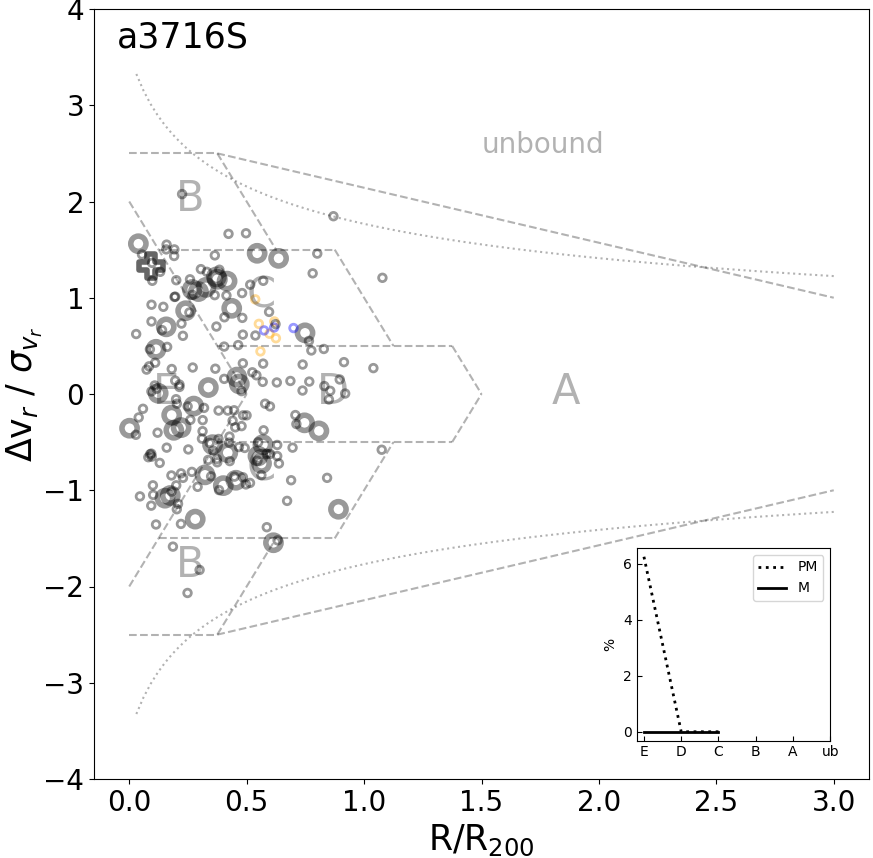}{0.45\textwidth}{}
          }
\caption{continued from Figure~\ref{fig:rosat_dsp1}, but the X-ray data of A3716 is from XMM-Newton.\label{fig:rosat_dsp2}}
\end{figure*}

\bibliography{ms}{}

\begin{thebibliography}{}
\expandafter\ifx\csname natexlab\endcsname\relax\def\natexlab#1{#1}\fi
\providecommand{\url}[1]{\href{#1}{#1}}
\providecommand{\dodoi}[1]{doi:~\href{http://doi.org/#1}{\nolinkurl{#1}}}
\providecommand{\doeprint}[1]{\href{http://ascl.net/#1}{\nolinkurl{http://ascl.net/#1}}}
\providecommand{\doarXiv}[1]{\href{https://arxiv.org/abs/#1}{\nolinkurl{https://arxiv.org/abs/#1}}}

\bibitem[{{Ahumada} {et~al.}(2020){Ahumada}, {Allende Prieto}, {Almeida}, {Anders}, {Anderson}, {Andrews}, {Anguiano}, {Arcodia}, {Armengaud}, {Aubert}, {Avila}, {Avila-Reese}, {Badenes}, {Balland}, {Barger}, {Barrera-Ballesteros}, {Basu}, {Bautista}, {Beaton}, {Beers}, {Benavides}, {Bender}, {Bernardi}, {Bershady}, {Beutler}, {Bidin}, {Bird}, {Bizyaev}, {Blanc}, {Blanton}, {Boquien}, {Borissova}, {Bovy}, {Brandt}, {Brinkmann}, {Brownstein}, {Bundy}, {Bureau}, {Burgasser}, {Burtin}, {Cano-D{\'\i}az}, {Capasso}, {Cappellari}, {Carrera}, {Chabanier}, {Chaplin}, {Chapman}, {Cherinka}, {Chiappini}, {Doohyun Choi}, {Chojnowski}, {Chung}, {Clerc}, {Coffey}, {Comerford}, {Comparat}, {da Costa}, {Cousinou}, {Covey}, {Crane}, {Cunha}, {Ilha}, {Dai}, {Damsted}, {Darling}, {Davidson}, {Davies}, {Dawson}, {De}, {de la Macorra}, {De Lee}, {Queiroz}, {Deconto Machado}, {de la Torre}, {Dell'Agli}, {du Mas des Bourboux}, {Diamond-Stanic}, {Dillon}, {Donor}, {Drory}, {Duckworth}, {Dwelly}, {Ebelke}, {Eftekharzadeh}, {Davis
  Eigenbrot}, {Elsworth}, {Eracleous}, {Erfanianfar}, {Escoffier}, {Fan}, {Farr}, {Fern{\'a}ndez-Trincado}, {Feuillet}, {Finoguenov}, {Fofie}, {Fraser-McKelvie}, {Frinchaboy}, {Fromenteau}, {Fu}, {Galbany}, {Garcia}, {Garc{\'\i}a-Hern{\'a}ndez}, {Garma Oehmichen}, {Ge}, {Geimba Maia}, {Geisler}, {Gelfand}, {Goddy}, {Gonzalez-Perez}, {Grabowski}, {Green}, {Grier}, {Guo}, {Guy}, {Harding}, {Hasselquist}, {Hawken}, {Hayes}, {Hearty}, {Hekker}, {Hogg}, {Holtzman}, {Horta}, {Hou}, {Hsieh}, {Huber}, {Hunt}, {Ider Chitham}, {Imig}, {Jaber}, {Jimenez Angel}, {Johnson}, {Jones}, {J{\"o}nsson}, {Jullo}, {Kim}, {Kinemuchi}, {Kirkpatrick}, {Kite}, {Klaene}, {Kneib}, {Kollmeier}, {Kong}, {Kounkel}, {Krishnarao}, {Lacerna}, {Lan}, {Lane}, {Law}, {Le Goff}, {Leung}, {Lewis}, {Li}, {Lian}, {Lin}, {Long}, {Longa-Pe{\~n}a}, {Lundgren}, {Lyke}, {Mackereth}, {MacLeod}, {Majewski}, {Manchado}, {Maraston}, {Martini}, {Masseron}, {Masters}, {Mathur}, {McDermid}, {Merloni}, {Merrifield}, {M{\'e}sz{\'a}ros}, {Miglio}, {Minniti},
  {Minsley}, {Miyaji}, {Mohammad}, {Mosser}, {Mueller}, {Muna}, {Mu{\~n}oz-Guti{\'e}rrez}, {Myers}, {Nadathur}, {Nair}, {Nandra}, {Correa do Nascimento}, {Nevin}, {Newman}, {Nidever}, {Nitschelm}, {Noterdaeme}, {O'Connell}, {Olmstead}, {Oravetz}, {Oravetz}, {Osorio}, {Pace}, {Padilla}, {Palanque-Delabrouille}, {Palicio}, {Pan}, {Pan}, {Parker}, {Paviot}, {Peirani}, {Ram{\'r}ez}, {Penny}, {Percival}, {Perez-Fournon}, {P{\'e}rez-R{\`a}fols}, {Petitjean}, {Pieri}, {Pinsonneault}, {Poovelil}, {Povick}, {Prakash}, {Price-Whelan}, {Raddick}, {Raichoor}, {Ray}, {Rembold}, {Rezaie}, {Riffel}, {Riffel}, {Rix}, {Robin}, {Roman-Lopes}, {Rom{\'a}n-Z{\'u}{\~n}iga}, {Rose}, {Ross}, {Rossi}, {Rowlands}, {Rubin}, {Salvato}, {S{\'a}nchez}, {S{\'a}nchez-Menguiano}, {S{\'a}nchez-Gallego}, {Sayres}, {Schaefer}, {Schiavon}, {Schimoia}, {Schlafly}, {Schlegel}, {Schneider}, {Schultheis}, {Schwope}, {Seo}, {Serenelli}, {Shafieloo}, {Shamsi}, {Shao}, {Shen}, {Shetrone}, {Shirley}, {Silva Aguirre}, {Simon}, {Skrutskie}, {Slosar},
  {Smethurst}, {Sobeck}, {Sodi}, {Souto}, {Stark}, {Stassun}, {Steinmetz}, {Stello}, {Stermer}, {Storchi-Bergmann}, {Streblyanska}, {Stringfellow}, {Stutz}, {Su{\'a}rez}, {Sun}, {Taghizadeh-Popp}, {Talbot}, {Tayar}, {Thakar}, {Theriault}, {Thomas}, {Thomas}, {Tinker}, {Tojeiro}, {Toledo}, {Tremonti}, {Troup}, {Tuttle}, {Unda-Sanzana}, {Valentini}, {Vargas-Gonz{\'a}lez}, {Vargas-Maga{\~n}a}, {V{\'a}zquez-Mata}, {Vivek}, {Wake}, {Wang}, {Weaver}, {Weijmans}, {Wild}, {Wilson}, {Wilson}, {Wolthuis}, {Wood-Vasey}, {Yan}, {Yang}, {Y{\`e}che}, {Zamora}, {Zarrouk}, {Zasowski}, {Zhang}, {Zhao}, {Zhao}, {Zheng}, {Zheng}, {Zhu}, \& {Zou}}]{sdssdr16}
{Ahumada}, R., {Allende Prieto}, C., {Almeida}, A., {et~al.} 2020, \apjs, 249, 3, \dodoi{10.3847/1538-4365/ab929e}

\bibitem[{{Alam} {et~al.}(2015){Alam}, {Albareti}, {Allende Prieto}, {Anders}, {Anderson}, {Anderton}, {Andrews}, {Armengaud}, {Aubourg}, {Bailey}, {Basu}, {Bautista}, {Beaton}, {Beers}, {Bender}, {Berlind}, {Beutler}, {Bhardwaj}, {Bird}, {Bizyaev}, {Blake}, {Blanton}, {Blomqvist}, {Bochanski}, {Bolton}, {Bovy}, {Shelden Bradley}, {Brandt}, {Brauer}, {Brinkmann}, {Brown}, {Brownstein}, {Burden}, {Burtin}, {Busca}, {Cai}, {Capozzi}, {Carnero Rosell}, {Carr}, {Carrera}, {Chambers}, {Chaplin}, {Chen}, {Chiappini}, {Chojnowski}, {Chuang}, {Clerc}, {Comparat}, {Covey}, {Croft}, {Cuesta}, {Cunha}, {da Costa}, {Da Rio}, {Davenport}, {Dawson}, {De Lee}, {Delubac}, {Deshpande}, {Dhital}, {Dutra-Ferreira}, {Dwelly}, {Ealet}, {Ebelke}, {Edmondson}, {Eisenstein}, {Ellsworth}, {Elsworth}, {Epstein}, {Eracleous}, {Escoffier}, {Esposito}, {Evans}, {Fan}, {Fern{\'a}ndez-Alvar}, {Feuillet}, {Filiz Ak}, {Finley}, {Finoguenov}, {Flaherty}, {Fleming}, {Font-Ribera}, {Foster}, {Frinchaboy}, {Galbraith-Frew}, {Garc{\'\i}a},
  {Garc{\'\i}a-Hern{\'a}ndez}, {Garc{\'\i}a P{\'e}rez}, {Gaulme}, {Ge}, {G{\'e}nova-Santos}, {Georgakakis}, {Ghezzi}, {Gillespie}, {Girardi}, {Goddard}, {Gontcho}, {Gonz{\'a}lez Hern{\'a}ndez}, {Grebel}, {Green}, {Grieb}, {Grieves}, {Gunn}, {Guo}, {Harding}, {Hasselquist}, {Hawley}, {Hayden}, {Hearty}, {Hekker}, {Ho}, {Hogg}, {Holley-Bockelmann}, {Holtzman}, {Honscheid}, {Huber}, {Huehnerhoff}, {Ivans}, {Jiang}, {Johnson}, {Kinemuchi}, {Kirkby}, {Kitaura}, {Klaene}, {Knapp}, {Kneib}, {Koenig}, {Lam}, {Lan}, {Lang}, {Laurent}, {Le Goff}, {Leauthaud}, {Lee}, {Lee}, {Licquia}, {Liu}, {Long}, {L{\'o}pez-Corredoira}, {Lorenzo-Oliveira}, {Lucatello}, {Lundgren}, {Lupton}, {Mack}, {Mahadevan}, {Maia}, {Majewski}, {Malanushenko}, {Malanushenko}, {Manchado}, {Manera}, {Mao}, {Maraston}, {Marchwinski}, {Margala}, {Martell}, {Martig}, {Masters}, {Mathur}, {McBride}, {McGehee}, {McGreer}, {McMahon}, {M{\'e}nard}, {Menzel}, {Merloni}, {M{\'e}sz{\'a}ros}, {Miller}, {Miralda-Escud{\'e}}, {Miyatake}, {Montero-Dorta}, {More},
  {Morganson}, {Morice-Atkinson}, {Morrison}, {Mosser}, {Muna}, {Myers}, {Nandra}, {Newman}, {Neyrinck}, {Nguyen}, {Nichol}, {Nidever}, {Noterdaeme}, {Nuza}, {O'Connell}, {O'Connell}, {O'Connell}, {Ogando}, {Olmstead}, {Oravetz}, {Oravetz}, {Osumi}, {Owen}, {Padgett}, {Padmanabhan}, {Paegert}, {Palanque-Delabrouille}, {Pan}, {Parejko}, {P{\^a}ris}, {Park}, {Pattarakijwanich}, {Pellejero-Ibanez}, {Pepper}, {Percival}, {P{\'e}rez-Fournon}, {P{\'e}rez-R{\`a}fols}, {Petitjean}, {Pieri}, {Pinsonneault}, {Porto de Mello}, {Prada}, {Prakash}, {Price-Whelan}, {Protopapas}, {Raddick}, {Rahman}, {Reid}, {Rich}, {Rix}, {Robin}, {Rockosi}, {Rodrigues}, {Rodr{\'\i}guez-Torres}, {Roe}, {Ross}, {Ross}, {Rossi}, {Ruan}, {Rubi{\~n}o-Mart{\'\i}n}, {Rykoff}, {Salazar-Albornoz}, {Salvato}, {Samushia}, {S{\'a}nchez}, {Santiago}, {Sayres}, {Schiavon}, {Schlegel}, {Schmidt}, {Schneider}, {Schultheis}, {Schwope}, {Sc{\'o}ccola}, {Scott}, {Sellgren}, {Seo}, {Serenelli}, {Shane}, {Shen}, {Shetrone}, {Shu}, {Silva Aguirre}, {Sivarani},
  {Skrutskie}, {Slosar}, {Smith}, {Sobreira}, {Souto}, {Stassun}, {Steinmetz}, {Stello}, {Strauss}, {Streblyanska}, {Suzuki}, {Swanson}, {Tan}, {Tayar}, {Terrien}, {Thakar}, {Thomas}, {Thomas}, {Thompson}, {Tinker}, {Tojeiro}, {Troup}, {Vargas-Maga{\~n}a}, {Vazquez}, {Verde}, {Viel}, {Vogt}, {Wake}, {Wang}, {Weaver}, {Weinberg}, {Weiner}, {White}, {Wilson}, {Wisniewski}, {Wood-Vasey}, {Ye`che}, {York}, {Zakamska}, {Zamora}, {Zasowski}, {Zehavi}, {Zhao}, {Zheng}, {Zhou}, {Zhou}, {Zou}, \& {Zhu}}]{alam15}
{Alam}, S., {Albareti}, F.~D., {Allende Prieto}, C., {et~al.} 2015, \apjs, 219, 12, \dodoi{10.1088/0067-0049/219/1/12}

\bibitem[{{Astropy Collaboration} {et~al.}(2013){Astropy Collaboration}, {Robitaille}, {Tollerud}, {Greenfield}, {Droettboom}, {Bray}, {Aldcroft}, {Davis}, {Ginsburg}, {Price-Whelan}, {Kerzendorf}, {Conley}, {Crighton}, {Barbary}, {Muna}, {Ferguson}, {Grollier}, {Parikh}, {Nair}, {Unther}, {Deil}, {Woillez}, {Conseil}, {Kramer}, {Turner}, {Singer}, {Fox}, {Weaver}, {Zabalza}, {Edwards}, {Azalee Bostroem}, {Burke}, {Casey}, {Crawford}, {Dencheva}, {Ely}, {Jenness}, {Labrie}, {Lim}, {Pierfederici}, {Pontzen}, {Ptak}, {Refsdal}, {Servillat}, \& {Streicher}}]{astropy}
{Astropy Collaboration}, {Robitaille}, T.~P., {Tollerud}, E.~J., {et~al.} 2013, \aap, 558, A33, \dodoi{10.1051/0004-6361/201322068}

\bibitem[{{Benavides} {et~al.}(2023){Benavides}, {Biviano}, \& {Abadi}}]{DS+}
{Benavides}, J.~A., {Biviano}, A., \& {Abadi}, M.~G. 2023, \aap, 669, A147, \dodoi{10.1051/0004-6361/202245422}

\bibitem[{{Bertin} \& {Arnouts}(1996)}]{sextractor}
{Bertin}, E., \& {Arnouts}, S. 1996, \aaps, 117, 393, \dodoi{10.1051/aas:1996164}

\bibitem[{{B{\'\i}lek} {et~al.}(2020){B{\'\i}lek}, {Duc}, {Cuillandre}, {Gwyn}, {Cappellari}, {Bekaert}, {Bonfini}, {Bitsakis}, {Paudel}, {Krajnovi{\'c}}, {Durrell}, \& {Marleau}}]{Bilek2020}
{B{\'\i}lek}, M., {Duc}, P.-A., {Cuillandre}, J.-C., {et~al.} 2020, \mnras, 498, 2138, \dodoi{10.1093/mnras/staa2248}

\bibitem[{{Binney} \& {Tremaine}(1987)}]{binney&tremaine87}
{Binney}, J., \& {Tremaine}, S. 1987, {Galactic dynamics}

\bibitem[{{Bundy} {et~al.}(2009){Bundy}, {Fukugita}, {Ellis}, {Targett}, {Belli}, \& {Kodama}}]{Bundy2009}
{Bundy}, K., {Fukugita}, M., {Ellis}, R.~S., {et~al.} 2009, \apj, 697, 1369, \dodoi{10.1088/0004-637X/697/2/1369}

\bibitem[{{Cava} {et~al.}(2009){Cava}, {Bettoni}, {Poggianti}, {Couch}, {Moles}, {Varela}, {Biviano}, {D'Onofrio}, {Dressler}, {Fasano}, {Fritz}, {Kj{\ae}rgaard}, {Ramella}, \& {Valentinuzzi}}]{cava09}
{Cava}, A., {Bettoni}, D., {Poggianti}, B.~M., {et~al.} 2009, \aap, 495, 707, \dodoi{10.1051/0004-6361:200810997}

\bibitem[{{Chilingarian} {et~al.}(2010){Chilingarian}, {Melchior}, \& {Zolotukhin}}]{kcorr1}
{Chilingarian}, I.~V., {Melchior}, A.-L., \& {Zolotukhin}, I.~Y. 2010, \mnras, 405, 1409, \dodoi{10.1111/j.1365-2966.2010.16506.x}

\bibitem[{{Chilingarian} \& {Zolotukhin}(2012)}]{kcorr2}
{Chilingarian}, I.~V., \& {Zolotukhin}, I.~Y. 2012, \mnras, 419, 1727, \dodoi{10.1111/j.1365-2966.2011.19837.x}

\bibitem[{{Conselice} {et~al.}(2009){Conselice}, {Yang}, \& {Bluck}}]{Conselice2009}
{Conselice}, C.~J., {Yang}, C., \& {Bluck}, A.~F.~L. 2009, \mnras, 394, 1956, \dodoi{10.1111/j.1365-2966.2009.14396.x}

\bibitem[{{Das} {et~al.}(2023){Das}, {Pandey}, \& {Sarkar}}]{das23}
{Das}, A., {Pandey}, B., \& {Sarkar}, S. 2023, Research in Astronomy and Astrophysics, 23, 025016, \dodoi{10.1088/1674-4527/acab44}

\bibitem[{{Dekel} {et~al.}(2009){Dekel}, {Sari}, \& {Ceverino}}]{Dekel2009}
{Dekel}, A., {Sari}, R., \& {Ceverino}, D. 2009, \apj, 703, 785, \dodoi{10.1088/0004-637X/703/1/785}

\bibitem[{{DePoy} {et~al.}(2008){DePoy}, {Abbott}, {Annis}, {Antonik}, {Barcel{\'o}}, {Bernstein}, {Bigelow}, {Brooks}, {Buckley-Geer}, {Campa}, {Cardiel}, {Castander}, {Castilla}, {Cease}, {Chappa}, {Dede}, {Derylo}, {Diehl}, {Doel}, {DeVicente}, {Estrada}, {Finley}, {Flaugher}, {Gaztanaga}, {Gerdes}, {Gladders}, {Guarino}, {Gutierrez}, {Hamilton}, {Haney}, {Holland}, {Honscheid}, {Huffman}, {Karliner}, {Kau}, {Kent}, {Kozlovsky}, {Kubik}, {Kuehn}, {Kuhlmann}, {Kuk}, {Leger}, {Lin}, {Martinez}, {Martinez}, {Merritt}, {Mohr}, {Moore}, {Moore}, {Nord}, {Ogando}, {Olsen}, {Onal}, {Peoples}, {Qian}, {Roe}, {Sanchez}, {Scarpine}, {Schmidt}, {Schmitt}, {Schubnell}, {Schultz}, {Selen}, {Shaw}, {Simaitis}, {Slaughter}, {Smith}, {Spinka}, {Stefanik}, {Stuermer}, {Talaga}, {Tarle}, {Thaler}, {Tucker}, {Walker}, {Worswick}, \& {Zhao}}]{DePoy2008}
{DePoy}, D.~L., {Abbott}, T., {Annis}, J., {et~al.} 2008, in Society of Photo-Optical Instrumentation Engineers (SPIE) Conference Series, Vol. 7014, Ground-based and Airborne Instrumentation for Astronomy II, ed. I.~S. {McLean} \& M.~M. {Casali}, 70140E, \dodoi{10.1117/12.789466}

\bibitem[{{Dey} {et~al.}(2019){Dey}, {Schlegel}, {Lang}, {Blum}, {Burleigh}, {Fan}, {Findlay}, {Finkbeiner}, {Herrera}, {Juneau}, {Landriau}, {Levi}, {McGreer}, {Meisner}, {Myers}, {Moustakas}, {Nugent}, {Patej}, {Schlafly}, {Walker}, {Valdes}, {Weaver}, {Y{\`e}che}, {Zou}, {Zhou}, {Abareshi}, {Abbott}, {Abolfathi}, {Aguilera}, {Alam}, {Allen}, {Alvarez}, {Annis}, {Ansarinejad}, {Aubert}, {Beechert}, {Bell}, {BenZvi}, {Beutler}, {Bielby}, {Bolton}, {Brice{\~n}o}, {Buckley-Geer}, {Butler}, {Calamida}, {Carlberg}, {Carter}, {Casas}, {Castander}, {Choi}, {Comparat}, {Cukanovaite}, {Delubac}, {DeVries}, {Dey}, {Dhungana}, {Dickinson}, {Ding}, {Donaldson}, {Duan}, {Duckworth}, {Eftekharzadeh}, {Eisenstein}, {Etourneau}, {Fagrelius}, {Farihi}, {Fitzpatrick}, {Font-Ribera}, {Fulmer}, {G{\"a}nsicke}, {Gaztanaga}, {George}, {Gerdes}, {Gontcho}, {Gorgoni}, {Green}, {Guy}, {Harmer}, {Hernandez}, {Honscheid}, {Huang}, {James}, {Jannuzi}, {Jiang}, {Joyce}, {Karcher}, {Karkar}, {Kehoe}, {Kneib}, {Kueter-Young}, {Lan},
  {Lauer}, {Le Guillou}, {Le Van Suu}, {Lee}, {Lesser}, {Perreault Levasseur}, {Li}, {Mann}, {Marshall}, {Mart{\'\i}nez-V{\'a}zquez}, {Martini}, {du Mas des Bourboux}, {McManus}, {Meier}, {M{\'e}nard}, {Metcalfe}, {Mu{\~n}oz-Guti{\'e}rrez}, {Najita}, {Napier}, {Narayan}, {Newman}, {Nie}, {Nord}, {Norman}, {Olsen}, {Paat}, {Palanque-Delabrouille}, {Peng}, {Poppett}, {Poremba}, {Prakash}, {Rabinowitz}, {Raichoor}, {Rezaie}, {Robertson}, {Roe}, {Ross}, {Ross}, {Rudnick}, {Safonova}, {Saha}, {S{\'a}nchez}, {Savary}, {Schweiker}, {Scott}, {Seo}, {Shan}, {Silva}, {Slepian}, {Soto}, {Sprayberry}, {Staten}, {Stillman}, {Stupak}, {Summers}, {Sien Tie}, {Tirado}, {Vargas-Maga{\~n}a}, {Vivas}, {Wechsler}, {Williams}, {Yang}, {Yang}, {Yapici}, {Zaritsky}, {Zenteno}, {Zhang}, {Zhang}, {Zhou}, \& {Zhou}}]{decals}
{Dey}, A., {Schlegel}, D.~J., {Lang}, D., {et~al.} 2019, \aj, 157, 168, \dodoi{10.3847/1538-3881/ab089d}

\bibitem[{{Dressler} \& {Shectman}(1988)}]{ds}
{Dressler}, A., \& {Shectman}, S.~A. 1988, \aj, 95, 985, \dodoi{10.1086/114694}

\bibitem[{{Fall} \& {Efstathiou}(1980)}]{Fall1980}
{Fall}, S.~M., \& {Efstathiou}, G. 1980, \mnras, 193, 189, \dodoi{10.1093/mnras/193.2.189}

\bibitem[{{Ferrari} {et~al.}(2006){Ferrari}, {Hunstead}, {Feretti}, {Maurogordato}, \& {Schindler}}]{Ferrari06}
{Ferrari}, C., {Hunstead}, R.~W., {Feretti}, L., {Maurogordato}, S., \& {Schindler}, S. 2006, \aap, 457, 21, \dodoi{10.1051/0004-6361:20065117}

\bibitem[{{Fiacconi} {et~al.}(2015){Fiacconi}, {Feldmann}, \& {Mayer}}]{Fiacconi2015}
{Fiacconi}, D., {Feldmann}, R., \& {Mayer}, L. 2015, \mnras, 446, 1957, \dodoi{10.1093/mnras/stu2228}

\bibitem[{{Flaugher} {et~al.}(2015){Flaugher}, {Diehl}, {Honscheid}, {Abbott}, {Alvarez}, {Angstadt}, {Annis}, {Antonik}, {Ballester}, {Beaufore}, {Bernstein}, {Bernstein}, {Bigelow}, {Bonati}, {Boprie}, {Brooks}, {Buckley-Geer}, {Campa}, {Cardiel-Sas}, {Castander}, {Castilla}, {Cease}, {Cela-Ruiz}, {Chappa}, {Chi}, {Cooper}, {da Costa}, {Dede}, {Derylo}, {DePoy}, {de Vicente}, {Doel}, {Drlica-Wagner}, {Eiting}, {Elliott}, {Emes}, {Estrada}, {Fausti Neto}, {Finley}, {Flores}, {Frieman}, {Gerdes}, {Gladders}, {Gregory}, {Gutierrez}, {Hao}, {Holland}, {Holm}, {Huffman}, {Jackson}, {James}, {Jonas}, {Karcher}, {Karliner}, {Kent}, {Kessler}, {Kozlovsky}, {Kron}, {Kubik}, {Kuehn}, {Kuhlmann}, {Kuk}, {Lahav}, {Lathrop}, {Lee}, {Levi}, {Lewis}, {Li}, {Mandrichenko}, {Marshall}, {Martinez}, {Merritt}, {Miquel}, {Mu{\~n}oz}, {Neilsen}, {Nichol}, {Nord}, {Ogando}, {Olsen}, {Palaio}, {Patton}, {Peoples}, {Plazas}, {Rauch}, {Reil}, {Rheault}, {Roe}, {Rogers}, {Roodman}, {Sanchez}, {Scarpine}, {Schindler}, {Schmidt},
  {Schmitt}, {Schubnell}, {Schultz}, {Schurter}, {Scott}, {Serrano}, {Shaw}, {Smith}, {Soares-Santos}, {Stefanik}, {Stuermer}, {Suchyta}, {Sypniewski}, {Tarle}, {Thaler}, {Tighe}, {Tran}, {Tucker}, {Walker}, {Wang}, {Watson}, {Weaverdyck}, {Wester}, {Woods}, {Yanny}, \& {DES Collaboration}}]{decam}
{Flaugher}, B., {Diehl}, H.~T., {Honscheid}, K., {et~al.} 2015, \aj, 150, 150, \dodoi{10.1088/0004-6256/150/5/150}

\bibitem[{{Flin} \& {Krywult}(2006)}]{flin06}
{Flin}, P., \& {Krywult}, J. 2006, \aap, 450, 9, \dodoi{10.1051/0004-6361:20041635}

\bibitem[{{Fujita}(2004)}]{Fujita2004}
{Fujita}, Y. 2004, \pasj, 56, 29, \dodoi{10.1093/pasj/56.1.29}

\bibitem[{{Gill} {et~al.}(2004){Gill}, {Knebe}, \& {Gibson}}]{Gill04}
{Gill}, S. P.~D., {Knebe}, A., \& {Gibson}, B.~K. 2004, \mnras, 351, 399, \dodoi{10.1111/j.1365-2966.2004.07786.x}

\bibitem[{{Gnedin}(2003)}]{Gnedin2003}
{Gnedin}, O.~Y. 2003, \apj, 589, 752, \dodoi{10.1086/374774}

\bibitem[{{G{\'o}mez} {et~al.}(2017){G{\'o}mez}, {Grand}, {Monachesi}, {White}, {Bustamante}, {Marinacci}, {Pakmor}, {Simpson}, {Springel}, \& {Frenk}}]{Gomez2017}
{G{\'o}mez}, F.~A., {Grand}, R. J.~J., {Monachesi}, A., {et~al.} 2017, \mnras, 472, 3722, \dodoi{10.1093/mnras/stx2149}

\bibitem[{{Gordon} {et~al.}(2019){Gordon}, {Pimbblet}, {Kaviraj}, {Owers}, {O'Dea}, {Walmsley}, {Baum}, {Crossett}, {Fraser-McKelvie}, {Lintott}, \& {Pierce}}]{Gordon19}
{Gordon}, Y.~A., {Pimbblet}, K.~A., {Kaviraj}, S., {et~al.} 2019, \apj, 878, 88, \dodoi{10.3847/1538-4357/ab203f}

\bibitem[{{Graham} {et~al.}(2015){Graham}, {Dullo}, \& {Savorgnan}}]{Graham2015}
{Graham}, A.~W., {Dullo}, B.~T., \& {Savorgnan}, G. A.~D. 2015, \apj, 804, 32, \dodoi{10.1088/0004-637X/804/1/32}

\bibitem[{{Gunn} \& {Gott}(1972)}]{Gunn1972}
{Gunn}, J.~E., \& {Gott}, J.~Richard, I. 1972, \apj, 176, 1, \dodoi{10.1086/151605}

\bibitem[{{Hanisch} {et~al.}(2001){Hanisch}, {Farris}, {Greisen}, {Pence}, {Schlesinger}, {Teuben}, {Thompson}, \& {Warnock}}]{FITS2}
{Hanisch}, R.~J., {Farris}, A., {Greisen}, E.~W., {et~al.} 2001, \aap, 376, 359, \dodoi{10.1051/0004-6361:20010923}

\bibitem[{{H{\"a}u{\ss}ler} {et~al.}(2013){H{\"a}u{\ss}ler}, {Bamford}, {Vika}, {Rojas}, {Barden}, {Kelvin}, {Alpaslan}, {Robotham}, {Driver}, {Baldry}, {Brough}, {Hopkins}, {Liske}, {Nichol}, {Popescu}, \& {Tuffs}}]{galapagos}
{H{\"a}u{\ss}ler}, B., {Bamford}, S.~P., {Vika}, M., {et~al.} 2013, \mnras, 430, 330, \dodoi{10.1093/mnras/sts633}

\bibitem[{Hwang \& Lee(2009)}]{hwang&lee09}
Hwang, H.~S., \& Lee, M.~G. 2009, Monthly Notices of the Royal Astronomical Society, 397, 2111, \dodoi{10.1111/j.1365-2966.2009.15100.x}

\bibitem[{{Jackson} {et~al.}(2022){Jackson}, {Kaviraj}, {Martin}, {Devriendt}, {Noakes-Kettel}, {Silk}, {Ogle}, \& {Dubois}}]{Jackson2022}
{Jackson}, R.~A., {Kaviraj}, S., {Martin}, G., {et~al.} 2022, \mnras, 511, 607, \dodoi{10.1093/mnras/stac058}

\bibitem[{{Ji} {et~al.}(2014){Ji}, {Peirani}, \& {Yi}}]{ji14}
{Ji}, I., {Peirani}, S., \& {Yi}, S.~K. 2014, \aap, 566, A97, \dodoi{10.1051/0004-6361/201423530}

\bibitem[{{Johnston} {et~al.}(2008){Johnston}, {Bullock}, {Sharma}, {Font}, {Robertson}, \& {Leitner}}]{Johnston2008}
{Johnston}, K.~V., {Bullock}, J.~S., {Sharma}, S., {et~al.} 2008, \apj, 689, 936, \dodoi{10.1086/592228}

\bibitem[{{Johnston} {et~al.}(1999){Johnston}, {Sigurdsson}, \& {Hernquist}}]{Johnston1999}
{Johnston}, K.~V., {Sigurdsson}, S., \& {Hernquist}, L. 1999, \mnras, 302, 771, \dodoi{10.1046/j.1365-8711.1999.02200.x}

\bibitem[{{Jones} {et~al.}(2000){Jones}, {Smail}, \& {Couch}}]{Jones2000}
{Jones}, L., {Smail}, I., \& {Couch}, W.~J. 2000, \apj, 528, 118, \dodoi{10.1086/308183}

\bibitem[{{Kado-Fong} {et~al.}(2018){Kado-Fong}, {Greene}, {Hendel}, {Price-Whelan}, {Greco}, {Goulding}, {Huang}, {Johnston}, {Komiyama}, {Lee}, {Lust}, {Strauss}, \& {Tanaka}}]{Kado-Fong2018}
{Kado-Fong}, E., {Greene}, J.~E., {Hendel}, D., {et~al.} 2018, \apj, 866, 103, \dodoi{10.3847/1538-4357/aae0f0}

\bibitem[{{Kaviraj} {et~al.}(2015){Kaviraj}, {Devriendt}, {Dubois}, {Slyz}, {Welker}, {Pichon}, {Peirani}, \& {Le Borgne}}]{Kaviraj2015}
{Kaviraj}, S., {Devriendt}, J., {Dubois}, Y., {et~al.} 2015, \mnras, 452, 2845, \dodoi{10.1093/mnras/stv1500}

\bibitem[{Kleiner {et~al.}(2014)Kleiner, Pimbblet, Owers, Jones, \& Stephenson}]{Kleiner2014}
Kleiner, D., Pimbblet, K.~A., Owers, M.~S., Jones, D.~H., \& Stephenson, A.~P. 2014, Monthly Notices of the Royal Astronomical Society, 439, 2755, \dodoi{10.1093/mnras/stu131}

\bibitem[{{Koulouridis} \& {Bartalucci}(2019)}]{Koulouridis2019}
{Koulouridis}, E., \& {Bartalucci}, I. 2019, \aap, 623, L10, \dodoi{10.1051/0004-6361/201935082}

\bibitem[{{Koulouridis} {et~al.}(2024){Koulouridis}, {Gkini}, \& {Drigga}}]{Koulouridis2024}
{Koulouridis}, E., {Gkini}, A., \& {Drigga}, E. 2024, \aap, 684, A111, \dodoi{10.1051/0004-6361/202348212}

\bibitem[{{Koulouridis} {et~al.}(2018){Koulouridis}, {Ricci}, {Giles}, {Adami}, {Ramos-Ceja}, {Pierre}, {Plionis}, {Lidman}, {Georgantopoulos}, {Chiappetti}, {Elyiv}, {Ettori}, {Faccioli}, {Fotopoulou}, {Gastaldello}, {Pacaud}, {Paltani}, \& {Vignali}}]{Koulouridis2018}
{Koulouridis}, E., {Ricci}, M., {Giles}, P., {et~al.} 2018, \aap, 620, A20, \dodoi{10.1051/0004-6361/201832974}

\bibitem[{{Lauer}(1988)}]{Lauer1988}
{Lauer}, T.~R. 1988, \apj, 325, 49, \dodoi{10.1086/165982}

\bibitem[{Lotz {et~al.}(2008)Lotz, Jonsson, Cox, \& Primack}]{Lotz08}
Lotz, J.~M., Jonsson, P., Cox, T.~J., \& Primack, J.~R. 2008, Monthly Notices of the Royal Astronomical Society, 391, 1137, \dodoi{10.1111/j.1365-2966.2008.14004.x}

\bibitem[{Lourenço {et~al.}(2020)Lourenço, Lopes, Laganá, Nascimento, Machado, Moura, Jaffé, Ribeiro, Vulcani, Moretti, \& Riguccini}]{lourenco20}
Lourenço, A. C.~C., Lopes, P. A.~A., Laganá, T.~F., {et~al.} 2020, Monthly Notices of the Royal Astronomical Society, 498, 835, \dodoi{10.1093/mnras/staa2464}

\bibitem[{{Macario} {et~al.}(2011){Macario}, {Markevitch}, {Giacintucci}, {Brunetti}, {Venturi}, \& {Murray}}]{macario11}
{Macario}, G., {Markevitch}, M., {Giacintucci}, S., {et~al.} 2011, \apj, 728, 82, \dodoi{10.1088/0004-637X/728/2/82}

\bibitem[{{Makino} \& {Hut}(1997)}]{makino97}
{Makino}, J., \& {Hut}, P. 1997, \apj, 481, 83, \dodoi{10.1086/304013}

\bibitem[{{Malin} \& {Carter}(1983)}]{Malin1983}
{Malin}, D.~F., \& {Carter}, D. 1983, \apj, 274, 534, \dodoi{10.1086/161467}

\bibitem[{{Martin} {et~al.}(2017){Martin}, {Kaviraj}, {Devriendt}, {Dubois}, {Laigle}, \& {Pichon}}]{Martin2017}
{Martin}, G., {Kaviraj}, S., {Devriendt}, J.~E.~G., {et~al.} 2017, \mnras, 472, L50, \dodoi{10.1093/mnrasl/slx136}

\bibitem[{{Martin} {et~al.}(2018){Martin}, {Kaviraj}, {Devriendt}, {Dubois}, \& {Pichon}}]{Martin2018}
{Martin}, G., {Kaviraj}, S., {Devriendt}, J.~E.~G., {Dubois}, Y., \& {Pichon}, C. 2018, \mnras, 1855, \dodoi{10.1093/mnras/sty1936}

\bibitem[{{Martin} {et~al.}(2021){Martin}, {Jackson}, {Kaviraj}, {Choi}, {Devriendt}, {Dubois}, {Kimm}, {Kraljic}, {Peirani}, {Pichon}, {Volonteri}, \& {Yi}}]{Martin2021}
{Martin}, G., {Jackson}, R.~A., {Kaviraj}, S., {et~al.} 2021, \mnras, 500, 4937, \dodoi{10.1093/mnras/staa3443}

\bibitem[{{Martin} {et~al.}(2022){Martin}, {Bazkiaei}, {Spavone}, {Iodice}, {Mihos}, {Montes}, {Benavides}, {Brough}, {Carlin}, {Collins}, {Duc}, {G{\'o}mez}, {Galaz}, {Hern{\'a}ndez-Toledo}, {Jackson}, {Kaviraj}, {Knapen}, {Mart{\'\i}nez-Lombilla}, {McGee}, {O'Ryan}, {Prole}, {Rich}, {Rom{\'a}n}, {Shah}, {Starkenburg}, {Watkins}, {Zaritsky}, {Pichon}, {Armus}, {Bianconi}, {Buitrago}, {Bus{\'a}}, {Davis}, {Demarco}, {Desmons}, {Garc{\'\i}a}, {Graham}, {Holwerda}, {Hon}, {Khalid}, {Klehammer}, {Klutse}, {Lazar}, {Nair}, {Noakes-Kettel}, {Rutkowski}, {Saha}, {Sahu}, {Sola}, {V{\'a}zquez-Mata}, {Vera-Casanova}, \& {Yoon}}]{Martin2022}
{Martin}, G., {Bazkiaei}, A.~E., {Spavone}, M., {et~al.} 2022, \mnras, 513, 1459, \dodoi{10.1093/mnras/stac1003}

\bibitem[{{Mart{\'\i}nez-Delgado} {et~al.}(2009){Mart{\'\i}nez-Delgado}, {Pohlen}, {Gabany}, {Majewski}, {Pe{\~n}arrubia}, \& {Palma}}]{Martinez-Delgado2009}
{Mart{\'\i}nez-Delgado}, D., {Pohlen}, M., {Gabany}, R.~J., {et~al.} 2009, \apj, 692, 955, \dodoi{10.1088/0004-637X/692/2/955}

\bibitem[{McIntosh {et~al.}(2008)McIntosh, Guo, Hertzberg, Katz, Mo, Van Den~Bosch, \& Yang}]{McIntosh2008}
McIntosh, D.~H., Guo, Y., Hertzberg, J., {et~al.} 2008, Monthly Notices of the Royal Astronomical Society, 388, 1537, \dodoi{10.1111/j.1365-2966.2008.13531.x}

\bibitem[{{Mihos}(2003)}]{mihos03}
{Mihos}, C. 2003, arXiv e-prints, astro.
\newblock \doarXiv{astro-ph/0305512}

\bibitem[{{Mihos} \& {Hernquist}(1996)}]{Mihos1996}
{Mihos}, J.~C., \& {Hernquist}, L. 1996, \apj, 464, 641, \dodoi{10.1086/177353}

\bibitem[{{Moore} {et~al.}(1999){Moore}, {Ghigna}, {Governato}, {Lake}, {Quinn}, {Stadel}, \& {Tozzi}}]{Moore99}
{Moore}, B., {Ghigna}, S., {Governato}, F., {et~al.} 1999, \apjl, 524, L19, \dodoi{10.1086/312287}

\bibitem[{{Moore} {et~al.}(1998){Moore}, {Governato}, {Quinn}, {Stadel}, \& {Lake}}]{Moore98}
{Moore}, B., {Governato}, F., {Quinn}, T., {Stadel}, J., \& {Lake}, G. 1998, \apjl, 499, L5, \dodoi{10.1086/311333}

\bibitem[{Moore {et~al.}(1996)Moore, Katz, Lake, Dressler, \& Oemler}]{Moore96}
Moore, B., Katz, N., Lake, G., Dressler, A., \& Oemler, A. 1996, Nature, 379, 613, \dodoi{10.1038/379613a0}

\bibitem[{{Moretti} {et~al.}(2017){Moretti}, {Gullieuszik}, {Poggianti}, {Paccagnella}, {Couch}, {Vulcani}, {Bettoni}, {Fritz}, {Cava}, {Fasano}, {D'Onofrio}, \& {Omizzolo}}]{moretti17}
{Moretti}, A., {Gullieuszik}, M., {Poggianti}, B., {et~al.} 2017, \aap, 599, A81, \dodoi{10.1051/0004-6361/201630030}

\bibitem[{{Naab} {et~al.}(2014){Naab}, {Oser}, {Emsellem}, {Cappellari}, {Krajnovi{\'c}}, {McDermid}, {Alatalo}, {Bayet}, {Blitz}, {Bois}, {Bournaud}, {Bureau}, {Crocker}, {Davies}, {Davis}, {de Zeeuw}, {Duc}, {Hirschmann}, {Johansson}, {Khochfar}, {Kuntschner}, {Morganti}, {Oosterloo}, {Sarzi}, {Scott}, {Serra}, {van de Ven}, {Weijmans}, \& {Young}}]{Naab2014}
{Naab}, T., {Oser}, L., {Emsellem}, E., {et~al.} 2014, \mnras, 444, 3357, \dodoi{10.1093/mnras/stt1919}

\bibitem[{Oh {et~al.}(2018)Oh, Kim, Lee, Sheen, Kim, Ree, Ho, Kyeong, Sung, Park, \& Yi}]{oh18}
Oh, S., Kim, K., Lee, J.~H., {et~al.} 2018, The Astrophysical Journal Supplement Series, 237, 14, \dodoi{10.3847/1538-4365/aacd47}

\bibitem[{{Oke}(1974)}]{Oke74}
{Oke}, J.~B. 1974, \apjs, 27, 21, \dodoi{10.1086/190287}

\bibitem[{{Oke} \& {Gunn}(1983)}]{Oke83}
{Oke}, J.~B., \& {Gunn}, J.~E. 1983, \apj, 266, 713, \dodoi{10.1086/160817}

\bibitem[{{Olave-Rojas} {et~al.}(2018){Olave-Rojas}, {Cerulo}, {Demarco}, {Jaff{\'e}}, {Mercurio}, {Rosati}, {Balestra}, \& {Nonino}}]{Olave-Rojas2018}
{Olave-Rojas}, D., {Cerulo}, P., {Demarco}, R., {et~al.} 2018, \mnras, 479, 2328, \dodoi{10.1093/mnras/sty1669}

\bibitem[{{Olave-Rojas} {et~al.}(2023){Olave-Rojas}, {Cerulo}, {Araya-Araya}, \& {Olave-Rojas}}]{Olave-Rojas2023}
{Olave-Rojas}, D.~E., {Cerulo}, P., {Araya-Araya}, P., \& {Olave-Rojas}, D.~A. 2023, \mnras, 519, 4171, \dodoi{10.1093/mnras/stac3762}

\bibitem[{{Pasquali} {et~al.}(2019){Pasquali}, {Smith}, {Gallazzi}, {De Lucia}, {Zibetti}, {Hirschmann}, \& {Yi}}]{pasquali19}
{Pasquali}, A., {Smith}, R., {Gallazzi}, A., {et~al.} 2019, \mnras, 484, 1702, \dodoi{10.1093/mnras/sty3530}

\bibitem[{{Peng} {et~al.}(2002){Peng}, {Ho}, {Impey}, \& {Rix}}]{galfit}
{Peng}, C.~Y., {Ho}, L.~C., {Impey}, C.~D., \& {Rix}, H.-W. 2002, \aj, 124, 266, \dodoi{10.1086/340952}

\bibitem[{{Pfleiderer}(1963)}]{Pfleiderer1963}
{Pfleiderer}, J. 1963, \zap, 58, 12

\bibitem[{Piraino-Cerda {et~al.}(2023)Piraino-Cerda, Jaffé, Lourenço, Crossett, Salinas, Kim, Sheen, Kelkar, Pallero, \& Bravo-Alfaro}]{piraino23}
Piraino-Cerda, F., Jaffé, Y.~L., Lourenço, A.~C., {et~al.} 2023, Monthly Notices of the Royal Astronomical Society, 528, 919, \dodoi{10.1093/mnras/stad3957}

\bibitem[{{Planck Collaboration} {et~al.}(2020){Planck Collaboration}, {Aghanim}, {Akrami}, {Ashdown}, {Aumont}, {Baccigalupi}, {Ballardini}, {Banday}, {Barreiro}, {Bartolo}, {Basak}, {Battye}, {Benabed}, {Bernard}, {Bersanelli}, {Bielewicz}, {Bock}, {Bond}, {Borrill}, {Bouchet}, {Boulanger}, {Bucher}, {Burigana}, {Butler}, {Calabrese}, {Cardoso}, {Carron}, {Challinor}, {Chiang}, {Chluba}, {Colombo}, {Combet}, {Contreras}, {Crill}, {Cuttaia}, {de Bernardis}, {de Zotti}, {Delabrouille}, {Delouis}, {Di Valentino}, {Diego}, {Dor{\'e}}, {Douspis}, {Ducout}, {Dupac}, {Dusini}, {Efstathiou}, {Elsner}, {En{\ss}lin}, {Eriksen}, {Fantaye}, {Farhang}, {Fergusson}, {Fernandez-Cobos}, {Finelli}, {Forastieri}, {Frailis}, {Fraisse}, {Franceschi}, {Frolov}, {Galeotta}, {Galli}, {Ganga}, {G{\'e}nova-Santos}, {Gerbino}, {Ghosh}, {Gonz{\'a}lez-Nuevo}, {G{\'o}rski}, {Gratton}, {Gruppuso}, {Gudmundsson}, {Hamann}, {Handley}, {Hansen}, {Herranz}, {Hildebrandt}, {Hivon}, {Huang}, {Jaffe}, {Jones}, {Karakci}, {Keih{\"a}nen},
  {Keskitalo}, {Kiiveri}, {Kim}, {Kisner}, {Knox}, {Krachmalnicoff}, {Kunz}, {Kurki-Suonio}, {Lagache}, {Lamarre}, {Lasenby}, {Lattanzi}, {Lawrence}, {Le Jeune}, {Lemos}, {Lesgourgues}, {Levrier}, {Lewis}, {Liguori}, {Lilje}, {Lilley}, {Lindholm}, {L{\'o}pez-Caniego}, {Lubin}, {Ma}, {Mac{\'\i}as-P{\'e}rez}, {Maggio}, {Maino}, {Mandolesi}, {Mangilli}, {Marcos-Caballero}, {Maris}, {Martin}, {Martinelli}, {Mart{\'\i}nez-Gonz{\'a}lez}, {Matarrese}, {Mauri}, {McEwen}, {Meinhold}, {Melchiorri}, {Mennella}, {Migliaccio}, {Millea}, {Mitra}, {Miville-Desch{\^e}nes}, {Molinari}, {Montier}, {Morgante}, {Moss}, {Natoli}, {N{\o}rgaard-Nielsen}, {Pagano}, {Paoletti}, {Partridge}, {Patanchon}, {Peiris}, {Perrotta}, {Pettorino}, {Piacentini}, {Polastri}, {Polenta}, {Puget}, {Rachen}, {Reinecke}, {Remazeilles}, {Renzi}, {Rocha}, {Rosset}, {Roudier}, {Rubi{\~n}o-Mart{\'\i}n}, {Ruiz-Granados}, {Salvati}, {Sandri}, {Savelainen}, {Scott}, {Shellard}, {Sirignano}, {Sirri}, {Spencer}, {Sunyaev}, {Suur-Uski}, {Tauber}, {Tavagnacco},
  {Tenti}, {Toffolatti}, {Tomasi}, {Trombetti}, {Valenziano}, {Valiviita}, {Van Tent}, {Vibert}, {Vielva}, {Villa}, {Vittorio}, {Wandelt}, {Wehus}, {White}, {White}, {Zacchei}, \& {Zonca}}]{planck18}
{Planck Collaboration}, {Aghanim}, N., {Akrami}, Y., {et~al.} 2020, \aap, 641, A6, \dodoi{10.1051/0004-6361/201833910}

\bibitem[{{Pontzen} {et~al.}(2017){Pontzen}, {Tremmel}, {Roth}, {Peiris}, {Saintonge}, {Volonteri}, {Quinn}, \& {Governato}}]{Pontzen2017}
{Pontzen}, A., {Tremmel}, M., {Roth}, N., {et~al.} 2017, \mnras, 465, 547, \dodoi{10.1093/mnras/stw2627}

\bibitem[{{Quinn}(1984)}]{Quinn1984}
{Quinn}, P.~J. 1984, \apj, 279, 596, \dodoi{10.1086/161924}

\bibitem[{{Quintana} {et~al.}(2020){Quintana}, {Proust}, {D{\"u}nner}, {Carrasco}, \& {Reisenegger}}]{Quintana2020}
{Quintana}, H., {Proust}, D., {D{\"u}nner}, R., {Carrasco}, E.~R., \& {Reisenegger}, A. 2020, \aap, 638, A27, \dodoi{10.1051/0004-6361/202037726}

\bibitem[{{Raychaudhury}(1990)}]{Raychaudhury90}
{Raychaudhury}, S. 1990, in Bulletin of the American Astronomical Society, Vol.~22, 1331

\bibitem[{{Rhee} {et~al.}(2017){Rhee}, {Smith}, {Choi}, {Yi}, {Jaff{\'e}}, {Candlish}, \& {S{\'a}nchez-J{\'a}nssen}}]{Rhee2017}
{Rhee}, J., {Smith}, R., {Choi}, H., {et~al.} 2017, \apj, 843, 128, \dodoi{10.3847/1538-4357/aa6d6c}

\bibitem[{{Roettiger} {et~al.}(1998){Roettiger}, {Stone}, \& {Mushotzky}}]{roettiger98}
{Roettiger}, K., {Stone}, J.~M., \& {Mushotzky}, R.~F. 1998, \apj, 493, 62, \dodoi{10.1086/305102}

\bibitem[{{Schawinski} {et~al.}(2014){Schawinski}, {Urry}, {Simmons}, {Fortson}, {Kaviraj}, {Keel}, {Lintott}, {Masters}, {Nichol}, {Sarzi}, {Skibba}, {Treister}, {Willett}, {Wong}, \& {Yi}}]{Schawinski2014}
{Schawinski}, K., {Urry}, C.~M., {Simmons}, B.~D., {et~al.} 2014, \mnras, 440, 889, \dodoi{10.1093/mnras/stu327}

\bibitem[{{Schlafly} \& {Finkbeiner}(2011)}]{mwext}
{Schlafly}, E.~F., \& {Finkbeiner}, D.~P. 2011, \apj, 737, 103, \dodoi{10.1088/0004-637X/737/2/103}

\bibitem[{{Schweizer}(1982)}]{Schweizer1982}
{Schweizer}, F. 1982, \apj, 252, 455, \dodoi{10.1086/159573}

\bibitem[{Scrucca {et~al.}(2016)Scrucca, Fop, Murphy, \& Raftery}]{mclust}
Scrucca, L., Fop, M., Murphy, T.~B., \& Raftery, A.~E. 2016, The {R} Journal, 8, 289.
\newblock \url{https://doi.org/10.32614/RJ-2016-021}

\bibitem[{{Shaw} \& Swaters(2015)}]{noao}
{Shaw}, R.~A., \& Swaters, R. 2015, {NOAO Data Handbook (Version 2.2; Tucson: National Optical Astronomy Observatory)}

\bibitem[{Sheen {et~al.}(2012)Sheen, Yi, Ree, \& Lee}]{Sheen_2012}
Sheen, Y.-K., Yi, S.~K., Ree, C.~H., \& Lee, J. 2012, The Astrophysical Journal Supplement Series, 202, 8, \dodoi{10.1088/0067-0049/202/1/8}

\bibitem[{Sobral {et~al.}(2015)Sobral, Stroe, Dawson, Wittman, Jee, Röttgering, van Weeren, \& Brüggen}]{Sobral15}
Sobral, D., Stroe, A., Dawson, W.~A., {et~al.} 2015, Monthly Notices of the Royal Astronomical Society, 450, 630, \dodoi{10.1093/mnras/stv521}

\bibitem[{{Sola} {et~al.}(2022){Sola}, {Duc}, {Richards}, {Paiement}, {Urbano}, {Klehammer}, {B{\'\i}lek}, {Cuillandre}, {Gwyn}, \& {McConnachie}}]{Sola2022}
{Sola}, E., {Duc}, P.-A., {Richards}, F., {et~al.} 2022, arXiv e-prints, arXiv:2203.03973.
\newblock \doarXiv{2203.03973}

\bibitem[{{Spavone} {et~al.}(2020){Spavone}, {Iodice}, {van de Ven}, {Falc{\'o}n-Barroso}, {Raj}, {Hilker}, {Peletier}, {Capaccioli}, {Mieske}, {Venhola}, {Napolitano}, {Cantiello}, {Paolillo}, \& {Schipani}}]{Spavone2020}
{Spavone}, M., {Iodice}, E., {van de Ven}, G., {et~al.} 2020, \aap, 639, A14, \dodoi{10.1051/0004-6361/202038015}

\bibitem[{{Stroe} \& {Sobral}(2021)}]{Stroe21}
{Stroe}, A., \& {Sobral}, D. 2021, \apj, 912, 55, \dodoi{10.3847/1538-4357/abe7f8}

\bibitem[{Stroe {et~al.}(2015)Stroe, Sobral, Dawson, Jee, Hoekstra, Wittman, van Weeren, Brüggen, \& Röttgering}]{Stroe15}
Stroe, A., Sobral, D., Dawson, W., {et~al.} 2015, Monthly Notices of the Royal Astronomical Society, 450, 646, \dodoi{10.1093/mnras/stu2519}

\bibitem[{Tanaka {et~al.}(2023)Tanaka, Koike, Naito, Shibata, Usuda-Sato, Yamaoka, Ando, Ito, Kobayashi, Kofuji, Kuwata, Nakano, Shimakawa, ichi Tadaki, Takebayashi, Tsuchiya, Umemoto, \& Bottrell}]{tanaka2023}
Tanaka, M., Koike, M., Naito, S., {et~al.} 2023, GALAXY CRUISE: Deep Insights into Interacting Galaxies in the Local Universe.
\newblock \doarXiv{2309.14710}

\bibitem[{{Toomre}(1977)}]{Toomre1977}
{Toomre}, A. 1977, in Evolution of Galaxies and Stellar Populations, ed. B.~M. {Tinsley} \& R.~B.~G. {Larson}, D.~Campbell, 401

\bibitem[{{Toomre} \& {Toomre}(1972)}]{Toomre1972}
{Toomre}, A., \& {Toomre}, J. 1972, \apj, 178, 623, \dodoi{10.1086/151823}

\bibitem[{{Trujillo} {et~al.}(2021){Trujillo}, {D'Onofrio}, {Zaritsky}, {Madrigal-Aguado}, {Chamba}, {Golini}, {Akhlaghi}, {Sharbaf}, {Infante-Sainz}, {Rom{\'a}n}, {Morales-Socorro}, {Sand}, \& {Martin}}]{Trujillo2022}
{Trujillo}, I., {D'Onofrio}, M., {Zaritsky}, D., {et~al.} 2021, \aap, 654, A40, \dodoi{10.1051/0004-6361/202141603}

\bibitem[{{Valenzuela} \& {Remus}(2022)}]{Valenzuela2022}
{Valenzuela}, L.~M., \& {Remus}, R.-S. 2022, arXiv e-prints, arXiv:2208.08443.
\newblock \doarXiv{2208.08443}

\bibitem[{{van Dokkum}(2005)}]{vandokkum05}
{van Dokkum}, P.~G. 2005, \aj, 130, 2647, \dodoi{10.1086/497593}

\bibitem[{{van Dokkum} {et~al.}(1999){van Dokkum}, {Franx}, {Fabricant}, {Kelson}, \& {Illingworth}}]{vandokkum99}
{van Dokkum}, P.~G., {Franx}, M., {Fabricant}, D., {Kelson}, D.~D., \& {Illingworth}, G.~D. 1999, \apjl, 520, L95, \dodoi{10.1086/312154}

\bibitem[{{Vera-Casanova} {et~al.}(2021){Vera-Casanova}, {G{\'o}mez}, {Monachesi}, {Gargiulo}, {Pallero}, {Grand}, {Marinacci}, {Pakmor}, {Simpson}, {Frenk}, \& {Morales}}]{VeraCasanova2021}
{Vera-Casanova}, A., {G{\'o}mez}, F.~A., {Monachesi}, A., {et~al.} 2021, arXiv e-prints, arXiv:2105.06467.
\newblock \doarXiv{2105.06467}

\bibitem[{{Vijayaraghavan} \& {Ricker}(2013)}]{Vijayaraghavan2013}
{Vijayaraghavan}, R., \& {Ricker}, P.~M. 2013, \mnras, 435, 2713, \dodoi{10.1093/mnras/stt1485}

\bibitem[{{Voges} {et~al.}(1999){Voges}, {Aschenbach}, {Boller}, {Br{\"a}uninger}, {Briel}, {Burkert}, {Dennerl}, {Englhauser}, {Gruber}, {Haberl}, {Hartner}, {Hasinger}, {K{\"u}rster}, {Pfeffermann}, {Pietsch}, {Predehl}, {Rosso}, {Schmitt}, {Tr{\"u}mper}, \& {Zimmermann}}]{rosat}
{Voges}, W., {Aschenbach}, B., {Boller}, T., {et~al.} 1999, \aap, 349, 389, \dodoi{10.48550/arXiv.astro-ph/9909315}

\bibitem[{{Wells} {et~al.}(1981){Wells}, {Greisen}, \& {Harten}}]{FITS1}
{Wells}, D.~C., {Greisen}, E.~W., \& {Harten}, R.~H. 1981, \aaps, 44, 363

\bibitem[{{White} \& {Rees}(1978)}]{white&rees78}
{White}, S.~D.~M., \& {Rees}, M.~J. 1978, \mnras, 183, 341, \dodoi{10.1093/mnras/183.3.341}

\bibitem[{Wittman {et~al.}(2024)Wittman, Imani, Olden, \& Golovich}]{Wittman24}
Wittman, D., Imani, D., Olden, R.~H., \& Golovich, N. 2024, The Astronomical Journal, 167, 49, \dodoi{10.3847/1538-3881/ad110c}

\bibitem[{{Yi} {et~al.}(2013){Yi}, {Lee}, {Jung}, {Ji}, \& {Sheen}}]{yi13}
{Yi}, S.~K., {Lee}, J., {Jung}, I., {Ji}, I., \& {Sheen}, Y.~K. 2013, \aap, 554, A122, \dodoi{10.1051/0004-6361/201321369}

\bibitem[{{Yuan} \& {Han}(2020)}]{yuan20}
{Yuan}, Z.~S., \& {Han}, J.~L. 2020, \mnras, 497, 5485, \dodoi{10.1093/mnras/staa2363}

\bibitem[{Yuan {et~al.}(2022)Yuan, Han, \& Wen}]{yuan22}
Yuan, Z.~S., Han, J.~L., \& Wen, Z.~L. 2022, Monthly Notices of the Royal Astronomical Society, 513, 3013, \dodoi{10.1093/mnras/stac1037}

\bibitem[{Zabludoff \& Zaritsky(1995)}]{Zabludoff1995}
Zabludoff, A.~I., \& Zaritsky, D. 1995, The Astrophysical Journal, 447, L21, \dodoi{10.1086/309552}

\bibitem[{Łokas(2023)}]{Lokas2023}
Łokas, E.~L. 2023, \aap, 678, A147, \dodoi{10.1051/0004-6361/202347735}

\end{thebibliography}
\bibliographystyle{aasjournal}



\end{document}